\documentclass[11pt,a4paper]{article}
\pdfoutput=1
\usepackage{jheppub}

\usepackage{amsmath,epsfig}
\usepackage{amssymb,amsfonts}
\usepackage{latexsym}
\usepackage[latin1]{inputenc}
\usepackage{slashed}
\usepackage{empheq}
\numberwithin{equation}{section}
\usepackage{dsfont}
\usepackage{cancel}
\usepackage{overpic}
\usepackage{subcaption}
\usepackage{subeqnarray}
\usepackage{xcolor}

\usepackage{graphicx}
\usepackage{longtable}
\usepackage{color}

\newcommand{\exclude}[1]{}
\def\<{\langle}
\def\>{\rangle}

\def\a#1{\alpha_{#1}}

\def\beq{\begin{equation}}
\def\eeq{\end{equation}}
\def\be{\begin{equation}}
\def\ee{\end{equation}}
\def\bea{\begin{eqnarray}}
\def\eea{\end{eqnarray}}
\def\bal{\begin{align}}
\def\eal{\end{align}}

\def\2b2[#1,#2][#3,#4]{\left( \begin{array}{cc} #1 & #2 \\ #3 & #4 \end{array}
\right)}
\def\3b3[#1,#2,#3][#4,#5,#6][#7,#8,#9]{\left( \begin{array}{ccc} #1 & #2 #3 \\
#4 & #5 & #6\\#7&#8&#9\end{array} \right)}

\newcommand\fverb{\setbox\pippobox=\hbox\bgroup\verb}
\newcommand\fverbdo{\egroup\medskip\noindent%
                        \fbox{\unhbox\pippobox}\ }
\newcommand\fverbit{\egroup\item[\fbox{\unhbox\pippobox}]}

\newcommand{\bear}{\begin{eqnarray}}

\newcommand{\eear}{\end{eqnarray}}

\newcommand{\bsea}{\begin{subeqnarray}}
\newcommand{\esea}{\end{subeqnarray}}
\newbox\pippobox

\def\f{\varphi}

\def\6{\partial}
\def\a{\alpha}

\def\e{\epsilon}

\def\sp{\;\;\;,\;\;\;}

\def\sq
\def\a{\alpha}

\def\e{\epsilon}

\title{Holographic RG flows on curved manifolds and the $F$-theorem}

\author[\natural]{J.~K.~Ghosh,}
\author[\flat, \natural]{E.~Kiritsis,}
\author[\natural]{F.~Nitti,}
\author[\natural]{L.~T.~Witkowski}

\affiliation[\natural]{\href{http://www.apc.univ-paris7.fr}{APC, AstroParticule et Cosmologie}, Universit\'e Paris Diderot, CNRS/IN2P3, CEA/IRFU, \\
Observatoire de Paris, Sorbonne Paris Cit\'e,\\
 10, rue Alice Domon et L\'eonie Duquet, 75205 Paris
Cedex 13, France}
\affiliation[\flat]{\href{http://hep.physics.uoc.gr}{Crete Center for Theoretical Physics},
Department of Physics,\\
University of Crete, 71003 Heraklion, Greece}

\preprint{CCTP-2018-11\\
\hphantom{AAAAAAAAAAAAAAAAAAAAAAAAAAAAAAAAAAAAAAAAllll} ITCP-IPP 2018/8}

\abstract{We study $F$-functions in the context of field theories on
  $S^3$ using gauge-gravity duality, with the radius of $S^3$ playing
  the role of RG scale. We show that the on-shell action, evaluated
  over a set of holographic RG flow solutions, can be used to define
  good $F$-functions, which decrease monotonically along the RG flow
  from the UV to the IR for a wide range of examples. If the operator
  perturbing the UV CFT has dimension $\Delta > 3/2$ these
  $F$-functions correspond to an appropriately renormalized free
  energy. If instead  the perturbing operator has dimension $\Delta <
  3/2$ it is the quantum effective potential, i.e.~the Legendre
  transform of the free energy, which gives rise to good
  $F$-functions. We check that these observations hold beyond
  holography for the case of a free fermion on $S^3$ ($\Delta=2$) and
  the free boson on $S^3$ ($\Delta=1$),  resolving a long-standing problem regarding the non-monotonicity of the free energy for the free massive scalar. 
  We also show that for a particular choice of entangling surface, we can define good $F$-functions from an entanglement entropy, which coincide with certain $F$-functions obtained from the on-shell action.}

\begin{document}
\maketitle

\section{Introduction and summary}

A fundamental property of Quantum Field Theory is that the number of degrees of freedom decreases under renormalization group (RG) flow. The quantitative description of this phenomenon is the subject of the so called `$c$-theorems'. The first component of any $c$-theorem is to identify a $c$-quantity that measures the number of degrees of freedom of the QFTs at the UV and IR fixed points of the RG flow. The second ingredient is a $c$-function with the property that it interpolates monotonically between the UV and IR values of the $c$-quantity along the flow.

For field theories in an even number of space-time dimensions suitable $c$-quantities can be identified with coefficients of the Weyl anomaly. In $d=2$ Zamolodchikov proposed a suitable $c$-function, which at the fixed points reduces to the Weyl anomaly coefficient $c$ \cite{Zamolodchikov:1986gt}. In $d=4$ it is the anomaly coefficient $a$ which plays the role of the $c$-quantity \cite{Cardy:1988cwa}. A proof of monotonicity under RG flow was presented in \cite{1107.3987}, therefore establishing the $a$-theorem in $d=4$. In odd space-time dimensions the Weyl anomaly is absent, and hence a different approach to the $c$-theorem is required.

Progress in this direction was made by relating the $c$-theorem to entropic considerations. In \cite{0405111} it was shown that the $c$-theorem in $d=2$ can be derived from strong subadditivity of an entanglement entropy. In \cite{1006.1263,1011.5819} it was observed that the $c$-quantity in any dimension (even and odd) can be defined as the universal contribution to an entanglement entropy across a suitably chosen surface.\footnote{For a recent review of entanglement entropy in holography and its application to RG flows and $c$-theorems see \cite{1801.10352}.}

The case of the $c$-theorem $d=3$, also referred to as the $F$-theorem, received particular attention. It was suggested in \cite{1012.3210, 1103.1181,1105.4598} that the role of the $c$-quantity can be played by (the finite part of) the free energy of the theory on the 3-sphere, $F= - \ln |Z_{S^3}|$. For CFTs the free energy on $S^3$ coincides with the entanglement entropy across a spherical surface \cite{1102.0440}, hence providing a link to the entropic formulation of \cite{1006.1263,1011.5819}. The free energy on $S^3$ was also proposed as a possible $c$-function (henceforth $F$-function) in $d=3$, with the radius of the sphere as the parameter along the RG flow \cite{1012.3210, 1103.1181,1105.4598}. A generalisation of the sphere partition function beyond $d=3$ was suggested as a definition for a $c$-function in general $d$ \cite{1409.1937} (see also \cite{1410.5973}). Evidence for the $F$-theorem in both supersymmetric and non-supersymmetric theories can be found in \cite{1103.1181,1105.4598,1011.5487,1012.3210,1105.2817,1302.7310}. 

However, there exist problems with the identification of the free energy with the $F$-function. In \cite{1105.4598} one curious observation was that for the simple case of a free massive scalar on $S^3$ the free energy failed to interpolate monotonically between UV and IR. Only by performing an ad-hoc subtraction of a suitably chosen function could it be made monotonic, suggesting that the free energy on $S^3$ fails to be a universally valid $F$-function.

A more successful definition of the $F$-function employs an appropriately defined entanglement entropy. In \cite{1202.2070} Liu and Mezei constructed a quantity termed `Renormalized Entanglement Entropy' (REE), whose functional dependence on the size of the entangling surface is interpreted as describing the RG flow of the entanglement entropy with distance scale. At the fixed points of a flow the REE reduces to the central charge of the corresponding CFT. For \emph{Poincar\'e-invariant} field theories in $d=3$ space-time dimensions the REE was proven to decrease monotonically from UV to IR in \cite{1202.5650}, suggesting that the REE can play the role of the $F$-function in $d=3$.
A different approach for isolating the finite contribution to the entanglement entropy based on mutual information was proposed in \cite{1506.06195}. 

The study of $c$-theorems and in particular the $F$-theorem remains an
active field with many directions for further study. For example, the
question of stationarity of the $F$-function at fixed points is currently unresolved with evidence against stationarity found in \cite{1207.3360}. For a recent work on the construction of $c$-functions in defect CFTs see \cite{1810.06995}.

In this work we will address open questions regarding the
$F$-theorem, both in its
formulation in terms of the free energy on $S^3$, and  in its entropic
formulation. Although the (UV-finite part of the) free energy on a
sphere  and the REE coincide at fixed points, the formulation of the
$F$-theorem in terms of the free energy on $S^3$
\cite{1012.3210, 1103.1181,1105.4598} seems problematic. As stated above, the free
energy on $S^3$ fails to be monotonic even for the case of a free massive scalar
\cite{1105.4598}, thus calling into question its identification as a
universally valid $F$-function. The same conclusion was reached  in
the context of  holographic RG flows in  \cite{Taylor}.

Another open question concerns the entropic formulation. In this case an $F$-function can be defined as the REE
 across a spherical surface \cite{1202.2070}. While this was proven to
 be monotonic under RG flow for Poincar\'e invariant theories, the
 status of the REE as an $F$-function beyond Poincar\'e-invariant
 theories is unclear. For example, in \cite{1504.00913} the behaviour
 of the REE under renormalization group flow was examined for the
 theory of a conformally coupled scalar on dS$_3$. In this case the
 REE fails to exhibit monotonicity and it is hence not a good $F$-function on dS$_3$.

Finally, it is not clear if and how
the two formulations of the $F$-theorem are related. In particular,
while the two definitions in terms of the entanglement entropy and
free energy coincide in the UV and IR \cite{1102.0440}, it is not
known to what extent this relation should persist along the RG
flow. While the $F$-theorem in three dimensions is by now well
established in terms of the entanglement entropy, an alternative
formulation directly in terms of the sphere partition function may still be
desirable as this quantity may be easier to compute in practice for
non-conformal field theories and it may evade some of the
difficulties outlined in \cite{1506.06195} in the computation of the
entanglement entropy in the presence of a  regulator.

Motivated by the discussion above, in this work we address the
following questions:
\begin{enumerate}
\item Can a true $F$-function be constructed from the free energy on $S^3$?
\item How can a good $F$-function be constructed from an entanglement entropy for theories on dS$_3$?
\item How are the formulations of the $F$-function in terms of the free energy on $S^3$ and in terms of an entanglement entropy related? In particular, under what circumstances do they coincide along the whole flow rather than only at the UV and IR end points?
\end{enumerate}
Here we will use holography to address these questions. In particular,
we will propose candidate $F$-functions constructed from the free
energy. We will test their monotonicity both in simple holographic
examples and in free theories. Also, we will elucidate how these
$F$-functions  are related to entanglement entropy on de Sitter space.
Throughout this work we will make use of  recent advances in the
understanding of holographic RG flows for field theories on curved
manifolds \cite{curvedRG}.

The rest of this  introduction summarises in a self-contained
  manner our setup, our results and concluding remarks.

\subsection{Setup and summary of results}
\label{sec:setupandsumm}
Our objects of study are QFTs on $S^d$ which can be defined as CFTs perturbed in the UV by a relevant scalar operator $\mathcal{O}$ of dimension $\Delta$. In the end, we will be mainly interested in $d=3$, but we will work in general $d$ whenever possible. The corresponding action for such theories is schematically given by
\begin{align}
\label{eq:QFT1} S_{\textrm{QFT}}= S_{\textrm{CFT}} + j \int d^d x \sqrt{\gamma^{(0)}} \, \mathcal{O} \, ,
\end{align}
where $j$ is a (constant) source for the operator $\mathcal{O}$ and
$\gamma^{(0)}_{\mu \nu}$ is the metric on $S^d$. To be specific, we
denote the scalar curvature of $S^d$ by $R$.

In this work we will study the response of theories of this kind to a change in $R$, which we will interpret as a response to the change of RG scale. We will refer to this as \emph{curvature-RG flow}. Following the curvature-RG flow from UV to IR corresponds to varying $R$ from $R \rightarrow \infty$ (UV) to $R \rightarrow 0$ (IR).

\vspace{0.3cm}

\noindent \textbf{Holographic setup.}
We will use holography to study such theories. In this context the
relevant physical system is $(d+1)$-dimensional Einstein-dilaton gravity with a potential. The action (for Lorentzian signature) is given by
\begin{align}
\label{eq:introS} S = M^{d-1} \int d^{d+1} x \sqrt{|g|} \left( R^{(g)} - \frac{1}{2} \partial_a \f \partial^a \f - V(\f) \right) + S_{\textrm{GHY}} \, .
\end{align}
Most of the time, we will be interested in the corresponding Euclidean theory, whose action is $S_E= -S$.
We consider solutions with the following ansatz for $\f$ and the metric $g_{ab}$:
\begin{align}
\f = \f(u) \, , \qquad ds^2 = g_{ab} dx^a dx^b = d u^2 + e^{2 A(u)} \zeta_{\mu \nu} dx^{\mu} dx^{\nu} \, .
\end{align}
Here $\zeta_{\mu \nu}$ is a metric on $S^d$ (or, in the Lorentzian
case,  dS$_d$) with curvature $R$. We also restrict attention to
purely negative potentials, i.e.~$V < 0$, that possess at least one
maximum and one minimum. This setup was analysed in detail in
\cite{curvedRG}. The main features are reviewed in section 2 and briefly summarised below.

In holography, extrema of the potential are associated with CFTs. We identify the theory associated with a maximum with the UV CFT, while the theory at a minimum will be identified with the IR CFT:
\begin{align}
\label{eq:UVdef} \textrm{CFT}_{\textrm{UV}} \quad & \Leftrightarrow \quad \f(u) = \f_{\textrm{UV}} \equiv \f_{\textrm{max}}  \, , \\
\label{eq:IRdef} \textrm{CFT}_{\textrm{IR}} \quad & \Leftrightarrow \quad \f(u) = \f_{\textrm{IR}} \equiv \f_{\textrm{min}} \, .
\end{align}

More precisely, a maximum of $V$ corresponds to a UV CFT perturbed by
a relevant scalar operator, thus realising equation \eqref{eq:QFT1} holographically. In geometric terms the UV is identified with an AdS$_{d+1}$ boundary of the bulk space-time, which we choose to be located at $u \rightarrow -\infty$. We can then read off the source $j$ and the metric $\gamma^{(0)}_{\mu \nu}$ in \eqref{eq:QFT1} from the near-boundary expansion for $\f$ and the bulk metric:
\begin{align}
ds^2 &\underset{u \rightarrow - \infty}{=} du^2 +  e^{-2u / \ell}  \gamma_{\mu \nu}^{(0)} dx^{\mu} dx^{\nu} + \ldots \, , \\
\f(u) &\underset{u \rightarrow - \infty}{=} \f_{\textrm{max}} +  \f_- \, \ell^{\Delta_-} e^{\Delta_- u / \ell} +  \f_+ \, \ell^{\Delta_+} e^{\Delta_+ u / \ell} + \ldots \, ,
\end{align}
with $\ell$ the length scale associated with AdS$_{d+1}$ and
\begin{align}
\Delta_{\pm} = \frac{1}{2} \left(d \pm \sqrt{d^2 + 4 \ell^2 V''(\f_{\textrm{UV}})} \right) \, ,
\end{align}
In `standard quantisation' $\f_-$ corresponds to the source $j$ in \eqref{eq:QFT1} while $\f_+$ is proportional to the vev $\langle \mathcal{O} \rangle$. In `alternative quantisation' this identification is reversed, with $\f_+$ the source and $\f_-$ related to the vev. Also, in standard quantisation the dimension $\Delta$ of the perturbing operator is given by $\Delta_+$, while in alternative quantisation it corresponds to $\Delta_-$. In this work we will consider both cases.

It is convenient to introduce the dimensionless quantity $\mathcal{R}$, which will use as the parameter along curvature-RG flow. It is defined as
\begin{align}
\mathcal{R} \equiv R |\f_-|^{-2 / \Delta_-} \, .
\end{align}
In standard quantisation this corresponds to the curvature in units of the source while in alternative quantisation it becomes the curvature in units of the vev $\langle \mathcal{O} \rangle$.

For given values of $R$ and $\f_-$ one can then obtain solutions $A_{R, \f_-} (u), \f_{R,\f_-}(u)$, which have the interpretation of holographic RG flows, with $u$ a coordinate along the flow.\footnote{Note that this notion of holographic RG flow is different from the concept of curvature-RG flow introduced before.} These solutions have been described in detail in \cite{curvedRG} and will be an important ingredient in the construction of good $F$-functions. In particular, the $F$-functions proposed in this work will be defined as suitable functionals over a set of solutions $A_{R, \f_-} (u), \f_{R,\f_-}(u)$.

\begin{figure}[t]
\centering
\begin{overpic}
[width=0.65\textwidth]{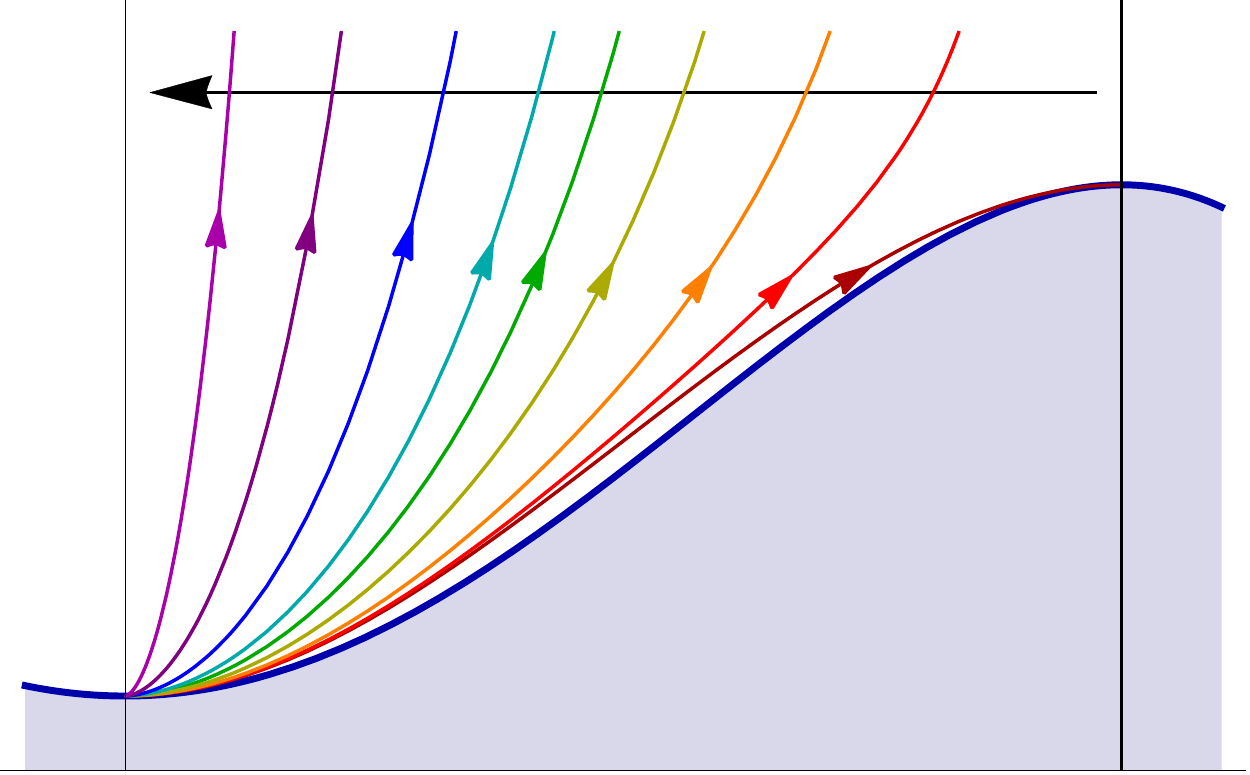}
\put(1,53.5){UV}
\put(92,53.5){IR}
\put(58,56){$\mathcal{R}$}
\put(99,3){$\f$}
\put(-2,62){$W(\f)$}
\put(61,25){$\sqrt{-\tfrac{4(d-1) V(\f)}{d}}$}
\end{overpic}
\caption{Family of holographic RG flow solutions $A_{R, \f_-}(u), \f_{R, \f_-}(u)$ with different values of dimensionless curvature $\mathcal{R}$. Here the different solutions are represented by plotting the corresponding function $W(\f) = - 2(d-1) \tfrac{dA}{du} \big(u(\f)\big)$. A flow with $\mathcal{R}=0$ has its end point $\f_0$ at the minimum of the potential. When $\mathcal{R}$ is increased the end point moves closer to the maximum of the potential with $\f_0 \rightarrow \f_{\textrm{max}}$ for $\mathcal{R} \rightarrow \infty$. On this family of holographic RG solutions parameterised by $\mathcal{R}$ we then define functionals $\mathcal{F}(\mathcal{R}) = \mathcal{F}[A_{R,\f_-}, \f_{R, \f_-}]$. For these functionals the notion of curvature-RG flow is then defined as the response to a change in $\mathcal{R}$ with the UV identified with $\mathcal{R} \rightarrow \infty$ and the IR given by $\mathcal{R} \rightarrow 0$. }
\label{fig:flowsintro}
\end{figure}
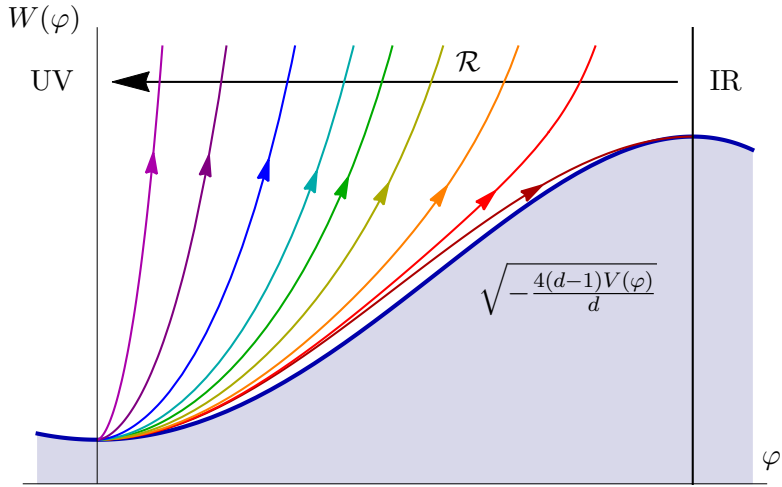

\vspace{0.3cm}

\noindent \textbf{The $F$-theorem and $F$-functions as functionals.}
In the following we specialise to $d=3$. As it has been known for some
time \cite{1012.3210, 1103.1181,1105.4598}, a good $F$-quantity is the finite part of the free
energy of the corresponding CFT, $F = - \ln
|Z_{S^3}|$. In holography, this can be calculated from the action
\eqref{eq:introS}, evaluated on the solutions \eqref{eq:UVdef} or
\eqref{eq:IRdef}, supplemented by appropriate counterterms (see
e.g.~\cite{Taylor}). In our conventions one finds
\begin{align}
F_{\textrm{UV}} &= 8 \pi^2 (M \ell_{\textrm{UV}})^2 \, , \quad \textrm{with} \quad \ell_{\textrm{UV}}^2 = - \frac{6}{V(\f_{\textrm{UV}})} \, \\
F_{\textrm{IR}} &= 8 \pi^2 (M \ell_{\textrm{IR}})^2 \, , \quad \ \textrm{with} \quad \ell_{\textrm{IR}}^2 \, = - \frac{6}{V(\f_{\textrm{IR}})} \, \\
\textrm{satisfying} \quad F_{\textrm{UV}} &\geq F_{\textrm{IR}} \, , \quad \textrm{in virtue of} \quad |V(\f_{\textrm{UV}})| \leq |V(\f_{\textrm{IR}})| \, .
\end{align}
For a  CFT, the finite part of the free energy is unambiguous and
independent of the subtraction scheme. Away from fixed points however,
these properties are lost, and the finite part of the free energy away
from fixed points becomes scheme-dependent.

The main goal of this paper is then to construct functions $\mathcal{F}(\mathcal{R})$ that interpolate monotonically between $F_{\textrm{UV}}$ and $F_{\textrm{IR}}$, i.e.~functions satisfying
\begin{align} \label{F-quant-intro}
\mathcal{F}(\mathcal{R}) \underset{\mathcal{R} \rightarrow \infty}{=} F_{\textrm{UV}} \, , \qquad \mathcal{F}(\mathcal{R}) \underset{\mathcal{R} \rightarrow 0}{=} F_{\textrm{IR}} \, , \qquad \mathcal{R} \frac{d}{d\mathcal{R}} \mathcal{F}(\mathcal{R}) \geq 0 \, .
\end{align}
We will do so by defining $\mathcal{F}(\mathcal{R})$ as a functional over a set of holographic RG flow solutions $A_{R, \f_-}(u), \f_{R, \f_-}(u)$,
\begin{align}
\mathcal{F}(\mathcal{R}) \equiv \mathcal{F}[A_{R, \f_-} (u), \f_{R,\f_-}(u)] \, .
\end{align}
Evaluating $\mathcal{F}(\mathcal{R})$  is therefore equivalent to evaluating the functional over a set of holographic RG flow solutions $A_{R, \f_-} (u), \f_{R,\f_-}(u)$ with $\mathcal{R}$ varying from 0 to $\infty$. An illustration of this is shown in figure \ref{fig:flowsintro}.

The discussion so far was equally valid for a boundary field theory defined on $S^3$ or dS$_3$. For example, the solutions $A_{R, \f_-}(u), \f_{R, \f_-}(u)$ are identical for these two cases. However, when defining $F$-functions we will  distinguish between theories on manifolds with Euclidean and Lorentzian signature.

\vspace{0.3cm}

\noindent \textbf{Candidate $F$-functions from the on-shell action.}
For the case of a boundary theory on $S^3$ we will construct $F$-functions from the on-shell action, that is the Euclidean form of the action \eqref{eq:introS} evaluated on solutions $A_{R, \f_-}(u), \f_{R, \f_-}(u)$. Note that in `standard quantisation' the Euclidean on-shell action corresponds to the free energy and, correspondingly, our $F$-functions can be understood as being constructed from the free energy on $S^3$ in this case.

The on-shell action suffers from both UV and IR divergences which need to be removed before it can act as a good $F$-function. In particular, introducing $\Lambda$ as an energy cutoff (in units of $\f_-$) the UV-divergent terms take the form
\begin{align}
S_{\textrm{on-shell}, E} \underset{\Lambda \rightarrow \infty}{\sim}
\mathcal{R}^{-3/2} \big[ \Lambda^3 + \mathcal{O} \big(\Lambda^{3-2
  \Delta_-} \big) \big] + \mathcal{R}^{-1/2} \big[\Lambda +
\mathcal{O} \big(\Lambda^{1-2 \Delta_-} \big) \big] \, . \label{UVdiv-intro}
\end{align}
Similarly, IR divergences take the form
\begin{align}
S_{\textrm{on-shell}, E} \underset{\mathcal{R} \rightarrow 0}{\sim} a
\mathcal{R}^{-3/2} + b \mathcal{R}^{-1/2} \, , \label{IRdiv-intro}
\end{align}
with $a$, $b$ numerical coefficients. 

In order to construct a function of ${\cal R}$ that matches the finite
$F$-quantities as in equation (\ref{F-quant-intro}) in the limits ${\cal R} \to 0, +\infty$, one has to eliminate both the  UV and IR divergences.  One key observation (see
sec.~\ref{sec:Ffunc}) is that both UV- and IR-divergences only come
with specific powers of $\mathcal{R}$. In particular, by removing
terms $\sim \mathcal{R}^{-3/2}$ and $\sim \mathcal{R}^{-1/2}$ all
divergent pieces can be eliminated. This is in contrast to the
structure of the power series in $\Lambda$ which, depending on  the
dimension $\Delta_-$, may display many subleading non-universal
divergent terms of the type $\Lambda^{3-2n\Delta_-}$, as
indicated in equation (\ref{UVdiv-intro}). This is one reason why it
is much more convenient to use ${\cal R}$, rather than $\Lambda$, to measure the response to a
change of scale.

One way of removing  divergences is to act on the on-shell action with appropriate differential operators, as has been done for the entanglement entropy across a spherical surface in \cite{1202.2070}. In particular, the operators
\begin{align}
\label{eq:Ddefintro} \mathcal{D}_{3/2} \equiv \frac{2}{3} \mathcal{R} \frac{d}{d \mathcal{R}} + 1 \, , \qquad \mathcal{D}_{1/2} \equiv 2 \mathcal{R} \frac{d}{d \mathcal{R}} + 1 \, ,
\end{align}
satisfy
\begin{align}
\mathcal{D}_{3/2} \mathcal{R}^{-3/2} = 0  \, , \qquad \mathcal{D}_{1/2} \mathcal{R}^{-1/2} =0 \, ,
\end{align}
which thus remove the divergent pieces while leaving the finite terms
intact. 

Acting   with the operators (\ref{eq:Ddefintro}) on  the regulated on-shell action $S_{\textrm{on-shell},E}
(\Lambda, \mathcal{R})$ yields our first candidate $F$-function:
\begin{align}
\label{eq:F1first} \mathcal{F}_1 (\mathcal{R}) & \equiv  \mathcal{D}_{1/2} \, \mathcal{D}_{3/2} \, S_{\textrm{on-shell},E} (\Lambda, \mathcal{R}) \, .
\end{align}

Another method of removing divergences is to employ holographic
renormalization and add appropriate counterterms to the action. As we
will see, in
$d=3$  holographic renormalization introduces two arbitrary constants,
which play the role of finite counterterms, and which we will denote
by $B_{ct}$ and $C_{ct}$. Picking different values for $B_{ct}$,
$C_{ct}$ corresponds to a choice of renormalization scheme. While
holographic renormalization successfully removes UV divergences for
any choice of these constants, the renormalized on-shell action will generically still contain IR divergences, which schematically are given by:
\begin{align}
S_{\textrm{on-shell}, E}^{\textrm{ren}} \underset{\mathcal{R} \rightarrow 0}{\sim} \mathcal{R}^{-3/2} (c_0 - C_{ct}) +  \mathcal{R}^{-1/2} (b_0 - B_{ct}) \, ,
\end{align}
where $c_0$ and $b_0$ are  numerical coefficients. These can again be
eliminated by acting with the differential operators
$\mathcal{D}_{3/2}$ and $\mathcal{D}_{1/2}$ on the renormalized
on-shell action, but this can be shown to reproduce the $F$-function
$\mathcal{F}_1(R)$ introduced in \eqref{eq:F1first}. However, the
IR-divergent terms can also be removed by an appropriate
renormalization scheme, i.e.~by choosing $C_{ct}= C_{ct,0}=c_0$ and
$B_{ct}= B_{ct,0}=b_0$.\footnote{The  constants $b_0$ and $c_0$ are
  well-defined concrete numbers in any given theory. In appendix
  \ref{app:renormalizationscheme} we relate this choice of finite counterterms  to a set of well-defined renormalization conditions on stress-tensor correlation functions.}

As we have two options (differentiation vs.~counterterm) for removing
each one of the terms $\sim \mathcal{R}^{-3/2}$ and $\sim
\mathcal{R}^{-1/2}$, we can propose four different $F$-functions from the renormalized on-shell action $S_{\textrm{on-shell},E}^{\textrm{ren}} (\mathcal{R} | B_{ct}, C_{ct})$:
\begin{align}
\label{eq:F1defintro} \mathcal{F}_1 (\mathcal{R}) & \equiv \mathcal{D}_{1/2} \, \mathcal{D}_{3/2} \, S_{\textrm{on-shell},E}^{\textrm{ren}} (\mathcal{R} | B_{ct}, C_{ct}) = \mathcal{D}_{1/2} \, \mathcal{D}_{3/2} \, S_{\textrm{on-shell},E} (\Lambda, \mathcal{R}) \, , \\
\label{eq:F2defintro} \mathcal{F}_2 (\mathcal{R}) & \equiv \mathcal{D}_{1/2} \, \hphantom{\mathcal{D}_{3/2}} \, S_{\textrm{on-shell},E}^{\textrm{ren}} (\mathcal{R} | B_{ct}, C_{ct,0}) \, , \\
\label{eq:F3defintro} \mathcal{F}_3 (\mathcal{R}) & \equiv \hphantom{\mathcal{D}_{1/2}} \, \mathcal{D}_{3/2} \, S_{\textrm{on-shell},E}^{\textrm{ren}} (\mathcal{R} | B_{ct,0}, C_{ct}) \, , \\
\label{eq:F4defintro} \mathcal{F}_4 (\mathcal{R}) & \equiv \hphantom{\mathcal{D}_{1/2} \, \mathcal{D}_{3/2}} \, S_{\textrm{on-shell},E}^{\textrm{ren}} (\mathcal{R} | B_{ct,0}, C_{ct,0}) \, .
\end{align}

\begin{figure}[t]
\centering
\begin{overpic}
[width=0.65\textwidth]{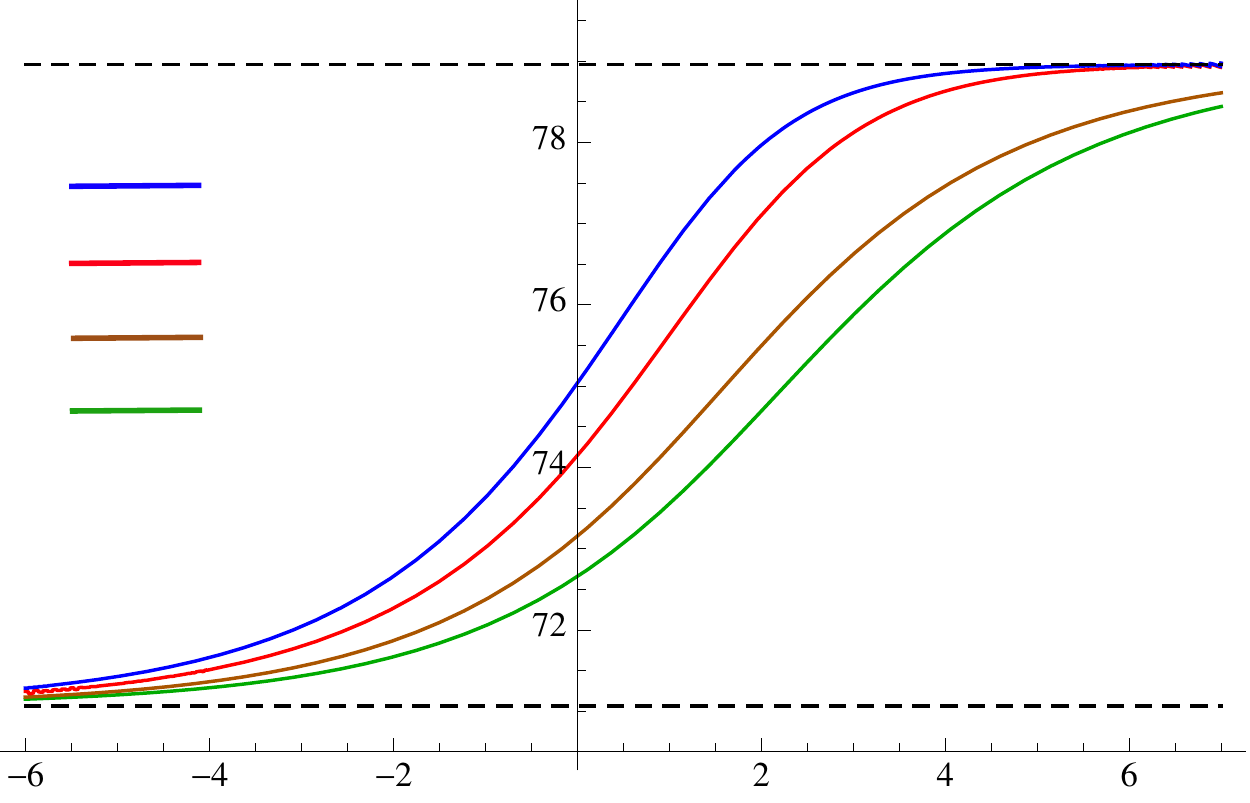}
\put(102,3){$\log(\mathcal{R})$}
\put(49,9.75){$8 \pi^2 (M \ell_{\textrm{IR}})^2$}
\put(49,61){$8 \pi^2 (M \ell_{\textrm{UV}})^2$}
\put(19,48.1){$\mathcal{F}_1$}
\put(19,42.1){$\mathcal{F}_2$}
\put(19,36.1){$\mathcal{F}_3$}
\put(19,30.1){$\mathcal{F}_4$}
\end{overpic}
\caption{$F$-functions
  $\mathcal{F}_{1,2,3,4}$ defined in \protect\eqref{eq:F1defintro}--\protect\eqref{eq:F4defintro} vs.~$\log(\mathcal{R})$ for a holographic model with dilaton potential \protect\eqref{eq:potrepeat} and $\Delta_-=1.2$.}
\label{fig:F1to4Delta1p2}
\end{figure}
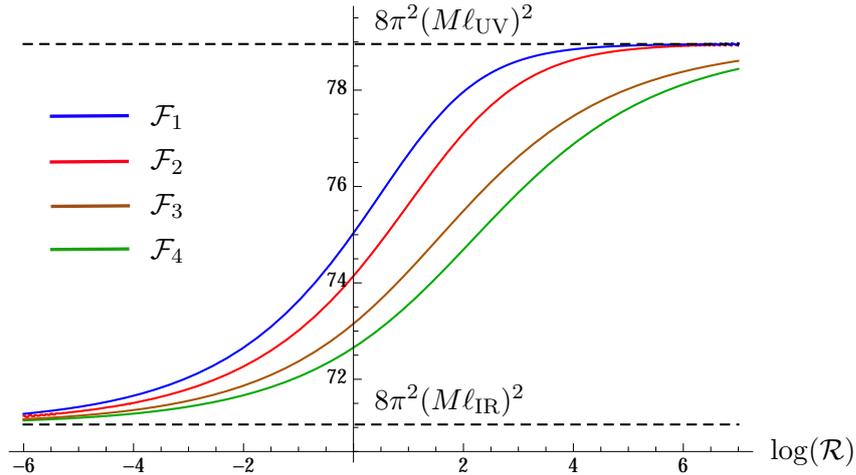

These are good candidate $F$-functions, as in the UV ($\mathcal{R} \rightarrow \infty$) and the IR ($\mathcal{R} \rightarrow 0$) they reduce to the corresponding $F$-quantity (see sec.~\ref{sec:Ffunc}):
\begin{align}
\mathcal{F}_i (R) & \underset{\mathcal{R} \rightarrow \infty}{\longrightarrow} F_{\textrm{UV}} = 8 \pi^2 (M \ell_{\textrm{UV}})^2 \, , \qquad \quad
\mathcal{F}_i (R)  \underset{\mathcal{R} \rightarrow 0}{\longrightarrow} F_{\textrm{IR}} = 8 \pi^2 (M \ell_{\textrm{IR}})^2 \, .
\end{align}
Proving monotonicity of these candidate $F$-functions is difficult,
as it requires a better understanding of the $R$-dependence of the bulk RG flow
solutions, and we will not attempt this in this
work. Rather, here we test our proposal in some concrete examples,
leaving the issue of a proof for future work. In particular,  we
confirmed monotonicity numerically for a wide range of potentials
$V(\f)$ which display two extrema and  allow for curvature RG-flow between them. For example, in figure \ref{fig:F1to4Delta1p2} we plot $\mathcal{F}_{1,2,3,4}(\mathcal{R})$ vs.~$\log(\mathcal{R})$ for the quadratic-quartic potential \eqref{eq:potrepeat} with $\Delta_-=1.2$. All four functions $\mathcal{F}_{1,2,3,4}(\mathcal{R})$ interpolate monotonically between $F_{\textrm{UV}}$ and $F_{\textrm{IR}}$.

Note that in `standard quantisation' the Euclidean on-shell action corresponds to the free energy. As a result, the definitions \eqref{eq:F1defintro}--\eqref{eq:F4defintro} meet one key objective of this paper: they provide a definition for $F$-functions obtained from the free energy.

\vspace{0.3 cm}

\noindent \textbf{The free energy vs.~the quantum effective potential as an $F$-function:}
The $F$-functions given in
\eqref{eq:F1defintro}--\eqref{eq:F4defintro} are well-defined for any
value $\tfrac{1}{2} < \Delta < 3$ of the dimension of the perturbing operator
$\mathcal{O}$, i.e.~for any relevant operator above the unitarity
bound.  However, as we will explain presently, their interpretation in terms of field-theoretic quantities is fundamentally different in the regimes $\Delta < \tfrac{3}{2}$ and $\Delta > \tfrac{3}{2} $.

When $\Delta > \tfrac{3}{2}$ we identify $\Delta= \Delta_+$, which is
equivalent to working in standard quantisation. In this case the
on-shell action corresponds to the \emph{free energy} of the dual QFT. The dimensionless parameter $\mathcal{R}$ along the curvature-RG flow is the curvature of $S^3$ in units of the source of the perturbing operator $\mathcal{O}$.

However, when $\Delta < \tfrac{3}{2}$ we need to identify $\Delta=
\Delta_-$, i.e.~we have to work in alternative quantisation. The
on-shell action is in this case identified with the \emph{quantum
  effective potential}, i.e.~the Legendre transform of the free energy
(see sec.~\ref{sec:altquant} for details). In the dual field theory,  the parameter $\mathcal{R}$ along the curvature-RG flow now corresponds to the curvature in units of the vev $\langle \mathcal{O} \rangle$.

With these interpretations in terms of field-theoretic quantities we
can then test our candidate $F$-functions beyond the framework of
holography. As a first step in this direction we consider the free
massive fermion and free massive boson on $S^3$. Note that, for the
free massive boson, the renormalized free energy on $S^3$ was found not to be
monotonic and hence not a good $F$-function as observed in
\cite{1105.4598}. Similarly, the universal contribution to the
entanglement entropy across a spherical surface in dS$_3$ also fails
to be monotonic, as shown in \cite{1504.00913}.

 In contrast we find that our proposals, when applied to the free
 massive boson do constitute good monotonic $F$-functions. Indeed, our holographic results suggest that the definition of the
 $F$-function in terms of field theory quantities depends on the dimension $\Delta$ of the perturbing operator in the UV and has to be adjusted accordingly. For the cases of the free massive boson and fermion this is given by:
\begin{align}
\textrm{Free massive fermion:} \quad & \Delta = [\psi^\dagger \psi] = 2 > \frac{3}{2} \, , \\
\textrm{Free massive boson:} \quad & \Delta = [\phi^2] \ \ = 1 < \frac{3}{2} \, .
\end{align}
Our holographic results then imply that for the free massive fermion the free energy on $S^3$ should be chosen to define a good $F$-function. However, it is the quantum effective potential which should be used to define an $F$-function for the massive boson, with $\mathcal{R} \sim R / \langle \phi^2 \rangle^2$ the dimensionless parameter along the curvature-RG flow.

In sec.~\ref{sec:freefields} we calculate the corresponding $F$-functions \eqref{eq:F1defintro}--\eqref{eq:F4defintro} explicitly and confirm that our proposals do indeed give rise to monotonic $F$-functions, both for the free massive fermion (sec.~\ref{sec:freefermion}) and boson (sec.~\ref{sec:freeboson}). Thus our $F$-function proposals successfully overcome a long-standing puzzle regarding $F$-functions for the free massive boson on $S^3$.\footnote{We also checked that the quantum effective potential does not give rise to a good $F$-function for the free massive fermion, leading to non-monotonicity. We did not include the calculation in this work, as it can be easily reproduced by modifying the analysis in sec.~\ref{sec:freeboson}.}

\vspace{0.3 cm}

\noindent \textbf{$F$-functions from the entanglement entropy:}
Another possibility for defining $F$-functions is to consider the
entanglement entropy across a spherical entangling surface. While this
has been proven to lead to a good $F$-function for theories in flat
space-time \cite{1202.2070,1202.5650}, here we consider a theory
defined on dS$_3$. In this case, the entanglement entropy across a
spherical surface which cuts in half the $t=0$ spatial
hypersurface can be shown to be related to the free energy on $S^3$, which in turn will give a relation between $F$-functions constructed from these quantities. This setup has been considered in field theory in \cite{1504.00913}, and here we study it holographically by calculating the corresponding Ryu-Takayanagi \cite{RT} entanglement entropy.

As we show in sec.~\ref{sec:ent1}, the entanglement entropy $S_{\textrm{EE}}$ is given by a functional of a holographic RG flow solution $A_{R, \f_-}(u), \f_{R, \f_-}(u)$:
\begin{align}
S_{\textrm{EE}}(\mathcal{R}) = S_{\textrm{EE}} \left[ A_{R, \f_-}(u), \f_{R, \f_-}(u) \right] \, .
\end{align}
Like the on-shell action, this quantity suffers from both UV and IR divergences,
which in this case depend on curvature as $ \sim
\mathcal{R}^{-1/2}$. As in the case of the on-shell action, the
divergences can be removed by acting with an appropriate differential
operator, or by renormalization with suitable renormalization
scheme. For one,  all divergent terms can be removed by acting with
$\mathcal{D}_{1/2}$ defined in \eqref{eq:Ddefintro}. If we choose
renormalization instead, this introduces one scheme-dependent
parameter $\tilde{B}_{ct}$. To define a good $F$-function this has to
be chosen such that IR divergence is cancelled. Overall, this gives
two candidate $F$-functions that can be defined from the de Sitter
entanglement entropy.
 
Notice that our holographic analysis implies that, in terms of the field theory
language, depending on the value of $\Delta$ the differential operator $\mathcal{D}_{1/2}$
has to be defined differently: 
for $\Delta >
 3/2$ the derivative with respect to $R$ is to be evaluated at a fixed value of the source of the operator perturbing the QFT in the UV, while for $\Delta <
 3/2$ it is the vev of this  operator that is to be kept constant. 
The non-monotonicity of the $F$-function constructed from the de Sitter entanglement entropy of a free massive scalar observed in \cite{1504.00913} can then be understood as originating from the use of an inappropriate differential operator (derivative at constant source rather than constant vev for a theory with $\Delta=1$).

Interestingly, for the setup considered here, i.e.~entanglement across a spherical surface in dS$_3$, the entanglement entropy also permits the interpretation as a thermal entropy, satisfying a corresponding thermodynamic identity (see sec.~\ref{sec:ent2} and \cite{1102.0440,1504.00913}). As  shown in appendix \ref{app:thermo0}, a consequence of this thermodynamic identity is that the renormalized entanglement entropy $S_{\textrm{EE}}(\mathcal{R} | \tilde{B}_{ct})$ is related to the on-shell action on $S^3$ as\footnote{The operator $\mathcal{D}_{3/2}$ eliminates any dependence on $C_{ct}$ as it only appears in the combination $C_{ct} \, \mathcal{R}^{-3/2}$.}
\begin{align}
\label{eq:SEESonshellrelationintro} S_{\textrm{EE}}^{\textrm{ren}}(\mathcal{R} | \tilde{B}_{ct}) = - \mathcal{D}_{3/2} \, S_{\textrm{on-shell}, E}^{\textrm{ren}} (\mathcal{R} | \tfrac{3}{2} \tilde{B}_{ct}, C_{ct}) \, ,
\end{align}
where $\mathcal{D}_{3/2}$ was defined in \eqref{eq:Ddefintro}. Using
this relation, we can show that the two $F$-functions that can be
defined from $S_{\textrm{EE}}$ coincide with two of the
$F$-functions derived from the on-shell action on $S^3$, namely 
 $\mathcal{F}_1(\mathcal{R})$ and $\mathcal{F}_3(\mathcal{R})$
in  equations  \eqref{eq:F1defintro} and \eqref{eq:F3defintro}.

Note that here the $F$-functions obtained from the entanglement
entropy and the on-shell action coincide over the whole course of the
RG flow and not only at the end points.
 In standard quantisation (applicable for $\Delta > 3/2$) the Euclidean on-shell action corresponds to the free energy. Our results then imply that the entanglement entropy across a spherical surface in dS$_3$ and the free energy on $S^3$ lead to equivalent definitions for $F$-functions. Making this connection precise was another key objective of this work. In alternative quantisation (applicable for $\Delta < 3/2$) our results show that the $F$-function constructed from the entanglement entropy is related to an $F$-function obtained from the quantum effective potential instead of the free energy.

\subsection{Conclusions, open questions and outlook}
We can now look back at the questions we asked at the beginning of
this introduction. In this paper we provide evidence, from both
holography and free theories, that for 3d QFTs it is indeed  possible
to construct monotonic $F$-functions starting from the free energy on $S^3$
and avoid the difficulties encountered in the past. As we show, this
is possible as long as one properly takes into account the following two considerations:
\begin{enumerate}
\item As the sphere path integral is UV divergent one has  to work
with renormalized quantities to define anything meaningful. Away from
the fixed points,  this
introduces a scheme dependence  of the resulting  finite functions. In
a generic scheme, the renormalized sphere free energy is finite but
non-monotonic, as  simple holographic examples have already shown \cite{Taylor}. The
key is to adopt renormalization schemes which, at the same time as the UV
divergences, remove also the IR divergences associated with taking
the volume of the $S^3$ to infinity.\footnote{In contrast, the value of the
  IR-finite part of the sphere free energy of a CFT at the fixed point
  is universal and scheme independent, which explains why the free
  energy on $S^3$ is always a good $F$-{\em quantity} for a CFT.} We have identified four different
schemes which accomplish this: three of them involve acting with
differential operators (in the curvature) on the regulated partition
function; the last one consists of a specific choice of counterterms, and in particular can
be related to specific renormalization conditions on correlators of
the {\em flat space} field theory.

\item Defining the subtraction procedure as outlined above is not
  enough, though: we find the definition of the $F$-function is
  different depending on the dimension $\Delta$ of the relevant operator which
  deforms the CFT away from the UV fixed point. If $\Delta>3/2$, then
  the $F$-functions are given by the (renormalized) free energy. If
  $\Delta <3/2$ on the other hand, they have to be defined with the
  same subtractions, but on its Legendre transform, i.e.~the quantum
  effective potential.\footnote{Note
    that one has to do this in the right order: taking the Legendre
    transform {\em after} the subtraction results in a function that
    is non-monotonic and sometimes not even IR-finite.}  This
  in particular is the reason why the renormalized free energy of a
  free scalar on $S^3$, {\em even in the schemes outlined in point 1},
  does not work as an $F$-function \cite{1105.4598}. It is here that
  holography has played a crucial role, as this would have been very
  hard to guess from purely field theoretical considerations. Instead, the natural quantity to work with on the gravity side is
  the on-shell action, and it is the same quantity which, depending on
  $\Delta$, plays the
  role of either the QFT free energy or its Legendre transform.
\end{enumerate}

Our other initial questions also concerned the entanglement entropy. We 
 find evidence that one can construct a good $F$-function from the de
 Sitter entanglement entropy. Like in the case of the free energy or its Legendre transform, an appropriate subtraction
 procedure has to be used, either by applying  a differential operator
 \`a la Liu and Mezei, or by a counterterm subtraction. We find that for $\Delta >
  3/2$ the corresponding
 $F$-functions coincide with a subset of those constructed from the
 free energy on $S^3$, while for $\Delta <
 3/2$ they correspond to a subset of those obtained from the Legendre transform of the free energy.

There remain many directions for further investigation. An open question concerns the relation of our proposed
$F$-functions to the one obtained from the Renormalized
Entanglement Entropy (REE) of a spherical region in the Minkowski QFT
\cite{1202.2070}. For a CFT at a fixed point, the corresponding  $F$-quantity
coincides with that defined from both the free energy and the de
Sitter entanglement entropy. This follows from the fact that, for a
CFT, one can use a conformal transformation to map the spherical
region in flat space to a similar region in de Sitter \cite{1102.0440}. Away from the
fixed point however this no longer holds, and we show explicitly that
the $F$-functions obtained from the free energy differ from the flat
space REE. It would be interesting to understand whether these are
related in some deeper way. 

In appendix \ref{mono} we also discuss another monotonic quantity in
holography which does not seem to obviously lead to a universal
$F$-function. It is obtained from the unrenormalized free energy density by varying the UV cut-off instead of
the curvature scale. It may be worth investigating whether a good
$F$-function may be obtained in this way. 

Another direction for further work is to gather
additional evidence that the functions
$\mathcal{F}_{1,2,3,4}(\mathcal{R})$ defined in
\eqref{eq:F1defintro}--\eqref{eq:F4defintro} are indeed good
$F$-functions. While we checked this explicitly for a wide range of
theories in holography and for free field theories, further
 tests are desirable.

In the context of holography, here we worked
 with bottom-up four-dimensional models, and it would be interesting
 to check whether our proposal holds in top-down models from string
 theories. For three-dimensional field theories, there are
 many  examples of gravity dual holographic RG flows  in
 gauged $N=8$, $d=4$ supergravity \cite{Ahn:2000aq,Ahn:2000mf,Ahn:2001by,Ahn:2002qga}, and their M-theory
 uplifts\cite{Corrado:2001nv,Bobev:2009ms}. One could also consider
 three-dimensional theories with flavor,  whose RG flows were studied
 e.g.~in \cite{Hohenegger:2009as,Hikida:2009tp,Ammon:2009wc} (in the quenched
 flavor limit)  and  \cite{Bea:2013jxa}. The quenched flavor case
 looks particularly treatable, as one does not need to find the
 curved domain wall solution in the
 full 11-dimensional bulk theory, but only  restrict to the
 contribution to the $F$-functions from the flavor degrees of freedom coming from the D6 branes
 wrapped on AdS$_4$$\times S^7$ using curved slicing of AdS$_4$, and
 use the flavor  mass as the deformation parameter.

Another interesting playground to test our proposal is in the context
of confining theories, which in holography may be realised both in bottom up models
(e.g.~setups with exponentially growing potentials,
like those studied in \cite{0707.1349} or the  AdS
soliton \cite{witten-thermo,Kuperstein:2004yf}) and in top-down 
 constructions like those studied in
\cite{witten-thermo,Csaki:1998qr}, dual to 3-dimensional confining
gauge theories (in this case the scale is provided by the radius of
an internal compact cycle). 

In \cite{curvedRG} it was observed that boundary field theories in
holography may undergo phase transitions when the boundary curvature
is adjusted (see also \cite{1108.6070}). The change of boundary
curvature is equivalent to the notion of curvature-RG flow explored in
this work. An interesting question to explore is whether the $F$-theorem holds across a curvature-induced phase transition. We leave this for future work.

Another open question is to what extent an $F$-theorem exists in higher odd dimensions, i.e.~$d=5, 7, \ldots$. Several proposals for $F$-functions in higher odd dimensions exist \cite{1006.1263, 1011.5819, 1105.4598, 1409.1937} and evidence for an $F$-theorem in $d=5$ can be found in \cite{1207.4359, 1404.1094}. However, so far there is no proof of monotonicity for odd dimensions with $d \geq 5$. The $F$-functions introduced in this work, while constructed for $d=3$, allow for a straightforward generalisation to any odd $d$ and are hence suitable for exploring the $F$-theorem beyond $d=3$. We hope to report on this in the future.

\section{Holographic space-times with curved slicing and RG flows}
\label{sec:ansatzreview}
 Our focus will be on  holographic RG flows of field theories on
 dS$_d$ and $S^d$ space-times, driven by a relevant deformation of a UV CFT.  These are dual to curved domain wall solutions in $d+1$
 dimensions. In this section we review these types of solutions in the
 context of Einstein gravity coupled to a scalar field. This section
 is in most part a review of \cite{curvedRG}, where a systematic study
 of curved domain walls with maximally-symmetric radial slices was
 performed, and the reader is referred to that work for more details.  While we present the setup for general $d$ we will ultimately be interested in odd $d$ and in particular $d=3$.

\subsection{Ansatz and conventions}
\label{sec:ansatz}
The relevant holographic framework is $(d+1)$-dimensional Einstein-dilaton gravity. In the Lorentzian case with signature $(-++\ldots+)$ the corresponding two-derivative action is given by
\begin{align}
\label{eq:action} S[g, \f]= M_p^{d-1} \int du \, d^ dx \, \sqrt{|g|} \left(R^{(g)} - \frac{1}{2} \partial_a \f \partial^a \f - V(\f) \right) + S_{GHY} \, ,
\end{align}
where we also included the Gibbons-Hawking-York term $S_{GHY}$. The Euclidean action can be obtained from \eqref{eq:action} by changing the overall sign, $S_E = - S$, and changing the
metric to positive signature.

To study holographic RG flows for field theories on dS$_d$ or $S^d$ we seek solutions of the following form for the $(d+1)$-dimensional metric and the scalar $\f$. In domain wall coordinates our ansatz is given by
\begin{align}
\label{eq:metric} ds^2 &= du^2 + e^{2 A(u)} \zeta_{\mu \nu} d x^{\mu} d x^{\nu} \, , \qquad \f = \f(u) \, ,
\end{align}
where $A(u)$ is a scale factor that depends on the coordinate $u$ only, and $\zeta_{\mu \nu}$ is a fiducial metric on dS$_d$ or $S^d$. To be specific, we label the radius associated with $\zeta_{\mu \nu}$ by $\alpha$. Note that $\alpha$ is unphysical as we can absorb a redefinition of $\alpha$ by an appropriate constant shift in $A(u)$.
Another useful quantity will be the induced metric $\gamma_{\mu \nu}(u)$ on a slice of fixed $u$, which is given by
\begin{align}
\label{eq:gammadef} \gamma_{\mu \nu}(u) \equiv e^{2 A(u)} \zeta_{\mu \nu} \, .
\end{align}

Throughout this paper, we will use the following shorthand notation. We will denote derivatives with respect to $u$ by a dot while derivatives with respect to $\f$ will be abbreviated to a prime, i.e.:
\begin{align}
\dot{f}(u) \equiv \frac{d f(u)}{du} \, , \qquad g'(\f) \equiv \frac{d g(\f)}{d \f} \, .
\end{align}

Given our ansatz \eqref{eq:metric} we obtain the following equations of motion:
\begin{align}
\label{eq:EOM1} 2(d-1) \ddot{A} + \dot{\f}^2 + \frac{2}{d} e^{-2A} R^{(\zeta)} &=0 \, , \\
\label{eq:EOM2} d(d-1) \dot{A}^2 - \frac{1}{2} \dot{\f}^2 + V - e^{-2A} R^{(\zeta)} &=0 \, , \\
\label{eq:EOM3} \ddot{\f} +d \dot{A} \dot{\f} - V' &= 0 \, ,
\end{align}
with $R^{(\zeta)}= d(d-1) / \alpha^2$ the scalar curvature associated with the metric $\zeta_{\mu \nu}$. Note that these equations remain unchanged when swapping between Lorentzian and Euclidean
signature.

To apply gauge-gravity duality, we will be exclusively interested in bulk solutions to \eqref{eq:EOM1}--\eqref{eq:EOM3} which are asymptotically AdS$_{d+1}$. In particular, we will seek bulk geometries with a boundary for $u \rightarrow - \infty$ at which the bulk space-times asymptotes to AdS$_{d+1}$. This implies that as we approach the boundary the scale factor has to take the form
\begin{align}
\nonumber A(u) & \underset{u \rightarrow - \infty}{\longrightarrow} \left. \ln \left( - \frac{\ell}{\alpha} \sinh \frac{u + c}{\ell} \right) \right|_{u \rightarrow - \infty} \, , \\
\label{eq:expAasymp} & = \ln \frac{\ell}{2 \alpha} - \frac{u + c}{\ell} - e^{2 (u + c) / \ell} + \mathcal{O}(e^{4 u / \ell}) \, ,
\end{align}
which is the scale factor corresponding to AdS$_{d+1}$ space-time in the coordinates of \eqref{eq:metric}. Here $\ell$ is the length scale associated with AdS$_{d+1}$, $\alpha$ is the length scale of the fiducial metric $\zeta_{\mu \nu}$ and $c$ is an integration constant.

It is this constant $c$ which sets the scalar curvature of the boundary manifold supporting the QFT. To see this,  we write the metric near the boundary in a Fefferman-Graham expansion:
\begin{align}
ds^2 = du^2 + \left(e^{-2u/ \ell} \gamma_{\mu \nu}^{(0)} + \ldots \right) d x^{\mu} d x^{\nu} \, .
\end{align}
It is the metric $\gamma_{\mu \nu}^{(0)}$ which describes the boundary space-time supporting the QFT. Given our ansatz, we can identify the boundary metric as
\begin{align}
\label{eq:boundarymetric} \gamma_{\mu \nu}^{(0)} = \left.{\left(e^{2u/\ell}  \, e^{2 A(u)} \, \right)}\right|_{u \rightarrow -\infty} \zeta_{\mu \nu} \, = \frac{\ell^2}{4 \alpha^2} \, e^{-2c/\ell}  \, \zeta_{\mu \nu} \, .
\end{align}
An important quantity in this work will be scalar curvature of the boundary space-time. Here and in the following this will be denoted by $R$. Given the expression for the boundary metric \eqref{eq:boundarymetric}, the asymptotic form of the scale factor \eqref{eq:expAasymp}, the boundary curvature is given by:
\begin{align}
\label{eq:boundaryR} R = \frac{4 \alpha^2}{\ell^2} \, e^{2c/ \ell} \, R^{(\zeta)} = \frac{4 d (d-1)}{\ell^2} \, e^{2c/ \ell} \, .
\end{align}
The boundary curvature is therefore set by the integration constant $c$ and the unphysical parameter $\alpha$ does not appear.

At this stage we introduce one further convention which we will adopt throughout this paper. As the radius $\alpha$ of the fiducial metric $\zeta_{\mu \nu}$ is unphysical, we can choose it freely without affecting any physical results. In the following, we will find it convenient to choose $\alpha$ to be identical to the physical radius associated with the boundary metric $\gamma_{\mu \nu}^{(0)}$, i.e.
\begin{align}
\alpha = \frac{\ell}{2} \, e^{-c/\ell} \, .
\end{align}
With this choice the following identities will be valid in the remainder of this paper.
\begin{align}
\gamma_{\mu \nu}^{(0)} = \zeta_{\mu \nu} \, , \qquad R = R^{(\zeta)} = \frac{4 d (d-1)}{\ell^2} \, e^{2c/ \ell} = \frac{d (d-1)}{\alpha^2} \, .
\end{align}
Also, given these conventions, the scale factor $A(u)$ in the vicinity of the boundary \eqref{eq:expAasymp} becomes:
\begin{align}
\label{eq:Aasymp} A(u) \underset{u \rightarrow - \infty}{=} -\frac{u}{\ell} - \frac{\ell^2 R}{4d(d-1)} \, e^{2u/ \ell} + \mathcal{O} \big(e^{4u/ \ell} \big) \, .
\end{align}

\subsection{Scalar functions formalism}
 Rather than working with the second-order equations
 \eqref{eq:EOM1}--\eqref{eq:EOM3}, it will be convenient to rewrite
 them as a system of first-order equations, by introducing an
 appropriate set of scalar functions of $\f$. This is always possible locally with the exception of special points at which $\dot{\f}=0$ and which we will refer
to as bounces, as discussed in \cite{exotic, curvedRG}. As long as
$\dot{\f}(u)\neq 0$, we can invert the relation between $u$ and
$\f(u)$ to define the following functions of $\f$:
\begin{align}
\label{eq:defWc} W(\f) & \equiv -2 (d-1) \dot{A} \, , \\
\label{eq:defSc} S(\f) & \equiv \dot{\f} \, , \\
\label{eq:defTc}  T(\f) & \equiv R \, e^{-2A} = \frac{d(d-1)}{\alpha^2} \, e^{-2A} \, .
\end{align}
where the expressions on the right hand side are evaluated at
$u=u(\f)$.

We can then rewrite \eqref{eq:EOM1}--\eqref{eq:EOM3} as a set of coordinate-independent, first-order non-linear differential equations in field space:
\begin{align}
\label{eq:EOM4} S^2 - SW' + \frac{2}{d} T &=0 \, , \\
\label{eq:EOM5} \frac{d}{2(d-1)} W^2 -S^2 -2 T +2V &=0 \, , \\
\label{eq:EOM6} SS' - \frac{d}{2(d-1)} SW - V' &= 0 \, .
\end{align}
As these equations are algebraic in $T$, we can partially solve this system by eliminating $T$. We are then left with the following system of equations:
\begin{align}
\label{eq:EOM7} \frac{d}{2(d-1)} W^2 +(d-1) S^2 - dSW' +2V &=0 \, , \\
\label{eq:EOM8} SS' - \frac{d}{2(d-1)} SW - V' &= 0 \, .
\end{align}

\subsection{Holographic RG flows for field theories on dS$_d$ or $S^d$}
\label{sec:holoRGreview}
Holographic RG flows correspond to solutions for $A(u)$ and $\f(u)$ where $e^{A(u)}$ evolves monotonically in $u$ from the UV fixed point at which $e^{A_{\textrm{UV}}} \rightarrow \infty$ to the IR end point at which $e^{A_{\textrm{IR}}} \rightarrow 0$. At the UV fixed point the scalar takes a value $\f_{\textrm{UV}}$ which coincides with a maximum of the potential $V$. The scalar then changes along the flow reaching a value $\f_0$ at the IR end point, with $\f_0$ some generic point. Note that, in contrast to the flow in $u$, the change in $\f(u)$  along the flow does not have to be monotonic. However, in this work it will be sufficient to consider solutions where $\f(u)$ is monotonic. In this case $\f$ can be used as a variable describing progress along the flow.\footnote{If $\f(u)$ reverses direction as a function of $u$, then $\f$ can still be used as a parameter along the flow, albeit piecewise along the various branches.} First, we will review the solutions in the vicinity of UV and IR loci before describing complete RG flows.

\vspace{0.3cm}
\noindent \textbf{UV fixed points.} UV fixed points are associated with extrema of the scalar potential $V(\f)$ with $V(\f_{\textrm{ext}}) <0$. For maxima of $V$ the RG flow solutions describe a relevant deformation away from the UV fixed point. In the vicinity of the UV, the potential can hence be expanded as\footnote{The parameter $\ell$ appearing in \eqref{eq:Vnearmax} is defined via $V(\f_{\textrm{UV}}) = - d(d-1) / \ell^2$. To avoid confusion with $\ell_{\textrm{IR}}$ defined later, we will sometimes also denote $\ell$ by $\ell_{\textrm{UV}}$. However, when possible, we will drop the subscript `UV' to remove clutter.}
\begin{align}
\label{eq:Vnearmax} V = - \frac{d(d-1)}{\ell^2} + \frac{1}{2} m^2 (\f-
\f_{\textrm{UV}})^2 + \mathcal{O}\big( (\f- \f_{\textrm{UV}})^3\big)
\, ,
\end{align}
For a UV maximum,\footnote{This is the generic situation for a UV fixed
point. A minimum of the potential may be a UV fixed point, but then
the RG flow is driven by an irrelevant operator acquiring a
vev. In this case, $m^2 >0$ in equation \eqref{eq:Vnearmax}.} we
require  $m^2$ to be negative and larger than the BF bound, $-d^2/4<
m^2 \ell^2 < 0 $.

The geometry in the UV takes the form of an asymptotically AdS$_{d+1}$ space-time with AdS length $\ell$. The solutions for $A(u)$ and $\f(u)$ in the vicinity of the UV will take the form of a near-boundary expansion for the scale factor and a scalar in AdS$_{d+1}$ space-time:
\begin{align}
\label{eq:AAdSasymp} A(u) &= -\frac{u}{\ell} - \frac{\ell^2 R }{4d (d-1)} \, e^{2u/\ell} + \ldots \, \\
\label{eq:phiAdSasymp} \f(u) &= \f_{\textrm{UV}} + \f_- \, \ell^{\Delta_-} \, e^{\Delta_- u / \ell} + \f_+ \, \ell^{\Delta_+} \, e^{\Delta_+ u / \ell} + \ldots \, ,
\end{align}
as $u\to -\infty$, where
\begin{align}
\label{eq:Deltapmdef} \Delta_{\pm} \equiv \frac{1}{2} \left(d \pm \sqrt{d^2 +4 \ell^2 m^2} \right) \, .
\end{align}
For a maximum, we have $0\leq \Delta_-\leq  d/2$ and $d/2\leq
\Delta_+ \leq d$.

The standard holographic dictionary assigns dimension  $\Delta_{\mathcal{O}} =
\Delta_+$ to the scalar  operator $\mathcal{O}$ dual to
the bulk field $\f$. The parameter $\f_-$ is
then identified with  the source $j$ of $\mathcal{O}$ which
parametrises the deformation of
the CFT away from the UV as in equation (\ref{eq:QFT1}),
while $\f_+$ is related to the vev of $\mathcal{O}$,
\begin{align} \label{source-vev}
j = \f_-, \qquad \langle \mathcal{O} \rangle = (2 \Delta_+ -d) \, \f_+ \, .
\end{align}

In the restricted range of parameters
\begin{align}
\label{eq:Deltarestriction} \frac{d-2}{2} < \Delta_- < \frac{d}{2} \, , \quad \textrm{or, equivalently} \quad \frac{d}{2} < \Delta_+ < \frac{d+2}{2} \, ,
\end{align}
one can use a different holographic dictionary, called `alternative
quantisation' \cite{9905104}, in which  the operator dimension is
identified with $\Delta_-$, and the source $j$ and the vev $\langle
\mathcal{O} \rangle$ in equation \eqref{source-vev} are interchanged. Generally,
 we will be using the standard dictionary, with $\Delta_{\mathcal{O}}
 =  \Delta_+$, unless explicitly stated.

We  now turn to the corresponding solutions in terms of the scalar
functions $W,S,T$. In the vicinity of a maximum of $V$ the functions
$W$ and $S$
can be expanded in powers of $(\f - \f_{\textrm{UV}})$. For
simplicity, and without loss of generality, from now on
we will set $\f_{\textrm{UV}}=0$. To leading order,
\begin{align}
\label{eq:WSboundaryleading} W(\f) =
\frac{2(d-1)}{\ell} + \frac{\Delta_-}{2 \ell} \f^2 + \mathcal{O} \big(
\f^3 \big) \, , \qquad S(\f) = {\Delta_- \over \ell} \f +  \mathcal{O} \big(
\f^2 \big) \, , \qquad \f \to 0
\end{align}
Solving for $W$ and $S$ introduces two (dimensionless) integration
constants which will be denoted by $C$ and $\mathcal{R}$,
respectively. In the near-boundary expansion around $\f=0$, they appear as
coefficients of  subleading, non-analytic terms:
\begin{align}
\nonumber W(\f) & 
  \supset \frac{\mathcal{R}}{d \ell} \, |\f |^\frac{2}{\Delta_-} \Big(1 + \mathcal{O}(\f) + \mathcal{O}(|\f|^{2/\Delta_-}\mathcal{R}) + \mathcal{O}(|\f|^{d/\Delta_-}C) \Big) \\
\label{eq:Wboundarynonanalytic} & \hphantom{AAA} + \frac{C}{\ell} \, |\f|^\frac{d}{\Delta_-}  \Big(1 + \mathcal{O}(\f) + \mathcal{O}(|\f|^{2/\Delta_-}\mathcal{R}) + \mathcal{O}(|\f|^{d/\Delta_-}C) \Big) \, , \\
\nonumber S(\f) & 
  \supset {d\over \Delta_-}\frac{C}{\ell} \, |\f|^{\frac{d}{\Delta_-}-1} \Big(1 + \mathcal{O}(\f) + \mathcal{O}(|\f|^{2/\Delta_-}\mathcal{R}) + \mathcal{O}(|\f|^{d/\Delta_-}C) \Big) \\
\label{eq:Sboundarynonanalytic} & \hphantom{AAA} + \mathcal{O} \Big( |\f|^{2/\Delta_-+1}\mathcal{R} \Big) \, .
\end{align}
For reference, the above near-boundary expansions for $W$ and $S$, but also that of $T$ are also collected in appendix \ref{app:nearmaximum}.

The upshot is that in the vicinity of a maximum, the solutions for $W$ and $S$ come as two-parameter families
\begin{align}
W_{C,\mathcal{R}} (\f) \, , \quad \textrm{and} \quad S_{C,\mathcal{R}}(\f) \, .
\end{align}
Given a set of solutions $W_{C,\mathcal{R}}$ and $S_{C,\mathcal{R}}$ one can integrate $W \sim \dot{A}$ and $S \sim \dot{\f}$ to obtain the corresponding solutions for $A(u)$ and $\f(u)$ (see appendix \ref{app:nearmaximum} for details). This introduces two further integration constants. The integration constant in $A$ is already fixed by our choice $\alpha = \frac{\ell}{2} e^{-c/ \ell}$. The other integration constant is physical and corresponds to $\f_-$ introduced above. Therefore, in standard quantization it is the source of the operator $\mathcal{O}$ dual to the bulk field $\f$.

We can then compare the solutions for $A(u)$ and $\f(u)$ (given in \eqref{eq:Amsol}, \eqref{eq:phimsol}) obtained from integrating $W_{C,\mathcal{R}}$ and $S_{C,\mathcal{R}}$ with the asymptotic form \eqref{eq:AAdSasymp}, \eqref{eq:phiAdSasymp}. One finds that the two match as long as one identifies
\begin{align}
\label{eq:dimlessRdef} \mathcal{R} &= R \, |\f_-|^{-2 / \Delta_-} \, , \\
C &= \frac{(2 \Delta_+ -d) \Delta_-}{d} \, \f_+ \, |\f_-|^{- \Delta_+ / \Delta_-} \, .
\end{align}
In standard quantization this establishes the identification of $C$ with the vev of the perturbing operator in units of its source:
\begin{align}
C & \underset{\textrm{std. quant.}}= \frac{\Delta_-}{d} \, \langle \mathcal{O} \rangle \, |\f_-|^{- \Delta_+ / \Delta_-} \, .
\end{align}
Similarly, $\mathcal{R}$ has the interpretation of the UV boundary
curvature $R$ in units of the operator source. Note that for a given
choice of  $(C, \mathcal{R})$ a set of solutions $W_{C,\mathcal{R}}$ and $S_{C,\mathcal{R}}$ does not describe a single flow, but a one-parameter family of flows labeled by the operator source $\f_-$. Only once we specify a value for the source $\f_-$ will a solution $W_{C,\mathcal{R}}$ and $S_{C,\mathcal{R}}$ correspond to a single flow.

Most importantly,   for a given bulk potential, only a
one-parameter subfamily of the solutions $W_{C,\mathcal{R}}$ and
$S_{C,\mathcal{R}}$ can be completed into an RG flow with a
regular interior. The remaining solutions correspond to geometries
that exhibit a singularity somewhere in the interior. In other words,
imposing regularity of the geometry results in a relation $C =
C(\mathcal{R})$. We will make this  point more explicit below, when
describing the IR geometry.

In addition to $W$ and $S$, another function, which we denote $U(\f)$,
will be useful in the following and we find it convenient to introduce
it at this stage. Given $W(\f)$ and $S(\f)$, we
define $U(\f)$ as one of the solution to the differential equation
\begin{align}
\label{eq:Uequation0} S U' - \frac{d-2}{2(d-1)} W U = -\frac{2}{d} \, .
\end{align}
As this is a first order differential equation, solving for $U$ introduces one further integration constant (in addition to $C$ and $\mathcal{R}$), which we will refer to as $B$. Using the near-boundary expansions of $W$ and $S$ given in \eqref{eq:WSboundaryleading}, \eqref{eq:Wboundarynonanalytic} and \eqref{eq:Sboundarynonanalytic}, we can derive the near-boundary behaviour of $U$ from \eqref{eq:Uequation0}:
\begin{align} \label{eq:Uuv}
U(\f) \underset{\f \rightarrow 0}{=} \ell \left[\frac{2}{d(d-2)} + B |\f|^{(d-2)/\Delta_-} + \mathcal{O} \big( \mathcal{R} |\f|^{2 / \Delta_-} \big) \right] \, .
\end{align}
Notice that the dependence on the integration constant $C$ through $W$
and $S$, only enters $U$ at subleading order in the near-boundary
expansion. Thus, in the vicinity of the UV, we find that there is a
family of solutions for $U$, which we will denote by $U_{B,
  \mathcal{R}}$. 

As for $W$ and $S$, there is a specific regularity condition  one
can impose (and which we will make explicit below when we describe the
IR)  which, given ${\cal R}$, completely fixes the
integration constant $B$. Therefore, equation (\ref{eq:Uequation0})
plus regularity give an unambiguous definition of the function
$U(\f)$, and one thus has $B =
B(\mathcal{R})$. While $C$ is related to the vev of the operator $\mathcal{O}$ in field theory, the constant $B$ computes an appropriately defined entanglement entropy. This will be explained in detail in sec.~\ref{sec:ent}.

The function $U(\f)$ defined as above has appeared before in the
literature, as the coefficient of the Einstein-Hilbert term in the
derivative expansion of the on-shell action evaluated on a RG flow solution  \cite{9912012,0703152,1106.4826,1401.0888}
\begin{align} \label{Einstein}
S_{\textrm{on-shell}} \supset \left. \int d^dx \sqrt{|\gamma|} \, \tfrac{d}{2} \, U(\f) \, R^{(\gamma)} \right|_{\textrm{UV}} \, ,
\end{align}
where $\gamma_{\mu \nu}$ is the induced metric defined in
\eqref{eq:gammadef}, $R^{(\gamma)}$ the corresponding curvature scalar
and the subscript UV implies that this is to be evaluated in the limit
$\f \rightarrow \f_{\textrm{UV}}$. The factor $d/2$ was included for
later convenience.\footnote{In fact, the function defined here differs
  by a factor $d/2$ from that entering the on-shell action as written
  in e.g. \cite{1401.0888}.} In this context, the regularity
condition simply means that the Einstein-Hilbert term in equation
(\ref{Einstein}) does not receive any contribution from the IR, and
that the derivative expansion of the corresponding first order flow
equations is self-consistent \cite{1401.0888}. Here  we will
see that the function $U$ has a more general meaning for any value of
the curvature, not
restricted to a derivative expansion of the on-shell action (which requires
small $R$) and it is related to the entanglement entropy.  

Having discussed maxima at length, note that minima of $V$ can also play the role of UV fixed points. In this case flows away from the UV fixed point are driven purely by a vev of an irrelevant operator (see e.g.~\cite{exotic, curvedRG}). Such flows can only be completed into globally regular solutions if the potential is tuned accordingly. We do not consider UV fixed points at minima of $V$ any further in this work.

\vspace{0.3cm}
\noindent \textbf{IR end points.} IR end points of holographic RG
flows for field theories on dS$_d$ or $S^d$ are associated with
generic points of the potential $V$. Most importantly, IR end points
cannot coincide with extrema of the potential (although they can be
arbitrarily close to an extremum). This is in contrast to flat domain
wall solutions, in which IR fixed points can only occur at minima of
the potential. Similarly, for flows with non-zero boundary curvature
$R$ the IR occurs at some finite value $u_0$ of the holographic
coordinate $u$,  with $u_0 \rightarrow + \infty$ for $R \rightarrow 0$.

At an IR end point at ($u_0, \f_0$) with $\f_0 \equiv \f(u_0)$ the flow stops, i.e.~$\dot{\f}(u_0)=0$ and  one has
\begin{align}
\label{eq:expAasympIR} e^{A(u)} & \underset{u \rightarrow u_0}{\longrightarrow} - \frac{u-u_0}{\alpha} + \mathcal{O} \big((u-u_0)^3 \big) \, , \\
\label{eq:phiasympIR} \f(u) & \underset{u \rightarrow u_0}{=}  \f_0 + \mathcal{O}\big((u-u_0)^2 \big) \, ,
\end{align}
The bulk geometry asymptotes to that of AdS$_{d+1}$ with AdS length
set by the value of the potential at $\f_0$, i.e. 
\be
R_{ab} = -{d\over \ell_0^2} g_{ab}  + \mathcal{O}\big((u-u_0)^2\big), \qquad \ell_0^2 \equiv
{d(d-1)\over |V(\f_0)|}
\ee
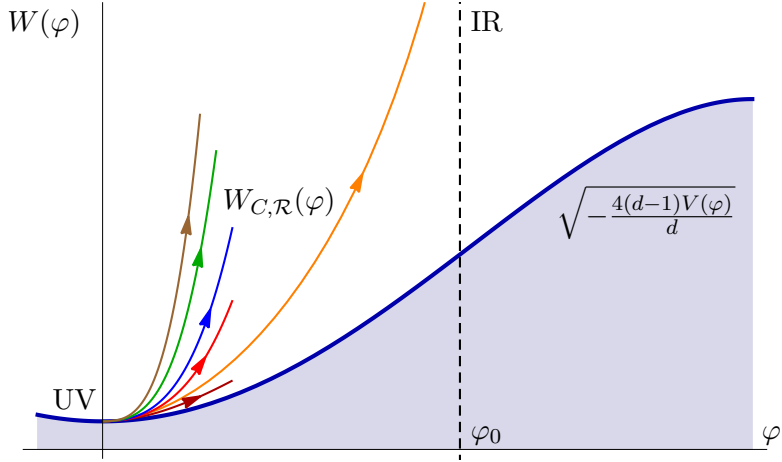
\begin{figure}[t]
\centering
\begin{overpic}
[width=0.65\textwidth]{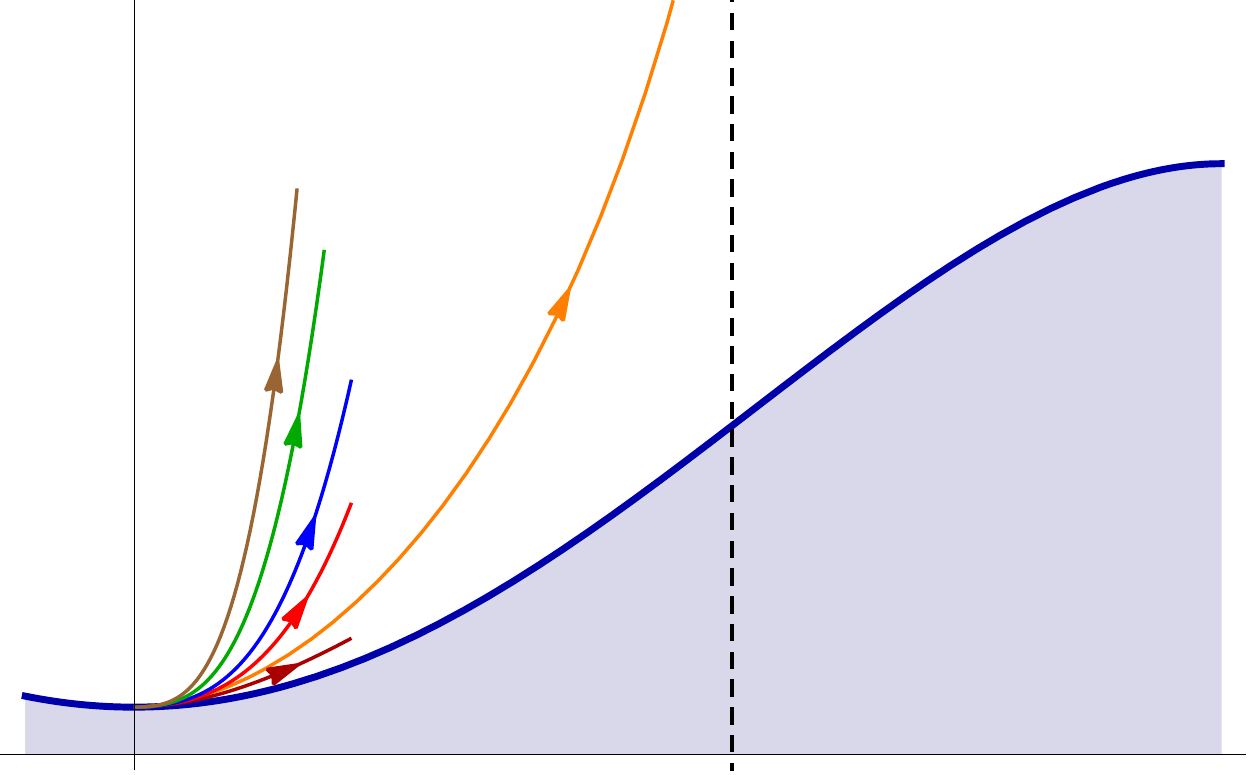}
\put(4,7){UV}
\put(60,3){$\f_0$}
\put(60,58){IR}
\put(99,3){$\f$}
\put(-2,58){$W(\f)$}
\put(26.8,34){$W_{C, \mathcal{R}}(\f)$}
\put(71.5,32){$\sqrt{-\tfrac{4(d-1) V(\f)}{d}}$}
\end{overpic}
\caption{Holographic RG flow solutions for the potential \protect\eqref{eq:pot}. Note the family of solutions $W_{C, \mathcal{R}}(\f)$ in the vicinity
  of a UV fixed point located at the maximum of the potential. For a given IR end point $\f_0$ only one of these solutions can be completed, resulting in a holographic RG flow for theories with a specific combination of $(C, \mathcal{R})$. The shaded region below the blue curve is not accessible for flows for field theories on dS$_d$ or $S^d$.}
\label{fig:Wex1}
\end{figure}

From the IR-asymptotic behavior in \eqref{eq:expAasympIR} and \eqref{eq:phiasympIR} we can derive the corresponding expressions for the functions $W(\f)$ and $S(\f)$ in the vicinity of the IR end point $\f_0$:
\begin{align}
\label{eq:WIR} W(\f) &= \frac{W_0}{\sqrt{|\f - \f_0|}} + \mathcal{O}\big({|\f - \f_0|}^0 \big) \, , \\
\nonumber S(\f) &= S_0 \sqrt{|\f - \f_0|} + \mathcal{O}\big(|\f - \f_0| \big) \, , \\
\nonumber \text{with} \quad S_0^2 &= \frac{2 |V'(\f_{0})|}{d+1}\sp W_0 = (d-1)S_0.
\end{align}
Note that the function $W$ diverges for $\f \rightarrow \f_0$, but this
does not imply that the geometry is singular there. As can be checked
from \eqref{eq:expAasympIR}, all curvature invariants remain
finite (see \cite{curvedRG} for details). 

The  behavior of $W$ and $S$ near $\f_0$ in equation (\ref{eq:WIR}), corresponding to IR-regular
solutions,  is completely determined by
$\f_0$, which appears as the only parameter. In contrast,  the {\em
  generic} solution for the scalar functions  depends on two independent
integration constants, which can be traced in  the UV to $C$ and
$\mathcal{R}$.

From \eqref{eq:Uequation0} and (\ref{eq:WIR})  one can derive the  behaviour
of $U(\f)$ in the IR, 
\begin{align}
\label{eq:UIR0} U(\f) \underset{\f \rightarrow \f_0}{=} {b \over |\f -
  \f_0|^{d-2 \over 2(d-1)}}  +
U_0 \, \sqrt{|\f - \f_0|} + \mathcal{O}\big(|\f - \f_0| \big) , \quad \textrm{with} \quad U_0 = \frac{4}{d(d-1) S_0}  \, ,
\end{align}
where $b$ is an integration constant. For generic values of $b$, $U(\f)$ diverges
as $\f\to \f_0$. We can therefore fix the integration constant of the
$U$-equation  by requiring that $U(\f)$ is finite at the IR
endpoint. This is our definition of ``regularity'' for the function
$U$, and it  fixes $b=0$ in equation (\ref{eq:UIR0}). In turn, this translates into some
($\f_0$-dependent) value $B=B(\f_0)$ of the integration appearing in the UV expansion (\ref{eq:Uuv}). 

For completeness, we also record the near IR behaviour of $T(\f)$. This can be deduced from \eqref{eq:WIR} with the help of \eqref{eq:EOM4}. One finds
\begin{align}
\label{eq:TIR} T= \frac{T_0}{|\f -\f_0|} + \mathcal{O}\big({|\f - \f_0|}^{-1/2} \big) \, , \quad \textrm{with} \quad T_0 = \frac{d(d-1)}{4} S_0^2 \, .
\end{align}

\vspace{0.3cm}
\noindent \textbf{Complete regular RG flows.}
Complete holographic RG flows correspond to regular solutions that
interpolate between the UV and IR behaviour as discussed
above. Consider a solution $W(\f)$, $S(\f)$, $U(\f)$ for a flow with UV fixed
point at an extremum and IR end point at some specific value $\f_0$
which is not an extremum. 
 As the solution near the IR is unique, the full solution
picks out one of the representatives of the family of solutions
$W_{C,\mathcal{R}}$, $S_{C,\mathcal{R}}$, $U_{B,\mathcal{R}}$ near the UV to form a
complete flow. Therefore, a solution with a given $\f_0$ corresponds to an
RG flow for a theory with specific boundary data, i.e.~we have a map
from end-point-value space  to solution space,
$$ \f_0 \to (B(\f_0), \, C(\f_0), \, \mathcal{R}(\f_0)).$$
This in turn can be used to obtain, by eliminating $\f_0$, the
functions $B(\mathcal{R})$ and $C(\mathcal{R})$, which however are not necessarily
single-valued. In other words, the parameter $\f_0$ is the most
convenient to scan the space of solutions, although it does not have a
direct interpretation in terms of boundary data.
The reason is that if a solution with a given end-point $\f_0$ exists, then {\em it is unique}.
We will see this in detail in an example, which will be discussed below.

Extracting $B(\mathcal{R})$ and $C(\mathcal{R})$ will be of great importance in this work. As will become apparent in section \ref{sec:collectF} and \ref{sec:ent}, the functions $B(\mathcal{R})$ and $C(\mathcal{R})$ contain the universal (i.e.~the UV-cutoff-independent) contributions to the free energy on $S^3$ and the entanglement entropy across a spherical surface in dS$_3$. As a result, $B(\mathcal{R})$ and $C(\mathcal{R})$ are the crucial building blocks out of which we will construct $F$-functions.

\begin{figure}[t]
\centering
\begin{subfigure}{.5\textwidth}
\centering
\begin{overpic}[width=1.0\textwidth,tics=10]{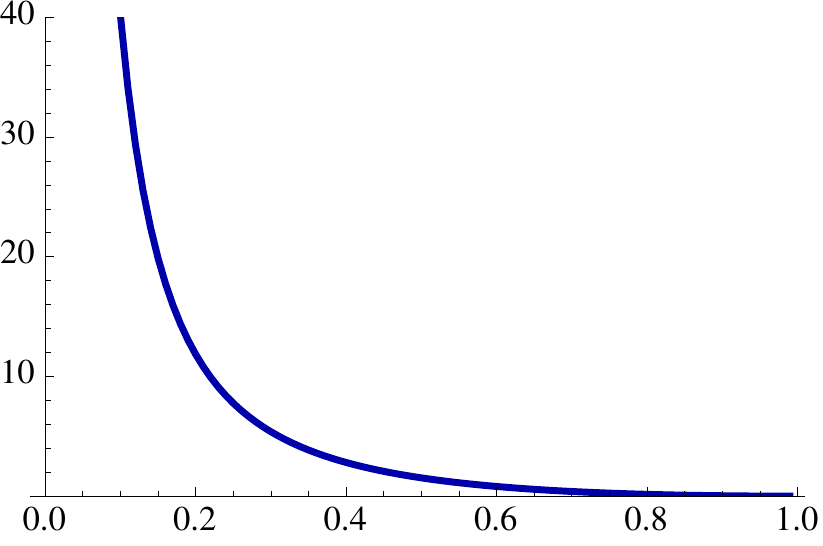}
\put (92,12) {$\f_0$} \put (6,68) {$\mathcal{R}$}
\end{overpic}
\caption{\hphantom{A}}
\label{fig:Rvsphi0}
\end{subfigure}%
\begin{subfigure}{.5\textwidth}
\centering
\begin{overpic}[width=1.0\textwidth,tics=10]{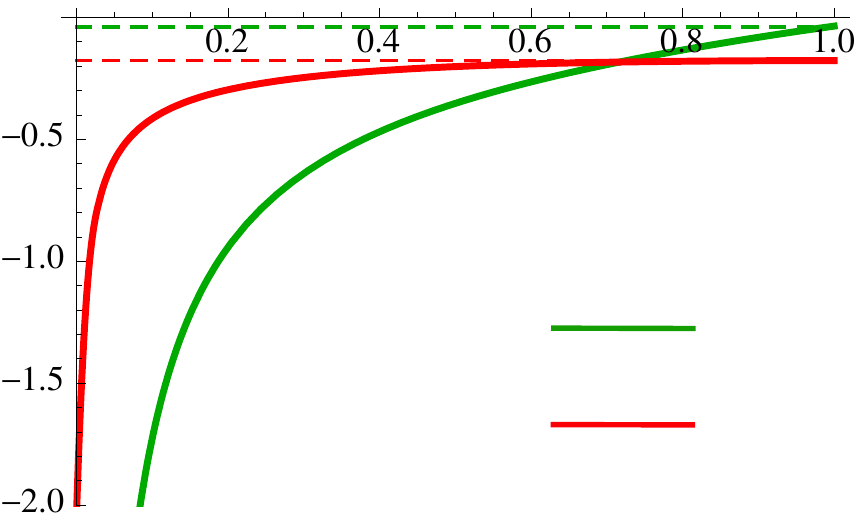}
\put (93,63) {$\f_0$} \put (83,21) {$B$} \put (83,9.5) {$C$}
\end{overpic}
\caption{\hphantom{A}}
\label{fig:BCvsphi0}
\end{subfigure}
\caption{\textbf{(a):} Dimensionless curvature $\mathcal{R}$ vs.~IR end point $\f_0$ for the potential \protect\eqref{eq:pot} with $d=3$, $\Delta_-=1.2$ and $\lambda$ chosen such that $\f_{\textrm{min}}=1$. Note that $\mathcal{R}$ increase as the IR end point $\f_0$ moves closer to the UV fixed point at $\f=0$. When $\f_0 \rightarrow \f_{\textrm{min}}=1$ the solutions asymptote to RG flows for a theory on flat space-time and hence $\mathcal{R}\rightarrow 0$. \textbf{(b):} $B$ and $C$ vs.~$\f_0$ for the same model parameters as in (a). Note that for $\f_0 \rightarrow \f_{\textrm{min}}=1$ both $B$ and $C$ attain a finite value indicated by the dashed green and red line, respectively. Here one finds that $B$ and $C$ decrease monotonically as $\f_0$ is decreased, ultimately diverging for $\f_0 \rightarrow \f_{\textrm{max}}=0$.}
\label{fig:RBCvsphi0}
\end{figure}
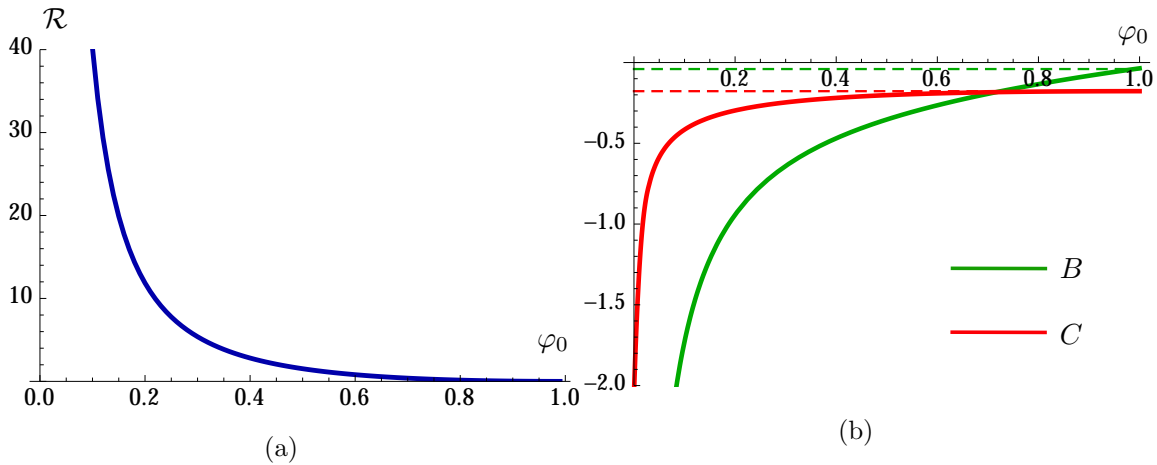

\vspace{0.3cm}
\noindent \textbf{Example.}
To conclude this section, we present an example for concreteness. We consider a potential with at least one maximum and minimum. To be specific, we will consider the following quadratic-quartic potential:
\begin{align}
\label{eq:pot} V(\f) = - \frac{d(d-1)}{\ell^2} - \frac{\Delta_- (d-\Delta_-)}{2 \ell^2} \f^2 + \frac{\lambda}{\ell^2} \f^4 \, ,
\end{align}
with $\lambda >0$ and $\tfrac{d-2}{2} < \Delta_- < \tfrac{d}{2}$. This potential has a maximum at $\f=\f_{\textrm{max}}=0$ which we identify with the UV fixed point. Here $\ell$ sets the AdS length in the UV. The parameter $\Delta_-$ is related to the scaling dimension $\Delta_+$ of the perturbing operator in the UV as $\Delta_- = d- \Delta_+$. Restricting attention to $\f \geq 0$ the potential also has a minimum at
\begin{align}
\f_{\textrm{min}}=\sqrt{\frac{\Delta_- (d-\Delta_-)}{4 \lambda}} \, .
\end{align}
In the following, we consider RG flows which are confined to the
interval $0 \leq \f \leq \f_{\textrm{min}}$. In addition, we restrict
our attention to solutions where $\f(u)$ evolves monotonically along a
flow,  as in
the present example. As a result, $\f$ is a suitable parameter along each flow. In this setting, RG flows for field theories on dS$_d$ or $S^d$ exhibit the following properties.
\begin{itemize}
\item In the vicinity of the UV fixed point at $\f=0$, we have families of solutions $W_{C,\mathcal{R}}$, $S_{C,\mathcal{R}}$ and $U_{B, \mathcal{R}}$ parametrised by $B$, $C$ and $\mathcal{R}$ as discussed above. For $W(\f)$ this is shown in fig.~\ref{fig:Wex1}. The family of solutions $W_{C, \mathcal{R}}(\f)$ is represented by the multitude of flows emanating from the UV in fig.~\ref{fig:Wex1}.
\item Any point $\f_0$ in the interval $0 < \f_0 < \f_{\textrm{min}}$ can act as an IR end point. In particular, for every such value $\f_0$ there exists a smooth solution interpolating between the UV at $\f=0$ and the IR end point at $\f=\f_0$. The solutions $W(\f)$ plotted in fig.~\ref{fig:flowsintro} (shown in the introduction) are solutions for $W(\f)$ for different values of $\f_0$ for the potential \eqref{eq:pot}. The dark red solution interpolating between the maximum and minimum of $V$ corresponds to the solution $W_{\textrm{flat}}(\f)$ and describes a holographic RG flow of a field theory on flat space-time.
\item However, only one particular representative of the family of
  solutions $W_{C,\mathcal{R}}$, $S_{C,\mathcal{R}}$ and $U_{B, \mathcal{R}}$ in the vicinity
  of the UV with a particular combination of $(B, C, \mathcal{R})$ is
  completed to a full flow solution with regular IR end point at a given
  $\f_0$ (see fig.~\ref{fig:Wex1}). Turning this around: choosing an
  IR end point $\f_0$ for a regular flow fixes the dimensionless curvature
  $\mathcal{R}$ and dimensionless parameters $B$ and $C$ associated with the solution.

\begin{figure}[t]
\centering

\end{figure}

\begin{figure}[t]
\centering
\begin{subfigure}{.5\textwidth}
 \centering
   \begin{overpic}
[width=1.0\textwidth]{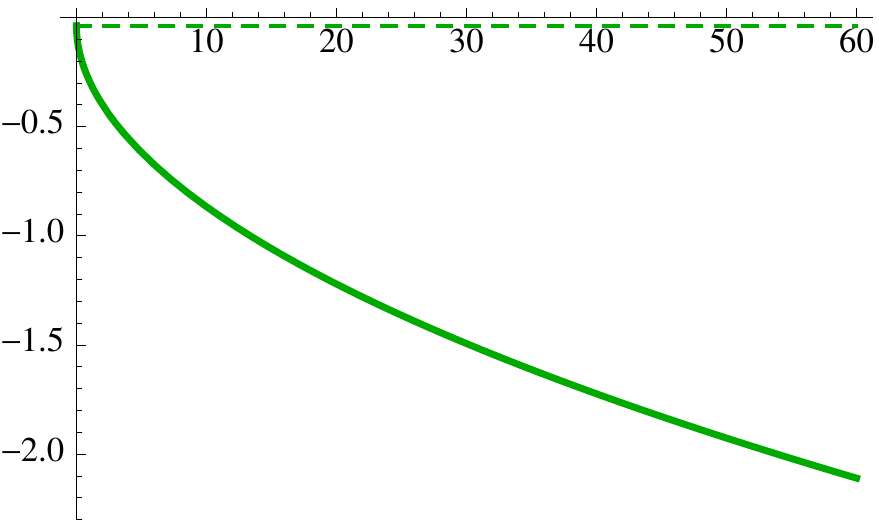}
\put(94,62){$\mathcal{R}$}
\put(12,3){$B(\mathcal{R})$}
\end{overpic}
\caption{\hphantom{A}}
\label{fig:BvsRDelta1p2}
\end{subfigure}%
\begin{subfigure}{.5\textwidth}
 \centering
   \begin{overpic}
[width=1.0\textwidth]{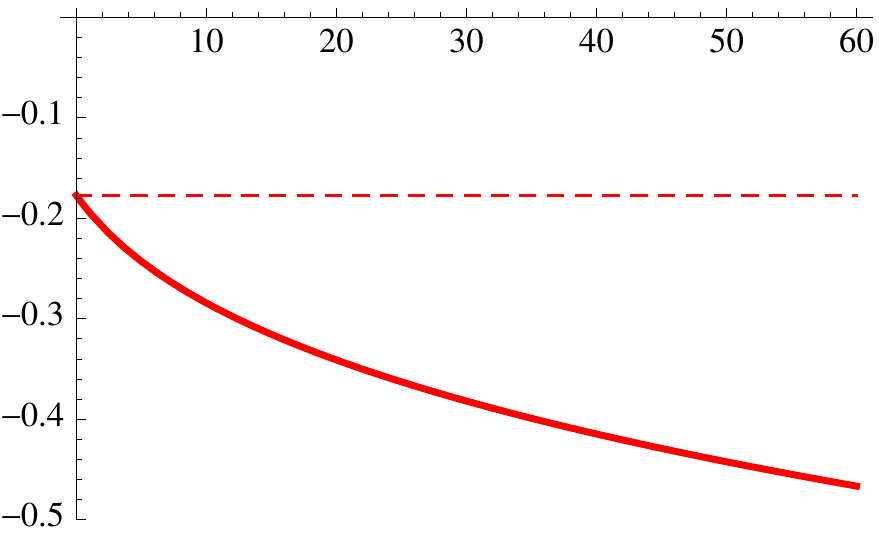}
\put(94,62){$\mathcal{R}$}
\put(12,3){$C(\mathcal{R})$}
\end{overpic}
\caption{\hphantom{A}}
\label{fig:CvsRDelta1p2}
\end{subfigure}%
\caption{$B$ vs.~$\mathcal{R}$ \textbf{(a)} and $C$ vs.~$\mathcal{R}$ \textbf{(b)} for RG flow solutions in the potential \protect\eqref{eq:pot} $d=3$, $\Delta_-=1.2$ and $\lambda$ chosen such that $\f_{\textrm{min}}=1$. For $\mathcal{R}\rightarrow 0$ both $B(\mathcal{R})$ and $C(\mathcal{R})$ attain a finite value indicated by the green and red dashed line, respectively.}
\label{fig:BCvsRDelta1p2}
\end{figure}

\item Once we have a complete solution, we can read off the corresponding values of $B$, $C$ and $\mathcal{R}$ from the asymptotic behaviour near the UV fixed point. In figure \ref{fig:RBCvsphi0} we display $\mathcal{R}$, $B$ and $C$ vs.~$\f_0$ for an example based on potential \eqref{eq:pot} with $d=3$ and $\Delta_-=1.2$ and $\lambda$ chosen such that $\f_{\textrm{min}}=1$. We can make the following observations. Note that for $\f_0 \rightarrow \f_{\textrm{min}}$ we find $\mathcal{R} \rightarrow 0$. This is consistent with the fact that RG flows for theories on flat manifolds (i.e.~manifolds with $\mathcal{R}=0$) have their IR end points at minima of $V$. Then, we find that $\mathcal{R}$ increases when the IR end point $\f_0$ is moved away from $\f_{\textrm{min}}$ and closer to $\f_{\textrm{max}}$, diverging as $\f_0 \rightarrow \f_{\textrm{max}}$. Similarly, $C$ diverges in magnitude as  $\f_0 \rightarrow \f_{\textrm{max}}$, while approaching a constant for $\f_0 \rightarrow \f_{\textrm{min}}$. Here, in this simple potential, the evolution of $\mathcal{R}$ (and $B$,$C$) with $\f_0$ is monotonic. Note that this is not generic and a much richer behaviour emerges in more complicated potentials. We refer readers to \cite{curvedRG} for further details.
\item An important observation is that flows with different values of $\mathcal{R}$ also exhibit different values of $B$ and $C$. Therefore, when evaluated on a continuum of RG flow solutions, $B$ and $C$ should be seen as a functions of $\mathcal{R}$. We can see this explicitly by plotting $B$ and $C$ vs.~$\mathcal{R}$ for RG flow solutions in the potential \eqref{eq:pot}. In figure  \ref{fig:BCvsRDelta1p2} this is shown for $d=3$ and $\Delta_-=1.2$. Note that the behaviour of $B(\mathcal{R})$ and $C(\mathcal{R})$ as a function of $\mathcal{R}$ can be determined analytically for $\mathcal{R} \rightarrow 0$ and $\mathcal{R} \rightarrow \infty$ (see appendix \ref{sec:largesmallR} for details). One can check that the numerical results are consistent with these analytic expectations.
\end{itemize}

To end this section, note that our review of holographic RG flows for field theories on dS$_d$ and $S^d$ has not been exhaustive. Phenomena we did not mention include flows which start at maxima of a potential and skip minima on their way to the IR. Similarly, we did not discuss flows along which $\f(u)$ changes direction. Related to this are interesting phenomena such as phase transitions triggered by space-time curvature. For all of this we refer readers to \cite{curvedRG}.

\section{On-shell action and free energy
}
\label{sec:collectF}
In this section we will present various quantities which will be
instrumental in the construction of the $F$-functions, as well as their
renormalization.

The basic quantity we consider in this section is  the (regularised or
renormalized) bulk on-shell
action, which is the action given in \eqref{eq:action} evaluated on solutions to the field equations
\eqref{eq:EOM1}--\eqref{eq:EOM3}. For the class of solutions
we are considering, this functional reduces to a function of boundary
values, $S_{\textrm{on-shell}} = S_{\textrm{on-shell}}(\f_-, R)$. However, the
need for regularisation and/or renormalization of this quantity
introduces a dependence on  extra parameters: a UV cut-off
$\Lambda$, or a choice of boundary counterterms.

In the dual field theory, the on-shell action has a different
interpretation depending whether one is using standard or alternative
holographic dictionary (see the discussion  in section  \ref{sec:holoRGreview}).

\begin{itemize}
\item In the  standard dictionary, the  on-shell action is  related to
the {\em free energy} of the theory as\footnote{In a thermodynamic interpretation of our system one may define the thermodynamic free energy $F_{\textrm{th}}$ as $\beta F_{\textrm{th}} = S_{\textrm{on-shell}, E}$ with $\beta$ the appropriate inverse temperature and $S_{\textrm{on-shell}, E}$ the Euclidean on-shell action. Note that this thermodynamic free energy $F_{\textrm{th}}$ differs from the free energy $F$ defined here.}
\begin{align} \label{F-S}
F (j,R) & \equiv - S_{\textrm{on-shell}} (\f_-, R) \, ,
\end{align}
with the source $j$ identified with $\f_-$.
\item
In the alternative dictionary, the on-shell action is identified,
rather than with the free energy, with its Legendre transform with
respect to the source $j$, i.e.  the {\em quantum
effective potential} $\Gamma (\langle \mathcal{O} \rangle, \mathcal{R})$ of the theory, which depends on the operator vev
$\langle \mathcal{O}  \rangle $ rather than the source $j$. That is, in alternative quantisation we have
\begin{align} \label{V-S}
\Gamma (\langle \mathcal{O} \rangle, R) & \equiv - S_{\textrm{on-shell}} (\f_-, R) \, ,
\end{align}
where $\f_-$ is now identified with $\langle \mathcal{O} \rangle$.
\end{itemize}

This distinction will be important later on. However, for simplicity,
we will be referring to the  on-shell action as {\em free energy},
thereby assuming the standard dictionary, unless stated
explicitly otherwise.

The appropriately renormalized free energy, which will be computed in
subsection \ref{sec:Sren} and will be the  starting
point for constructing our proposals for $F$-functions in section
\ref{sec:Ftheorem}.

\subsection{The free energy of a holographic RG flow}
\label{sec:freeenergyS3}
We begin by writing the  expression for the free energy $F$ for a
field theory on $S^d$, by using the definition (\ref{F-S}), where
$S_{\textrm{on-shell}}$ is given by \eqref{eq:action} evaluated on a
solution.

The on-shell action can be expressed as a functional over $A(u)$ (we refer
readers to appendix \ref{app:freeenergy} for details of the
derivation). This leads to the following expression for the free energy,
\begin{align}
\label{eq:Son} F = 2(d-1) M^{d-1} V_d \, {\big[e^{dA} \dot{A} \big]}_{\textrm{UV}} - \frac{2 M^{d-1} R}{d} V_d \int_{\textrm{UV}}^{\textrm{IR}} du \, e^{(d-2)A} \, ,
\end{align}
where we defined
\begin{align}
V_d \equiv \int \textrm{d}^dx \sqrt{|\zeta|} = \textrm{Vol}(S^d) \, .
\end{align}
The integration in \eqref{eq:Son} is over the entire geometry of the
holographic RG flow from the boundary at $u \rightarrow - \infty$
(referred to as UV) to the interior at some finite $u_0$ (referred to
as IR). Because  of  regularity of the IR end-point, the
IR gives a vanishing contribution to the first term in equation
(\ref{eq:Son}), as can be easily seen using equation (\ref{eq:expAasympIR}). 

Integrating up to the boundary gives rise to UV-divergences which we hence regulate by introducing a UV cutoff. In particular, we will work with a dimensionless cutoff defined as
\begin{align}
\label{eq:dimlesscutoff} \Lambda \equiv {\left. \frac{e^{A(u)} }{\ell \, |\f_-|^{1/\Delta_-}} \right|}_{u = \log \epsilon} \, ,
\end{align}
such that taking $\Lambda \rightarrow \infty$ is equivalent to letting $\epsilon \rightarrow 0$ corresponding to the boundary at $u \rightarrow - \infty$. Hence, whenever we write `UV' in the following, this implies that the corresponding quantity is to be evaluated at $u = \log \epsilon$ or $\f= \f (\log \epsilon)$.

It will be convenient to rewrite the free energy given in \eqref{eq:Son} in terms of the functions $W(\f)$, $S(\f)$ and $T(\f)$. The resulting expression will be more amenable to numerical analysis as well as analytical considerations compared to \eqref{eq:Son}. As a first step, we write
\begin{align}
\label{eq:Volsphere} V_d = \int \textrm{d}^dx \sqrt{|\zeta|} = \frac{2 \pi^{\frac{d+1}{2}}}{\Gamma(\frac{d+1}{2})} \, \alpha^{d} = \tilde{\Omega}_d \, R^{-\frac{d}{2}}  \, , \quad \textrm{with} \quad \tilde{\Omega}_d \equiv \frac{2 d^{\frac{d}{2}} (d-1)^{\frac{d}{2}} \pi^{\frac{d+1}{2}}}{\Gamma(\frac{d+1}{2})} \, .
\end{align}
Inserting this and using the definitions \eqref{eq:defWc}--\eqref{eq:defTc} the free energy \eqref{eq:Son} can be written as
\begin{align}
\label{eq:Son2} F =  - M^{d-1} \tilde{\Omega}_d \left( {\big[ T^{- \frac{d}{2}}(\f) W(\f) \big]}_{\textrm{UV}} + \frac{2}{d} \int_{\textrm{UV}}^{\textrm{IR}} d \f \, S^{-1}(\f) T^{-\frac{d}{2}+1}(\f) \right) \, .
\end{align}
We now  rewrite the second term. In particular note that we can express\footnote{To avoid confusion, in eq.~\eqref{eq:STintid} we denote the differential symbol by $\textrm{d}$, while the number of (boundary) space-time dimensions is written as $d$. In the remainder of this analysis there is little danger of confusion and so we revert to using $d$ for both differentials and the number of dimensions.}
\begin{align}
\label{eq:STintid} \frac{2}{d} \, S^{-1} (\f) T^{-\frac{d}{2}+1} (\f)   \, \textrm{d} \f = - \textrm{d} \left(T^{-\frac{d}{2}+1} (\f) U(\f) \right) \, ,
\end{align}
in terms of  the  function $U(\f)$ introduced in section
\ref{sec:holoRGreview} 
through equation (\ref{eq:Uequation0}) and specified by the regularity condition
\begin{align}
\label{eq:UIR} U(\f) \underset{\f \rightarrow \f_0}{=} U_0 \, \sqrt{|\f - \f_0|} + \mathcal{O}\big(|\f - \f_0| \big)  \, .
\end{align}
As we explained in the previous section, this regular IR expansion acts as a boundary condition for the
differential equation (\ref{eq:Uequation0})  and
fixes the integration constant $B$ appearing in $U$ uniquely in terms of
$\f_0$. Therefore, by introducing the function $U(\f)$ we did not
introduce any extra freedom or new parameter. As for the coefficient
$C(\mathcal{R})$, we can trade the dependence on $\f_0$ in $B(\f_0)$ with a
dependence on $\mathcal{R}$ which, contrary to $\f_0$, is one of the
boundary data, and write $B \equiv B(\mathcal{R})$. As shown in
sec.~\ref{sec:ent}, $B(\mathcal{R})$ computes an appropriately defined entanglement entropy.

Going back to the computation of the free energy, using \eqref{eq:STintid} the 2nd term in \eqref{eq:Son2} becomes:
\begin{align}
\label{eq:2ndterm} \frac{2}{d} \int_{\textrm{UV}}^{\textrm{IR}} d \f \, S^{-1}(\f) T^{-\frac{d}{2}+1}(\f) = - {\big[T^{-\frac{d}{2}+1}(\f) U(\f) \big]}_{\textrm{IR}} + {\big[T^{-\frac{d}{2}+1}(\f) U(\f) \big]}_{\textrm{UV}} \, .
\end{align}
One can check that the contribution from the IR to the above
expression vanishes, by inserting the corresponding near-IR expansions
for $T(\f)$ and $U(\f)$ given in \eqref{eq:TIR} and
\eqref{eq:UIR}. This is ensured by the regularity conditions on 
 $U(\f)$ at the IR end-point.\footnote{We can now better understand the meaning
of the regularity condition we imposed on $U$  by setting $b=0$ in
equation (\ref{eq:UIR0}).  While for $W$ IR regularity  is a
necessary condition  for the regularity of the bulk solution,
the regularity of  $U$  is a  choice, but a particularly convenient
one as it allows to write the free energy purely as a  UV
boundary term. Any other choice of the integration constant in the $U$-equation would
have given the same numerical result for $F$, but this would arise by
a combination of UV and IR terms \cite{1401.0888}. This goes against the spirit of
holography, in which one should be able to write the field theory partition
function purely in terms of UV boundary data.} Collecting all results, we arrive at an expression for
$F$ which is purely a UV boundary term:
\begin{align}
\label{eq:SonUV} F = - M^{d-1} \tilde{\Omega}_d \left({\big[ T^{- \frac{d}{2}}(\f) W(\f) \big]}_{\f(\log \epsilon)} +  {\big[T^{-\frac{d}{2}+1}(\f) U(\f) \big]}_{\f(\log \epsilon)} \right) \, ,
\end{align}
where we now display the UV cutoff explicitly. 

From \eqref{eq:SonUV} it is clear  that the on-shell action depends on
the UV curvature $R$ only through the dimensionless combination
$\mathcal{R} = R \, |\f_-|^{-2 / \Delta_-}$. The reason is that any
dependence on $R$ enters \eqref{eq:SonUV} only through $W$, $T$ and
$U$ which themselves are functions of $\mathcal{R}$ only.\footnote{The
  fact that $U$ only depends on $R$ through $\mathcal{R}$ follows from
  the fact $U$ can be determined entirely in terms of $W$ and $S$ from
  equation \eqref{eq:Uequation0}.} In addition, it depends on the
dimensionless cutoff $\Lambda$ through $\f(\log \epsilon)$. As the
free energy is dimensionless and there are no further dimensionful
parameters beyond $R$ and $\f_-$ in this problem, it is only  a function of the two dimensionless parameters $\Lambda$ and $\mathcal{R}$, i.e.~$F=F(\Lambda, \mathcal{R})$.

The structure of the UV-divergent terms as $\Lambda \to +\infty$
depends on the number of dimensions $d$, and  it is easiest to give results for a specific $d$. Therefore, in the remainder of this section, we turn to our main case of interest and work with $d=3$ in the following.

\subsubsection*{The unrenormalized free energy in $d=3$}
Starting with \eqref{eq:SonUV} we insert the near-boundary expansions
for $T$, $W$ and $U$ collected in appendix \ref{app:nearmaximum}. This
gives us $F$ in terms of an expansion in $\f(u)$, which we can turn into an expansion in powers of $\Lambda$ by trading
$\f(u)$ for $e^{A(u)}$ with the help of \eqref{eq:phimsol} and
\eqref{eq:Amsol}. After a lot  of algebra, in $d=3$ we find the
following result as $\Lambda \to +\infty$:
\begin{align}
\label{eq:Fd3cutoff} F^{d=3} (\Lambda,\mathcal{R}) = - (M \ell)^2
\tilde{\Omega}_3 \, \Bigg\{ & \hphantom{+} \ \mathcal{R}^{-3/2} \bigg[
4 \Lambda^3 \Big(1 + \mathcal{O} \big(\Lambda^{-2 \Delta_-} \big)
\Big) + C(\mathcal{R})  \bigg] \\
\nonumber & + \mathcal{R}^{-1/2} \bigg[ \Lambda \Big(1 + \mathcal{O}
\big(\Lambda^{-2 \Delta_-} \big) \Big) + B(\mathcal{R})  \bigg] + \mathcal{O}\Big(\mathcal{R}^{1/2} \Lambda^{-1} \Big)\Bigg\} \, .
\end{align}
We can then make the following observations:
\begin{itemize}
\item There is a leading divergence of the form $\sim \Lambda^3$ and a
  subleading divergence $\sim \Lambda$. Depending on the precise value
  of $\Delta_-$ there may be many further divergent terms. However, all
  UV-divergent terms either come with a curvature factor
  $\mathcal{R}^{-3/2}$ or $\mathcal{R}^{-1/2}$. Moreover, the integration constants $B(\mathcal{R})$ and
  $C(\mathcal{R})$ only contribute finite terms. 
    This observation will be important later. It
  is a manifestation of the well-known fact in holographic
  renormalization that UV divergences are universal and vevs only contribute to finite terms in the
  on-shell action.
\item The most important part of $F$ as far as the $F$-theorem is concerned is what we will refer to as the `universal contribution'. This is the $\Lambda$-independent piece of \eqref{eq:Fd3cutoff} and is given by
\begin{align} \label{Ffinite}
F^{d=3}_{\textrm{univ}} = - (M \ell)^2 \tilde{\Omega}_3 \left( \mathcal{R}^{-3/2} C(\mathcal{R}) + \mathcal{R}^{-1/2}  B(\mathcal{R})  \right) \, .
\end{align}
This depends on the boundary parameter $\mathcal{R}$ and on the (curvature-dependent)
parameters $C(\mathcal{R})$ and $B(\mathcal{R})$, which in turn are related to the following field theoretic
quantities: $C(\mathcal{R})$ is related to a vev in standard quantisation:
\be \label{vevs}
C(\mathcal{R}) = {\Delta_- \over d}\langle \mathcal{O} \rangle
  |\f_-|^{-{\Delta_+\over \Delta_-}} \, ,
\ee
while $B(\mathcal{R})$ computes an appropriately defined entanglement entropy (see sec.~\ref{sec:ent}). In addition, for $\mathcal{R} \rightarrow 0$ one can show (see app.~\ref{app:thermo}) that $B(\mathcal{R})$ is related to a derivative with respect to curvature of the vev $\langle \mathcal{O} \rangle$ as:
\begin{align}
B(\mathcal{R}) \big|_{\mathcal{R} = 0} = 2 \left. \frac{\partial C(\mathcal{R})}{\partial \mathcal{R}} \right|_{\mathcal{R} = 0} = \frac{2 \Delta_-}{d} |\f_-|^{- \frac{\Delta_+-2}{\Delta_-}} \left. \frac{\partial }{\partial R} \langle \mathcal{O} \rangle \right|_{R = 0} \, ,
\end{align}
where the last equality holds for fixed $\f_-$.
\end{itemize}

\subsection{The renormalized free energy}
\label{sec:Sren}
The universal contribution to the free energy written in equation (\ref{Ffinite}) can be affected by finite
local counterterms. Therefore, to obtain the finite part of the free
energy in a systematic and unambiguous way, we need to resort to
holographic renormalization.

In holographic renormalization of a general dilaton-gravity theory, the counterterms can be conveniently
organised in terms of  curvature invariants associated
with the induced metric $\gamma_{\mu \nu}$, multiplied by suitable
functions of the scalar field \cite{1106.4826}. The intrinsic
curvature appears up to a maximum power of $d/2$ (plus logarithmic
contributions associated to anomalies) for $d$ even and $(d-1)/2$ for $d$ odd. For example, the first two counterterms are given by
\begin{align}
\label{eq:Sct1} F^{(0)}_{ct} &= M^{d-1} \int_{\textrm{UV}} d^dx \sqrt{|\gamma|} \, W_{ct} (\f) \, , \\
\label{eq:Sct2} F^{(1)}_{ct} &= M^{d-1} \int_{\textrm{UV}} d^dx \sqrt{|\gamma|} \, R^{(\gamma)} U_{ct} (\f) \, , \\
\nonumber \vdots
\end{align}
and these are all the counterterms needed in $d=3$. The functions $W_{ct}$ and $U_{ct}$ satisfy the equations\footnote{The eq.~\eqref{eq:Wcteq} for $W_{ct}$ is equivalent to the EOM for $W$ in the case $R=0$.}
\begin{align}
\label{eq:Wcteq} \frac{d}{4 (d-1)} W_{ct}^2 - \frac{1}{2} {\big( W_{ct}' \big)}^2 & = - V \, , \\
\label{eq:Ucteq} W_{ct}' \, U_{ct}' - \frac{d-2}{2 (d-1)} \, W_{ct} \, U_{ct} &=-1 \, .
\end{align}
These are equivalent to the ``flat'' superpotential equation with
$T=0$ and to (a rescaled version of) the
``flat'' $U$-equation (\ref{eq:Uequation0}) with $S(\f)$ replaced by
$W'$. Therefore they track the flat space holographic RG flow.

As we will be exclusively interested in the case $d=3$ in this work, the two counterterms \eqref{eq:Sct1} and \eqref{eq:Sct2} are sufficient and we hence refrain from giving explicit expressions of counterterms at higher orders in $R^{(\gamma)}$, but they can be found in \cite{1106.4826}.

Equations \eqref{eq:Wcteq} and \eqref{eq:Ucteq} determine the functions $W_{ct}$ and $U_{ct}$ up to two integration constants, which we call $C_{ct}$ and $B_{ct}$, respectively and which we can choose at will. A particular choice of these constants corresponds to a choice of renormalization scheme. It will also be useful to record the expansion of the functions
$W_{ct}$ and $U_{ct}$ in the vicinity of the UV boundary. In
particular, close to a UV fixed point at $\f=0$ the functions $W_{ct}$ and $U_{ct}$ can be expanded in powers of $\f$ as follows, \cite{Myers,Kraus,HaroSkenderisSolodukhin}:
\begin{align}
W_{0,ct}(\f) &= \frac{2(d-1)}{\ell}+\frac{\Delta_- }{2\ell}\f^2+ \frac{C_{ct}}{\ell} |\f |^{d/\Delta_-}+\ldots , \label{W0ct} \\
U_{ct}(\f) &= \frac{\ell}{d-2} + \ell \, B_{ct} \, |\f |^{(d-2)/\Delta_-}+\ldots , \label{Uct}
\end{align}
where $C_{ct}$ and $B_{ct}$ now appear explicitly.

The renormalized free energy is then given by the free energy \eqref{eq:SonUV} with all necessary counterterms added. For $d=3$ this gives
\begin{align}
\label{eq:Sonren} F^{d=3, \textrm{ren}} (\mathcal{R}| B_{ct}, C_{ct}) = \lim_{\Lambda \to \infty} \left[
F^{d=3}(\Lambda, \mathcal{R}) + F_{ct}^{(0)} + F_{ct}^{(1)}  \right]\, .
\end{align}
where we explicitly emphasised the fact that the dependence on $\Lambda$ has been traded with
a dependence on counterterms.

\subsubsection*{Renormalized free energy for $d=3$}
As a first step, it will be convenient to rewrite the counterterms \eqref{eq:Sct1} and \eqref{eq:Sct2} as follows. Using the fact that
\begin{align}
R^{(\gamma)} =T \, , \qquad \textrm{and}  \qquad \sqrt{-\gamma}= R^{d/2} T^{-d/2} \, \sqrt{-\zeta} \, ,
\end{align}
the counterterms become
\begin{align}
\label{eq:Sct1b} F^{(0)}_{ct} &=  M^{d-1} \tilde{\Omega}_d \, {\big[ T^{-\frac{d}{2}} (\f) \, W_{ct} (\f) \big]}_{\textrm{UV}} \, , \\
\label{eq:Sct2b} F^{(1)}_{ct} &=  M^{d-1} \tilde{\Omega}_d \, {\big[ T^{-\frac{d}{2}+1}(\f) \, U_{ct} (\f) \big]}_{\textrm{UV}} \, ,
\end{align}
where $\tilde{\Omega}_d$ is defined in \eqref{eq:Volsphere}. Using this we find the following expression for the renormalized free energy in $d=3$:
\begin{align}
\label{eq:Fren1} F^{d=3, \textrm{ren}} = - M_p^{2} \tilde{\Omega}_3 \left({\big[ T^{- \frac{3}{2}}(\f) \big( W(\f) - W_{ct}(\f) \big) \big]}_{\textrm{UV}} +  {\big[T^{-\frac{1}{2}}(\f) \big( U(\f) - U_{ct}(\f) \big) \big]}_{\textrm{UV}} \right) \, .
\end{align}
Inserting the expressions for the UV expansions for $T$, $W$, $W_{ct}$, $U$ and $U_{ct}$ from \eqref{eq:Tnearmax}, \eqref{eq:Wmsol}, \eqref{W0ct}, \eqref{eq:Uuv}, \eqref{Uct} we finally arrive at
\begin{align}
\label{eq:Fren2} F^{d=3, \textrm{ren}} (\mathcal{R}| B_{ct}, C_{ct})= - (M \ell)^2 \tilde{\Omega}_3 \Big[ \mathcal{R}^{-3/2} \big( C(\mathcal{R}) - C_{ct}  \big) + \mathcal{R}^{-1/2} \big( B(\mathcal{R}) - B_{ct}  \big) \Big] \, .
\end{align}
This now shows the explicit dependence of the free energy on the two renormalization constants $C_{ct}$ and $B_{ct}$.

\subsection{Expressions at small and large curvature}
\label{sec:smalllargeR}
We begin by collecting the results from the previous sections. There we presented expressions for the free energy on $S^3$. In particular, we have the following expressions for the cutoff-regulated and renormalized quantities:
\begin{align}
\label{eq:Fsummary1} & F (\Lambda,\mathcal{R})  = - (M \ell)^2 \tilde{\Omega}_3 \, \Bigg\{ \hphantom{+} \ \mathcal{R}^{-3/2} \bigg[ 4 \Lambda^3 \Big(1 + \mathcal{O} \big(\Lambda^{-2 \Delta_-} \big) + C(\mathcal{R}) \Big) \bigg] \\
\nonumber & \ \qquad \qquad \qquad \qquad \qquad + \mathcal{R}^{-1/2} \bigg[ \Lambda \Big(1 + \mathcal{O} \big(\Lambda^{-2 \Delta_-} \big) + B(\mathcal{R}) \Big) \bigg] + \mathcal{O} \Big(\mathcal{R}^{1/2} \Lambda^{-1} \Big)\Bigg\}  \, , \\
\label{eq:Fsummary3} &F^{\textrm{ren}} (\mathcal{R}| B_{ct}, C_{ct}) = - (M \ell)^2 \tilde{\Omega}_3 \Big[ \mathcal{R}^{-3/2} \big( C(\mathcal{R}) - C_{ct}  \big) + \mathcal{R}^{-1/2} \big( B(\mathcal{R}) - B_{ct}  \big) \Big] \, ,
\end{align}
where we now suppress the superscript $d=3$ to remove clutter.

Generally, we will have to revert to numerical methods to evaluate these quantities. However, for both $\mathcal{R} \rightarrow \infty$ and $\mathcal{R} \rightarrow 0$ we can make analytical progress. Hence, in this section we will collect analytical results for $F$  for both $\mathcal{R} \rightarrow \infty$ and $\mathcal{R} \rightarrow 0$. The relevant calculations are shown in appendices \ref{app:largeR} and \ref{app:smallR}, respectively. Here we will only show the results which will be most relevant for our later study of the $F$-theorem. In particular, it will be most economical to record the expressions for $C(\mathcal{R})$ and $B(\mathcal{R})$, which we will do in the following.

\subsubsection*{Large curvature results: $\mathcal{R} \rightarrow \infty$}
From the calculations in Appendix \ref{app:largeR} we obtain:
\begin{align}
\label{eq:BClargeR} C(\mathcal{R}) \underset{\mathcal{R} \rightarrow \infty}{=} \mathcal{O} \big(\mathcal{R}^{3/2- \Delta_-} \big) \, , \qquad B(\mathcal{R}) \underset{\mathcal{R} \rightarrow \infty}{=} -8 \pi^2 \tilde{\Omega}_3^{-2} \, \mathcal{R}^{1/2} \left( 1+ \mathcal{O} \big(\mathcal{R}^{-\Delta_-} \big) \right) \, .
\end{align}
Inserting this into equation
\eqref{eq:Fsummary1}--\eqref{eq:Fsummary3} we obtain the
corresponding large-curvature asymptotics for the free energy. To be brief, we give explicit expressions for the renormalized quantities only. Therefore, for $\mathcal{R} \rightarrow \infty$ we obtain:
\begin{align}
\label{eq:FlargeR} F^{\textrm{ren}}  & \underset{\mathcal{R} \rightarrow \infty}{=} \hphantom{-} (M \ell)^2 \left(8 \pi^2 + \tilde{\Omega}_3 B_{ct} \mathcal{R}^{-1/2} + \tilde{\Omega}_3 C_{ct} \mathcal{R}^{-3/2} +\mathcal{O}(\mathcal{R}^{-\Delta_-}) \right) \, .
\end{align}
Therefore, $F^{\textrm{ren}}$ is finite for $\mathcal{R} \rightarrow \infty$ approaching the value $8 \pi^2 (M \ell)^2$.

We identify this value as the free energy (or central charge) of the UV CFT. The reason is as follows. For fixed $R$ taking $\mathcal{R} \rightarrow \infty$ corresponds to the limit of vanishing source, i.e.~$|\f_-| \rightarrow 0$. Hence the value of the (renormalized) free energy obtained for $\mathcal{R} \rightarrow \infty$ can be identified with that of the corresponding CFT. 

The above observation gives rise to the following general result. For a CFT associated with an extremum of the potential  at $\f_{\textrm{CFT}}$ the (renormalized) free energy is given by
\begin{align}
\label{eq:FrenCFTdef} F_{\textrm{CFT}} = 8 \pi^2 (M \ell_{\textrm{CFT}})^2 \, , \qquad \textrm{with} \qquad \ell_{\textrm{CFT}}^2 \equiv - \frac{6}{V(\f_{\textrm{CFT}})} \, .
\end{align}
This is valid regardless whether the extremum is a maximum or a minimum of the potential. Also note that the renomalized value of the free energy of a CFT is unambiguous, i.e.~there is no scheme-dependence.

\subsubsection*{Small curvature expansion: $\mathcal{R} \rightarrow 0$}
From the analysis in appendix \ref{app:smallR} one finds:
\begin{align}
\label{eq:CsmallR} C(\mathcal{R}) & \underset{\mathcal{R} \rightarrow 0}{=} C_0 + C_1 \mathcal{R} + \mathcal{O} \big(\mathcal{R}^{2} \big) + \mathcal{O} \big(\mathcal{R}^{3/2- \Delta_-^{\textrm{IR}}} \big) \, , \\
\label{eq:BsmallR} B(\mathcal{R}) & \underset{\mathcal{R} \rightarrow 0}{=} B_0 \left( 1+ \mathcal{O} \big(\mathcal{R} \big) \right) -8 \pi^2 \tilde{\Omega}_3^{-2} \frac{\ell_{\textrm{IR}}^2}{\ell^2} \, \mathcal{R}^{1/2} \left( 1+ \mathcal{O} \big(\mathcal{R}^{-\Delta_-^{\textrm{IR}}} \big) \right) \, ,
\end{align}
where we have  defined
\begin{align}
\label{eq:LIRdef} \ell_{\textrm{IR}}^2 \equiv -
\frac{d(d-1)}{V(\f_\textrm{IR})} \, , \qquad \Delta_-^{\textrm{IR}} = \frac{1}{2} (d - \sqrt{d^2 + 4 \ell_{\textrm{IR}}^2 V''(\f_{\textrm{IR}})} ) \, ,
\end{align}
where $\f_\textrm{IR}$ is the minimum of the potential. Note that
$\Delta_-^{\textrm{IR}} < 0$ since $V''(\f_{\textrm{IR}}) >0$.

The quantities $C_0$, $B_0$ and  $C_1$  appearing in equations \eqref{eq:CsmallR}--\eqref{eq:BsmallR} are numerical coefficients. The quantity $C_0$ is the flat-space value of the vev $\langle \mathcal{O} \rangle$ and $C_1$ is the
coefficient of the first curvature correction to $\langle \mathcal{O} \rangle$. The coefficient $B_0$ corresponds to the flat-space limit of an appropriately defined entanglement entropy (see sec.~\ref{sec:ent}).\footnote{In fact, for our setup one finds that $B_0=2 C_1$, This follows from a thermodynamic relation between the free energy on $S^3$ and an appropriately defined entanglement entropy and will be explained in sec.~\ref{sec:ent} and app.~\ref{app:thermo}.}

This leads to the following expressions for the renormalized free energy:
\begin{align}
\label{eq:SonsmallR} F^{\textrm{ren}} & \underset{\mathcal{R} \rightarrow 0}{=} - (M \ell)^2 \tilde{\Omega}_3 \big(C_0-C_{ct} \big) \mathcal{R}^{-3/2} - (M \ell)^2 \tilde{\Omega}_3 \big(B_0 + C_1-B_{ct} \big) \mathcal{R}^{-1/2} \\
\nonumber & \qquad + 8 \pi^2 (M \ell_{\textrm{IR}})^2 + \mathcal{O}(\mathcal{R}^{- \Delta_-^{\textrm{IR}}}) + \mathcal{O}(\mathcal{R}^{1/2}) \, .
\end{align}
Note that in general  $F^{\textrm{ren}}$ diverges for $\mathcal{R}
\rightarrow 0$. The leading divergence is of the form
$\mathcal{R}^{-3/2}$ and it can be understood as a volume
divergence. This is the statement that the free energy is an extensive
quantity and grows with the volume of the $S^3$, i.e.~$\textrm{Vol}(S^3) \sim \mathcal{R}^{-3/2}$ for fixed $\f_-$. The same IR divergence also occurs in the unrenormalized quantity $F(\Lambda, \mathcal{R})$.
The coefficient of that divergence is the free-energy density of the flat theory.

In addition to the divergent terms, the expression
\eqref{eq:SonsmallR} also exhibit a finite contribution. From \eqref{eq:FrenCFTdef} we identify this term as the
central charge of the IR CFT associated with the IR fixed point at the
minimum of the potential.  Interestingly, while this central charge is a property of the IR CFT only, here it emerges from the free energy of a holographic RG flow solution from the UV fixed point to this IR.

 These observations suggest that  we may be able  construct
$F$-functions (depending on ${\cal R}$) out of the free energy,  which interpolate between the
central charges of the UV and IR CFTs, if we can isolate a quantity
which generalises the finite contribution also away from the fixed
points.  This is what we will propose in the next section.

\section{Constructing  $F$-functions from the free energy}
\label{sec:Ftheorem}
Having collected all the necessary ingredients, we can finally turn to
constructing  candidate $F$-functions. We start with a definition of
the $F$-theorem. We then explain how the $F$-theorem is related  to
holographic RG flows\footnote{This idea was already explored in
  \cite{Taylor}. As we will see, the key reason that work gave a negative
  result lies in how  infrared divergences are treated}. Finally, using the results from section \ref{sec:collectF} we propose several candidate $F$-functions and check their viability numerically.
\subsection{Definitions and strategy}
\label{sec:Fdef}
The $F$-theorem \cite{1103.1181} is a statement about Lorentz-invariant quantum field theories in $d=3$ and their behaviour under renormalization group flow. In its minimal form, it is concerned with properties of two conformal field theories which are connected by an RG flow. It states that at both the UV and IR fixed points one can define a quantity $F$ such that
\begin{align}
F_{\textrm{UV}} \geq F_{\textrm{IR}} \, .
\end{align}
A refined version demands that there exists a function $\mathcal{F}(\mathcal{R})$, with $\mathcal{R}$ some parameter along the flow, which exhibits the following properties:
\begin{itemize}
\item At the fixed points of the flow, the function $\mathcal{F}(\mathcal{R})$ takes the values $\mathcal{F}_{\textrm{UV}}$ and $\mathcal{F}_{\textrm{IR}}$ respectively.
\item The function $\mathcal{F}(\mathcal{R})$ evolves monotonically along the flow, i.e.~
\begin{align}
\frac{d}{d \mathcal{R}} \mathcal{F} (\mathcal{R}) \geq 0 \, ,
\end{align}
for a parameter $\mathcal{R}$ that decreases monotonically when going from UV to IR.
\item It is also expected that $\mathcal{F}(\mathcal{R})$ should be stationary at the fixed points of the RG flow. This expectation arises from the observation that the Zamolodchikov $c$-function in $d=2$ is stationary at the fixed points. However, as pointed out in \cite{1207.3360}, there exist $F$-function candidates that satisfy the first two requirements (correct values in UV/IR, monotonicity), but violate stationarity. We hence leave it open whether the $F$-function should be stationary at the fixed points and focus on the other two conditions in this work.
\end{itemize}

Here we will explore several candidate $F$-functions in a holographic setting for a field theory on $S^3$. We lay out our strategy in the following:
\begin{enumerate}
\item In the holographic context, the UV theory is the CFT associated with a maximum $\f_{\textrm{max}}$ of the bulk potential $V$. The IR CFT will be associated with a neighbouring minimum $\f_{\textrm{min}}$:
\begin{align}
\nonumber \textrm{CFT}_{\textrm{UV}} \quad  \longleftrightarrow \quad & \f(u) = \f_{\textrm{UV}} = \f_{\textrm{max}} = const. \, , \\
\nonumber \textrm{CFT}_{\textrm{IR}} \quad \longleftrightarrow \quad & \f(u) = \f_{\textrm{IR}} = \f_{\textrm{min}} = const. \, .
\end{align}
\item We identify the $F$-quantity with the renormalized free energy of the corresponding CFT. This can be calculated as follows. Note that at an extremum of $V$ the bulk geometry is AdS$_{4}$. The scale factor is then given by
\be
A_{\textrm{UV}}(u) = \ln \left(- \frac{\ell_{\textrm{UV}}}{\alpha} \sinh \frac{u+c_{\textrm{UV}}}{\ell_{\textrm{UV}}} \right) \, , \quad \textrm{with} \quad \ell_{\textrm{UV}}^2 = -\frac{6}{V(\f_{\textrm{UV}})}
\ee
for the UV CFT and associated AdS space, and
\be
A_{\textrm{IR}}(u) = \ln \left(- \frac{\ell_{\textrm{IR}}}{\alpha} \sinh \frac{u+c_{\textrm{IR}}}{\ell_{\textrm{IR}}} \right) \, , \quad \, \, \, \textrm{with} \quad \ell_{\textrm{IR}}^2 = -\frac{6}{V(\f_{\textrm{IR}})} \, .
\ee
for the IR CFT and its associated AdS space.

The UV and IR values of the $F$-quantity are then given by the
renormalized action evaluated on these solutions. From \eqref{eq:FrenCFTdef} it follows that this is given by:\footnote{Note that for a CFT the value of the on-shell action is uniquely defined, i.e.~there is no scheme-dependence. Different schemes (in $d=3$) correspond to theories with different coefficients for the finite counterterms $\sim \int d^3 x \, \sqrt{\zeta} \, |\f_-|^{3 / \Delta_-}$ and $\sim \int d^3 x \, \sqrt{\zeta} \, R \, |\f_-|^{1 / \Delta_-}$. However, for a CFT $\f_-=0$ and no such finite counterterms exist.}
\begin{align}
F_{\textrm{UV}} = 8 \pi^2 (M \ell_{\textrm{UV}})^2 \, , \qquad F_{\textrm{IR}} = 8 \pi^2 (M \ell_{\textrm{IR}})^2 \, .
\end{align}
As $\ell_{\textrm{UV}} > \ell_{\textrm{IR}}$ these indeed satisfy $F_{\textrm{UV}} \geq F_{\textrm{IR}}$.
\item It is the dimensionless curvature $\mathcal{R}$ which will play the role of the parameter along the flow, which we will also refer to as curvature-RG flow to distinguish it from the holographic RG flow in $u$. The candidate $F$-functions will be specific functionals for a given holographic flow solution $A(u), \f(u)$. To be precise, we will consider $F$-functions constructed out of the (renormalized) free energy introduced in section \ref{sec:collectF}. As pointed out there, given a flow solution $A(u), \f(u)$ with UV data $R, \f_-$ these functionals only depend on the UV sources via the dimensionless combination $\mathcal{R}$. As a result, the candidate $F$-functions we will consider will only depend on UV sources via $\mathcal{R}$, i.e.
\begin{align}
\mathcal{F}(\mathcal{R}) \equiv \mathcal{F}\left[A_{R, \f_-} (u) , \, \f_{R, \f_-} (u) \right] \, .
\end{align}
\item In terms of curvature-RG flow in $\mathcal{R}$, the notion of UV and IR are defined as follows. By UV we refer to the limit  $\mathcal{R} \rightarrow \infty$. The corresponding UV value for the $F$-function is the functional $\mathcal{F}$ evaluated on a holographic RG flow solution $A(u)$, $\f(u)$ with end point $\f_0 \rightarrow \f_{\textrm{UV}}$. Moving away from the UV corresponds to letting $\mathcal{R}$ decrease, which is equivalent to evaluating $\mathcal{F}$ over solutions $A(u), \f(u)$ with end points that successively move away from $\f_{\textrm{UV}}$. The IR is defined as $\mathcal{R} \rightarrow 0$. The related solutions $A(u), \f(u)$ exhibit $\f_0 \rightarrow \f_{\textrm{IR}}$, i.e.~the end point approaches a minimum of the potential, and the corresponding value $\mathcal{F}(\mathcal{R})$ is the functional evaluated on this solution.\footnote{Note that this notion of UV and IR in terms of $\mathcal{R} \rightarrow \infty$ and $\mathcal{R} \rightarrow 0$ differs from the usual definition of UV fixed point and IR end point for a single holographic flow solution. For a single holographic RG flow the terms UV and IR refer to the limits $(u \rightarrow - \infty, \, \f \rightarrow \f_{\textrm{UV}})$ and $(u \rightarrow u_0, \, \f \rightarrow \f_{0})$, respectively.}
\end{enumerate}
The upshot is: To calculate $\mathcal{F}(\mathcal{R})$ along the RG flow defined by $\mathcal{R}$ we have to evaluate the functional $\mathcal{F}$ over a family of holographic RG flow solutions whose IR end points $\f_0$ move successively from $\f_{\textrm{UV}}$ to $\f_{\textrm{IR}}$.

\subsection{Candidate $F$-functions}
\label{sec:Ffunc}
In this section, we will propose several suitable candidate $F$-functions. The building blocks will be both the renormalized and unrenormalized free energy, with the relevant expressions collected in \eqref{eq:Fsummary1}--\eqref{eq:Fsummary3}. Our candidate $F$-functions will be constructed to satisfy two criteria:
\begin{enumerate}
\item Any candidate $F$-function has to be free of both UV and IR divergences.
\item In the UV $(\mathcal{R} \rightarrow \infty)$ and the IR $(\mathcal{R} \rightarrow 0)$ the $F$-functions should reproduce the free energy of the corresponding UV and IR CFTs, i.e.
\begin{align}
\mathcal{F} (\mathcal{R}) & \underset{\mathcal{R} \rightarrow \infty}{\longrightarrow} F_{\textrm{UV}} = 8 \pi^2 (M \ell_{\textrm{UV}})^2 \, , \\
\mathcal{F} (\mathcal{R}) & \underset{\mathcal{R} \rightarrow 0}{\longrightarrow} \ \, F_{\textrm{IR}} \, = 8 \pi^2 (M \ell_{\textrm{IR}})^2 \, .
\end{align}
\end{enumerate}

We begin by examining the various divergent pieces present in the free
energy. The unrenormalized free energy given in \eqref{eq:Fsummary1} exhibits UV-divergent terms, which in terms of the dimensionless cutoff $\Lambda$ defined in \eqref{eq:dimlesscutoff} take the following schematic form:
\begin{align}
\nonumber \Lambda\textrm{-dependent terms in } F(\Lambda, \mathcal{R}): \quad \sim \mathcal{R}^{-1/2} (\Lambda + \ldots)  \quad \textrm{and} \quad \sim \mathcal{R}^{-3/2} (\Lambda^3 + \ldots) \, .
\end{align}
In addition, the unrenormalized free energy contains terms which do
not depend on $\Lambda$, but diverge when $\mathcal{R} \rightarrow
0$. These are IR divergences associated with a diverging volume of
$S^3$.  The relevant terms take the following schematic form (see sec.~\ref{sec:smalllargeR} for details):
\begin{align}
\nonumber \textrm{IR-divergent terms in } F(\Lambda, \mathcal{R}): \quad \sim \mathcal{R}^{-1/2} (B_0 + C_1) \big|_{\mathcal{R} \rightarrow 0} \quad \textrm{and} \quad \sim \mathcal{R}^{-3/2} C_0 \big|_{\mathcal{R} \rightarrow 0} \,  ,
\end{align}
with $C_0$, $B_0$ and $C_1$ numerical constants.

Alternatively, we can work with the renormalized free energy $F^{\textrm{ren}}$.
While this is free of UV-divergences,
$F^{\textrm{ren}}$ still exhibits IR divergent terms. In this case, these are schematically given by (see \eqref{eq:SonsmallR}):
\begin{align}
\nonumber \textrm{IR-divergent terms in} & \\
\nonumber F^{\textrm{ren}}(\mathcal{R}| B_{ct}, C_{ct}): \quad & \sim
\mathcal{R}^{-1/2} (B_0 + C_1- B_{ct}) \big|_{\mathcal{R} \rightarrow
  0} \quad \textrm{and} \quad \sim \mathcal{R}^{-3/2} (C_0 -C_{ct})
\big|_{\mathcal{R} \rightarrow 0} \, . 
\end{align}

As one can observe from the above equations, both UV-divergent as well as IR-divergent
terms come with the same functional dependence on $\mathcal{R}$,
i.e.~the problematic terms come with a factor $\mathcal{R}^{-3/2}$
and/or $\mathcal{R}^{-1/2}$.  A similar observation, regarding the
entanglement entropy,  has also been made for theories in flat
space-time in \cite{1202.2070}. There it was shown that the divergent
contribution to the entanglement entropy across a scalable surface
(with, say, scale $a$) only come with several distinct powers of that
scale $a$. The same holds in our case for the free energy, with  the scale $a$ given
by the curvature $R$. The main difference is that here the field theory itself is defined on curved space-time with constant scalar curvature $R$.

The challenge for constructing a viable $F$-function now is to isolate the finite contributions to the free energy or the entanglement entropy, i.e.~we need to ensure that both UV-cutoff-dependent terms as well as the explicitly IR-divergent terms do not enter into the $F$-function. There are at least two ways of doing this:
\begin{enumerate}
\item For one, we can remove any contribution with curvature
  dependence $\mathcal{R}^{-3/2}$ and/or $\mathcal{R}^{-1/2}$ by
  acting on $F^{(\textrm{ren})}$ with an appropriate differential
  operator, similarly to what was done in \cite{1202.2070}. In particular, we define
\begin{align}
\label{eq:Ddef} \mathcal{D}_{3/2} \equiv \left(\frac{2}{3} \, \mathcal{R} \, \frac{\partial}{\partial \mathcal{R}} + 1 \right) \, , \quad \textrm{and} \quad \mathcal{D}_{1/2} \equiv \left(2 \, \mathcal{R} \, \frac{\partial}{\partial \mathcal{R}} + 1 \right)  \, .
\end{align}
These satisfy
\begin{align}
\mathcal{D}_{3/2} \mathcal{R}^{-3/2} =0 \, , \qquad \mathcal{D}_{1/2} \mathcal{R}^{-1/2} =0  \, ,
\end{align}
and hence remove the divergent contributions while leaving terms with any other power of $\mathcal{R}$ intact.
\item By working with the renormalized quantity $F^{\textrm{ren}}$ we
  can guarantee the absence of UV-divergences. The observation then is
  that we can remove the remaining IR-divergent pieces by choosing a
  suitable renormalization scheme. This amounts to an appropriate
  specific
  choice of renormalization parameters which we will call $B_{ct,0}$ and $C_{ct,0}$:
\begin{align}
\label{eq:Bct0Cct0def} B_{ct,0} &\equiv B(0) + \left. \tfrac{\partial C(\mathcal{R})}{\partial \mathcal{R}} \right|_{\mathcal{R}=0}= B_0 + C_1 \, , \qquad C_{ct,0} \equiv C(0) = C_0 \, .
\end{align}
In order to make clear that these renormalization conditions are
well-defined in terms of the dual field theory language, we show in appendix \ref{app:renormalizationscheme} that this choice of renormalization conditions is equivalent to requiring that certain correlation functions involving the renormalized stress-tensor $T_{\mu \nu}^{(\textrm{ren})}$ of the boundary theory vanish for $R \rightarrow 0$.
\end{enumerate}

We are now in a position to propose candidate $F$-functions. To be specific, we begin with $F$-functions constructed out of the renormalized free energy $F^{\textrm{ren}}$. This has two IR-divergent terms, one at order $\mathcal{R}^{-3/2}$ and one at order $\mathcal{R}^{-1/2}$. As stated above, each of these terms can be removed in two ways, either by differentiation or with the help of counterterms. This gives four possibilities for removing divergent pieces and we hence define the four divergence-free quantities $\mathcal{F}_{1,2,3,4}(\mathcal{R})$ constructed from $F^{\textrm{ren}}$:
\begin{align}
\label{eq:F1def} \mathcal{F}_1 (\mathcal{R}) & \equiv \mathcal{D}_{1/2} \, \mathcal{D}_{3/2} \, F^{\textrm{ren}} (\mathcal{R}| B_{ct}, C_{ct}) \, , \\
\label{eq:F2def} \mathcal{F}_2 (\mathcal{R}) & \equiv \mathcal{D}_{1/2} \, \hphantom{\mathcal{D}_{3/2}} \, F^{\textrm{ren}} (\mathcal{R}| B_{ct}, C_{ct,0}) \, , \\
\label{eq:F3def} \mathcal{F}_3 (\mathcal{R}) & \equiv \hphantom{\mathcal{D}_{1/2}} \, \mathcal{D}_{3/2} \, F^{\textrm{ren}} (\mathcal{R}| B_{ct,0}, C_{ct}) \, , \\
\label{eq:F4def} \mathcal{F}_4 (\mathcal{R}) & \equiv \hphantom{\mathcal{D}_{1/2} \, \mathcal{D}_{3/2}} \, F^{\textrm{ren}} (\mathcal{R}| B_{ct,0}, C_{ct,0}) \, .
\end{align}
with $\mathcal{D}_{1/2}$, $\mathcal{D}_{3/2}$ defined in \eqref{eq:Ddef} and $B_{ct,0}$, $C_{ct,0}$ given in \eqref{eq:Bct0Cct0def}. As we will check explicitly at the end of this section, for $\mathcal{R} \rightarrow \infty$ and $\mathcal{R} \rightarrow 0$ the functions $\mathcal{F}_{1,2,3,4}$ reduce to $F_{\textrm{UV}}$ and $F_{\textrm{IR}}$, respectively. As a result, they pass the minimum test and are four good candidate $F$-functions.

An important observation is that the functions $\mathcal{F}_{1,2,3,4}$ can be written entirely in terms of $\mathcal{R}$, $B(\mathcal{R})$ and $C(\mathcal{R})$ (as well as the counter-terms). In particular, inserting \eqref{eq:Fsummary3} into \eqref{eq:F1def}--\eqref{eq:F1def} one obtains
\begin{align}
\label{eq:F1def2} \mathcal{F}_1 (\mathcal{R}) & = -(M \ell)^2 \, \tilde{\Omega}_3 \ \frac{4}{3} \mathcal{R}^{1/2} \Big( 2 B' + C'' + \mathcal{R} B''  \Big) \, , \\
\label{eq:F2def2} \mathcal{F}_2 (\mathcal{R}) & = -(M \ell)^2 \, \tilde{\Omega}_3 \ 2 \mathcal{R}^{-3/2} \Big(  - (C - C_{ct,0}) + \mathcal{R} C' + \mathcal{R}^2 B' \Big) \, , \\
\label{eq:F3def2} \mathcal{F}_3 (\mathcal{R}) & = -(M \ell)^2 \, \tilde{\Omega}_3 \ \frac{2}{3} \mathcal{R}^{-1/2} \Big((B + C' -B_{ct,0}) + \mathcal{R} B' \Big) \, , \\
\label{eq:F4def2} \mathcal{F}_4 (\mathcal{R}) & = -(M \ell)^2 \, \tilde{\Omega}_3 \ \mathcal{R}^{-3/2} \Big( (C - C_{ct,0}) + \mathcal{R} (B - B_{ct,0}) \Big) \, ,
\end{align}
where $\tilde{\Omega}_3$ is defined in \eqref{eq:Volsphere}. Here we suppressed the argument of $B(\mathcal{R})$ and $C(\mathcal{R})$ to reduce clutter and ${}^{\prime}$ refers to a derivative with respect to $\mathcal{R}$.

Similarly, we can start with the unrenormalized free energy $F(\Lambda, \mathcal{R})$. This again has divergent terms at order $\mathcal{R}^{-3/2}$ and $\mathcal{R}^{-1/2}$. Here, from  the various options for cancelling these terms, we can only apply the method involving differentiation. However, one can check this does not give rise to a new $F$-function. From \eqref{eq:Fsummary1} one finds
\begin{align}
\label{eq:F1def3} \mathcal{D}_{1/2} \, \mathcal{D}_{3/2} \, F (\Lambda,\mathcal{R}) = -(M \ell)^2 \, \tilde{\Omega}_3 \ \frac{4}{3} \mathcal{R}^{1/2} \Big( 2 B' + C'' + \mathcal{R} B''  \Big) = \mathcal{F}_1(\mathcal{R}) \, .
\end{align}
As a result, the functions $\mathcal{F}_{1,2,3,4}$ defined in \eqref{eq:F1def}--\eqref{eq:F4def} exhaust the possibilities for candidate $F$-functions that can be constructed directly from the free energy on $S^3$ by the procedures mentioned above.

One can show explicitly that all our candidate $F$-functions reduce to $F_{\textrm{UV}}$ and $F_{\textrm{IR}}$ in the UV ($\mathcal{R} \rightarrow \infty$) and IR ($\mathcal{R} \rightarrow 0$), respectively. For example, using the results in sec.~\ref{sec:smalllargeR} we find the following behaviour for $\mathcal{R} \rightarrow \infty$:
\begin{align}
\mathcal{F}_{1} (\mathcal{R}) & \underset{\mathcal{R} \rightarrow
  \infty}{=}  F_{\textrm{UV}} + \mathcal{O}(\mathcal{R}^{-\Delta_-})
\, , \label{eq:F1largeR} \\
\mathcal{F}_{2} (\mathcal{R}) & \underset{\mathcal{R} \rightarrow \infty}{=} F_{\textrm{UV}} + \mathcal{O}(\mathcal{R}^{-\Delta_-}) + (M \ell_{\textrm{UV}})^2 \tilde{\Omega}_3 \, C_{ct,0} \, \mathcal{R}^{-3/2} \, , \label{eq:F2largeR}  \\
\mathcal{F}_{3} (\mathcal{R}) & \underset{\mathcal{R} \rightarrow \infty}{=} F_{\textrm{UV}} + \mathcal{O}(\mathcal{R}^{-\Delta_-})  + (M \ell_{\textrm{UV}})^2 \tilde{\Omega}_3 \, B_{ct,0} \, \mathcal{R}^{-1/2} \, ,  \label{eq:F3largeR} \\
\mathcal{F}_{4} (\mathcal{R}) & \underset{\mathcal{R} \rightarrow
  \infty}{=} F_{\textrm{UV}} + \mathcal{O}(\mathcal{R}^{-\Delta_-}) +
(M \ell_{\textrm{UV}})^2 \tilde{\Omega}_3 \, B_{ct,0} \,
\mathcal{R}^{-1/2} + (M \ell_{\textrm{UV}})^2 \tilde{\Omega}_3 \,
C_{ct,0} \, \mathcal{R}^{-3/2} \, ,  \label{eq:F4largeR}
\end{align}
We can also make another observation. From the above it follows that all our candidate $F$-functions are also stationary in the UV, i.e.~they obey $\partial_{\mathcal{R}} \mathcal{F}_i |_{\mathcal{R} \rightarrow \infty} =0$.

Similarly, using the results in sec.~\ref{sec:smalllargeR} the $F$-functions behave for $\mathcal{R} \rightarrow 0$ as
\begin{align}
\mathcal{F}_{i} (\mathcal{R}) & \underset{\mathcal{R} \rightarrow 0}{=}  F_{\textrm{IR}} + \mathcal{O}(\mathcal{R}^{-\Delta_-^{\textrm{IR}}}) + \mathcal{O}(\mathcal{R}^{1/2}) \, , \qquad i = 1\ldots4\, .
\end{align}

Finally, we need to check whether our functions
$\mathcal{F}_i(\mathcal{R})$ decrease monotonically with RG flow. Only
if this is the case we can declare success and present them as good
$F$-functions. This is a difficult task and an analytic
proof of monotonicity, if possible at all, is beyond the scope of this
paper. In section (\ref{sec:numerical}) we will test our proposal
numerically on some simple but generic examples and show that all the
proposed $F$-functions are indeed monotonic. This lends support to our
proposal, and further tests (and eventually a proof) will be left for
future work.

\subsection{An $F$-function from holographic RG flow in flat space-time}
Before testing our proposal on examples, in this section we will
review a different type of $F$-function which arises from
holographic RG flow in {\em flat} space-time. This was originally
constructed by Liu and Mezei \cite{1202.2070} starting from the
entanglement entropy across  a spherical region  in flat space.

The construction is as follows. We consider a ball  of radius
$\alpha$ in a flat space-time quantum field theory, and compute the
entanglement entropy between the points inside the ball and those outside the ball, which we will denote by $S_{\textrm{FEE}}(\alpha)$
  This quantity is both UV
divergent and, once UV-regulated, has a large-volume divergence as
$\alpha\to \infty$, just like the free energy we have been studying
in this section. Liu and Mezei proposed as an $F$-function the ``Renormalized
entanglement entropy'' (REE), which we will denote by $\mathcal{F}_0$, and which is defined as
 \begin{equation} \label{F0}
\mathcal{F}_0(\mathcal{\alpha})= \left(\alpha {d \over d\alpha} -1 \right)S_{\textrm{FEE}}(\alpha) \ ,
\end{equation}
It was subsequently proven to be a good $F$-function in \cite{1202.5650}, i.e.~it interpolates monotonically between the values of the CFT central charge at the end-points of the flat-space RG flow.

In a field theory with a holographic dual, the entanglement entropy
across a region of space is computed via the Ryu-Takayanagi
prescription \cite{RT}: one picks a $(d-2)$-dimensional surface on the AdS boundary which
coincides with the entangling surface and extends it to a geodesic
$(d-1)$-dimensional surface in the bulk. The entanglement entropy is then computed
in
terms of the minimal surface area ${\cal A}$ by
\begin{equation}
\label{eq:SFEEdef} S_{\textrm{FEE}}(\alpha)= {\cal  A}/4G_{d+1} \ ,
\end{equation}
where $G_{d+1}$ is Newton's constant in $(d+1)$ dimensions.

\begin{figure}[t]
\centering
\begin{overpic}
[width=0.65\textwidth]{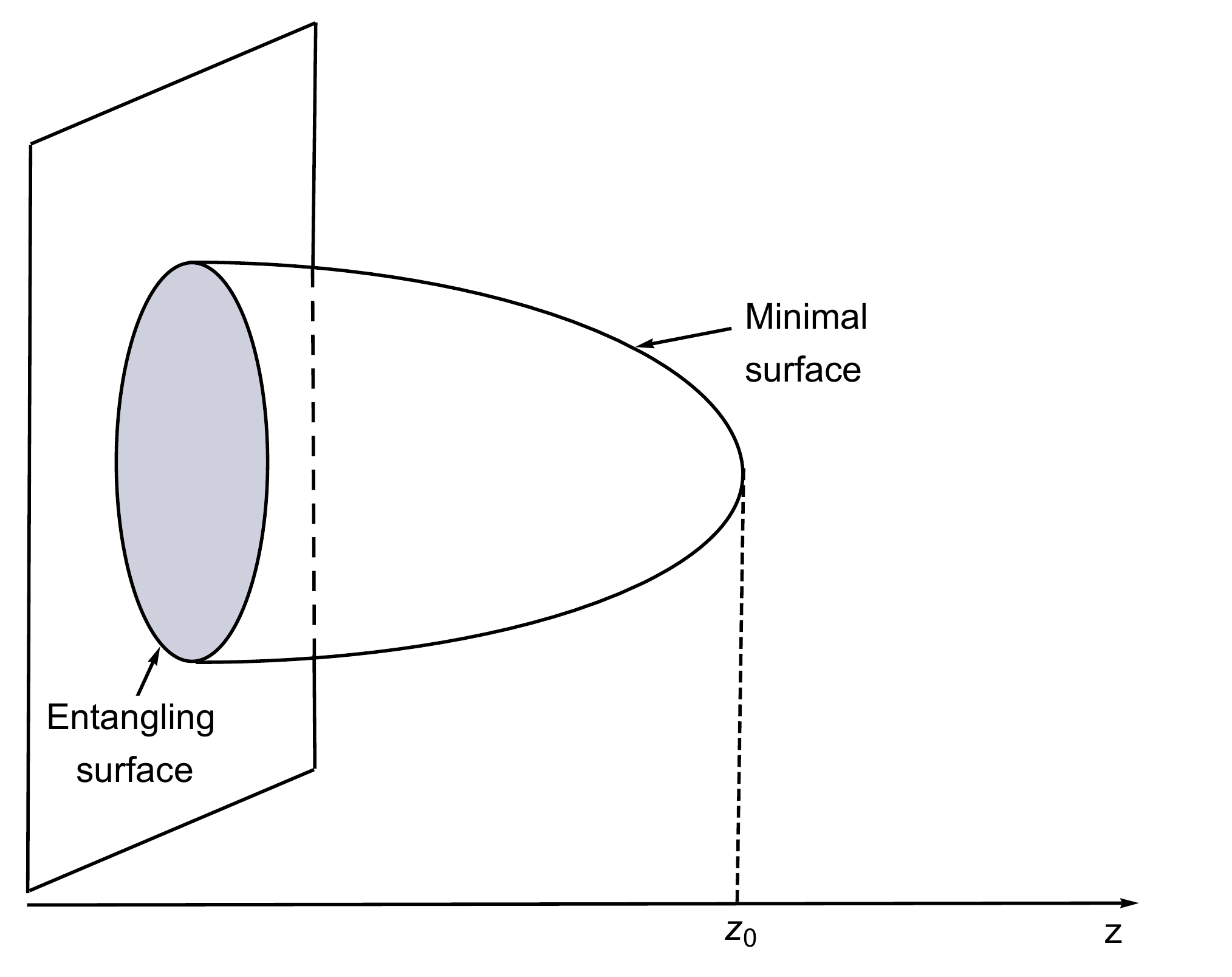}
\end{overpic}
\caption{Cartoon of the minimal surface to compute the entanglement entropy in flat space. The coordinate $z$ is the holographic direction in conformal coordinates. }
\label{flatee}
\end{figure}

In the case we are discussing, for $d=3$, the entangling surface is a
circle of radius $\alpha$, and the situation is described
schematically in Figure \ref{flatee}. The details of the calculation  are given in Appendix
\ref{FEE}. Notice that for this calculation we consider the vacuum, flat-space
RG flow theory, whose metric is (here it is convenient to use conformal
coordinates),
\be
ds^2 = \rho (z)\left(dz^2 + \eta_{\mu\nu}dx^\mu dx^\nu\right) \, .
\ee
where $\rho(z)$ is related to the scale factor introduced in \eqref{eq:metric} as $\rho(z)=e^{A\left(u (z)\right)}$.

To calculate the minimal surface $\mathcal{A}$ we need to choose a solution $\rho(z)$ relevant to the problem at hand. Here this is $\rho(z)=e^{A_{\textrm{flat}}\left(u (z)\right)}$ where $A_{\textrm{flat}}(u)$ is a holographic RG flow solution for a field theory in flat space-time. The minimal surface area is then calculated as
\begin{equation}
\mathcal{A}= 2\pi \int_{\epsilon}^{z_0} dz \ a^2(z)
r(z)\sqrt{1+(r'(z))^2},
\end{equation}
where $r(z)$ describes the embedding of the Ryu-Takayanagi minimal
surface, and $z_0$ is the point where the latter smoothly
caps-off. The geodesic equation and regularity conditions for $r(z)$
are described in appendix \ref{FEE}. The entanglement entropy $S_{\textrm{FEE}}$ then follows from \eqref{eq:SFEEdef}.

As in the case of the on-shell action studied in sec.~\ref{sec:collectF} we find that in holography the entanglement entropy $S_{\textrm{FEE}}$ is a function of the dimensionless combination $\mathcal{R} = R |\f_-|^{-2 / \Delta_-}$ introduced in eq.~\eqref{eq:dimlessRdef}. Then the REE can be written as
\begin{equation} \label{F0-ii}
\mathcal{F}_0(\mathcal{R})=-\mathcal{D}_{1/2} S_{\textrm{FEE}}(\mathcal{R}) \ .
\end{equation}
where $\mathcal{D}_{1/2}$ is the differential operator defined in
eq.~\eqref{eq:Ddef}.

As with the other $F$-functions, the REE of Liu and Mezei \eqref{F0-ii} can be computed numerically in specific
examples. A numerical comparison
between  our candidate $F$-functions and $\mathcal{F}_0$
will be performed in the next section.

\subsection{Numerical tests of monotonicity}
\label{sec:numerical}
To test our proposal, in this section we will evaluate the candidate $F$-functions $\mathcal{F}_i(\mathcal{R})$ with $i=1,2, \ldots 4$ for a set of example RG flows. In our holographic setting this amounts to choosing a dilaton potential. Here we will work with the potential \eqref{eq:pot}, which we reproduce below for $d=3$:
\begin{align}
\label{eq:potrepeat} V(\f) = - \frac{6}{\ell_{\textrm{UV}}^2} - \frac{\Delta_- (3-\Delta_-)}{2 \ell_{\textrm{UV}}^2} \, \f^2 + \frac{\lambda}{\ell_{\textrm{UV}}^2} \, \f^4 \, .
\end{align}
This potential has a maximum at $\f_{\textrm{UV}} =0$ and a minimum at $\f_{\textrm{IR}} = \sqrt{\tfrac{\Delta_- (3-\Delta_-)}{4 \lambda}}$. We can also define a quantity $\ell_{\textrm{IR}}$ as in \eqref{eq:LIRdef} which by definition satisfies $\ell_{\textrm{IR}} < \ell_{\textrm{UV}}$. For definiteness, in the following we will set
\begin{align}
\ell_{\textrm{IR}}^2 = b \, \ell_{\textrm{UV}}^2 \, \quad \textrm{with} \quad b = 0.9 \, ,
\end{align}
which we can do by choosing
\begin{align}
\lambda = \frac{\Delta_-^2 (3-\Delta_-)^2 \, b}{96 (1-b)} \, .
\end{align}
Then, in the following, we will also set $M \ell_{\textrm{UV}}=1$. With this potential we can vary the dimensions of perturbing operators and therefore we can check various different flows.

\begin{figure}[t]
\centering
\begin{overpic}
[width=0.65\textwidth]{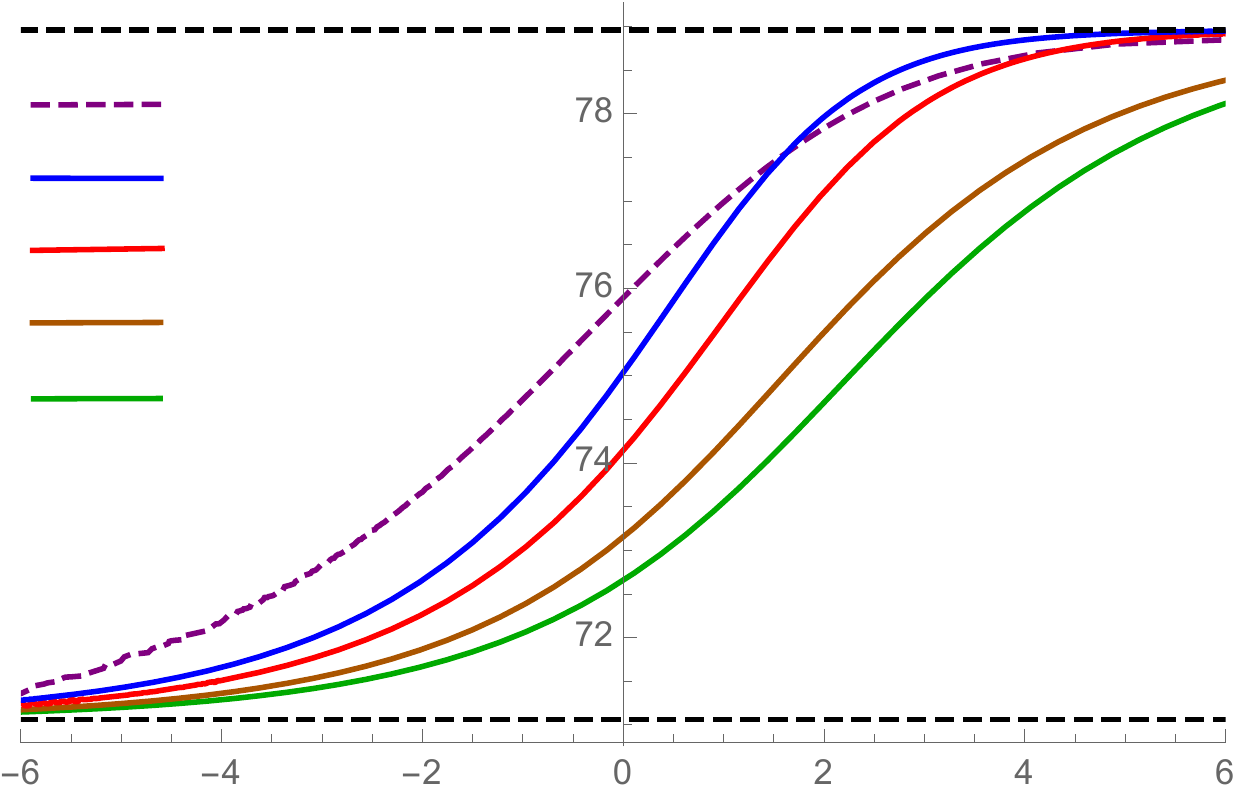}
\put(102,3){$\log(\mathcal{R})$}
\put(52,8){$8 \pi^2 (M \ell_{\textrm{IR}})^2$}
\put(52,64){$8 \pi^2 (M\ell_{\textrm{UV}})^2$}
\put(15,54.2){$\mathcal{F}_0$}
\put(15,48.3){$\mathcal{F}_1$}
\put(15,42.4){$\mathcal{F}_2$}
\put(15,36.5){$\mathcal{F}_3$}
\put(15,30.6){$\mathcal{F}_4$}
\end{overpic}
\caption{$F$-functions
  $\mathcal{F}_{1,2,3,4}$ defined in \protect\eqref{eq:F1def}--\protect\eqref{eq:F4def} and Liu-Mezei $F$-function ${\mathcal F}_0$ (i.e.~the REE of \cite{1202.2070}) vs.~$\log(\mathcal{R})$ for a holographic model with dilaton potential \protect\eqref{eq:potrepeat} and $\Delta_-=1.2$.}
\label{fig:F0to4Delta1p2}
\end{figure}
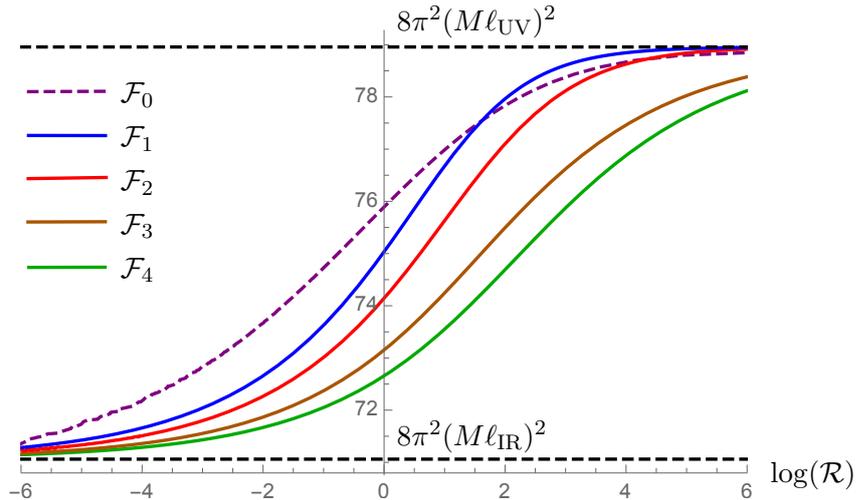

To construct solutions, and compute the corresponding on-shell action, in the
whole range of $\mathcal{R}$ we proceed as follows.  We pick a value $\f_0$ with $\f_{\textrm{max}} < \f_0 < \f_{\textrm{min}}$ and solve equations \eqref{eq:EOM7}, \eqref{eq:EOM8} and \eqref{eq:Uequation0} subject to the boundary conditions \eqref{eq:WIR} and \eqref{eq:UIR} to obtain a solution for the functions $W(\f)$, $S(\f)$ and $U(\f)$. From the near-boundary (i.e.~$\f \rightarrow \f_{\textrm{UV}} =0$) behaviour of $W(\f)$, $S(\f)$ and $U(\f)$ we can then extract the corresponding values of $\mathcal{R}_0$, $C(\mathcal{R}_0)$ and $B(\mathcal{R}_0)$, respectively.\footnote{See \eqref{eq:Wmsol}, \eqref{eq:Smsol} and \eqref{eq:Unearmax} for the near-boundary expansions of $W(\f)$, $S(\f)$ and $U(\f)$, respectively.} By varying the end point $\f_0$ from $\f_{\textrm{UV}}$ to $\f_{\textrm{IR}}$ and repeating the analysis we can hence extract $C(\mathcal{R})$ and $B(\mathcal{R})$ as functions of $\mathcal{R}$. This is summarised schematically  below:
\begin{align}
\nonumber \textrm{Choose a value for } \f_0 \longrightarrow \ & W(\f), \ S(\f),  \ U(\f) \longrightarrow \textrm{ choose new value for }  \f_0 \longrightarrow \ldots \\
\nonumber & \ \ \downarrow \quad \quad \ \ \downarrow \quad \quad \ \downarrow \\
\nonumber & \ \ \mathcal{R}_0 \quad C(\mathcal{R}_0) \ \, B(\mathcal{R}_0)
\end{align}
Once we have $C(\mathcal{R})$ and $B(\mathcal{R})$ as functions of
$\mathcal{R}$ we can determine the counterterms $B_{ct,0}$ and $C_{ct,0}$ from \eqref{eq:Bct0Cct0def}. Finally, using \eqref{eq:F1def2}--\eqref{eq:F4def2} we can compute the functions $\mathcal{F}_i(\mathcal{R})$.

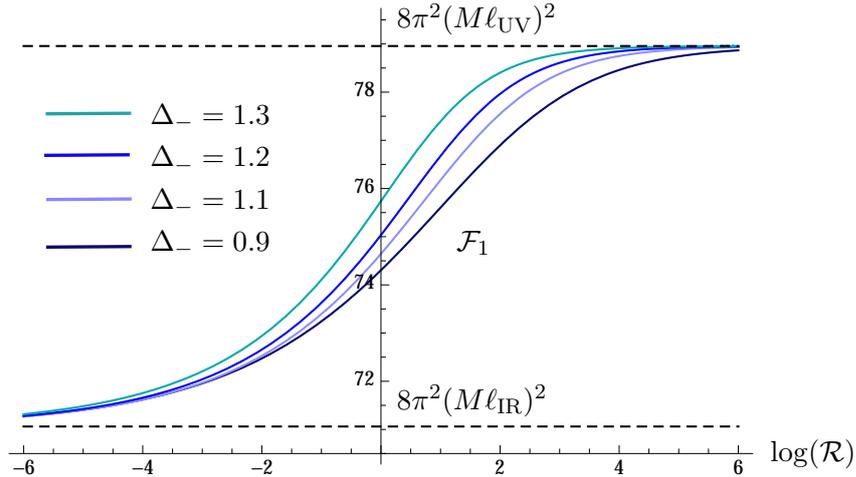
\begin{figure}[t]
\centering
\begin{overpic}
[width=0.65\textwidth]{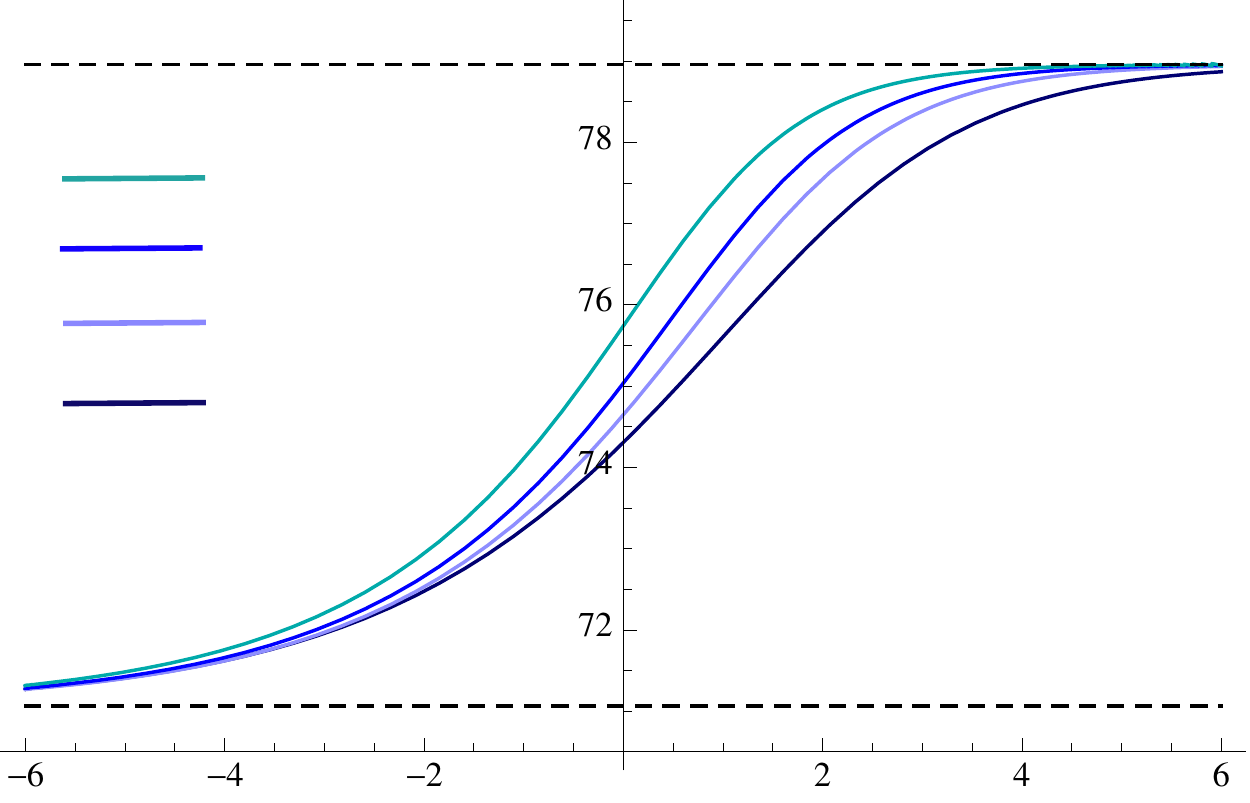}
\put(102,3){$\log(\mathcal{R})$}
\put(52,10){$8 \pi^2 (M \ell_{\textrm{IR}})^2$}
\put(52,61.5){$8 \pi^2 (M \ell_{\textrm{UV}})^2$}
\put(19,48.5){$\Delta_-=1.3$}
\put(19,42.8){$\Delta_-=1.2$}
\put(19,37.1){$\Delta_-=1.1$}
\put(19,31.4){$\Delta_-=0.9$}
\put(60,31.4){$\mathcal{F}_1$}
\end{overpic}
\caption{$\mathcal{F}_{1}$ vs.~$\log(\mathcal{R})$ for a holographic model with dilaton potential \protect\eqref{eq:potrepeat} and $\Delta_-=0.9$ (dark blue), $1.1$, (light blue), $1.2$ (blue) and $1.3$ (cyan).}
\label{fig:F1Delta0p9and1p1and1p2and1p3}
\end{figure}

In figure \ref{fig:F0to4Delta1p2} we plot $\mathcal{F}_{1,2,3,4}$ vs.~$\log(\mathcal{R})$ for the potential \eqref{eq:potrepeat} with $\Delta_-=1.2$. For comparison, we also display the Liu-Mezei $F$-function \cite{1202.2070} labelled by $\mathcal{F}_0(\mathcal{R})$ and given in \eqref{F0-ii} for a flat-space RG flow in the same potential \eqref{eq:potrepeat} with $\Delta_-=1.2$.\footnote{In the case of $\mathcal{F}_0(\mathcal{R})$, the quantity $\mathcal{R}$ refers to the curvature of the spherical entangling surface in units of $\f_-$ and \emph{not} the curvature of the background space-time, which is flat in this case.} We make the following observations.
\begin{itemize}
\item In the UV ($\mathcal{R} \rightarrow \infty$) all four candidate $F$-functions asymptote to $F_{\textrm{UV}}= 8 \pi^2 (M \ell_{\textrm{UV}})^2$ as expected. Similarly, in the IR ($\mathcal{R} \rightarrow 0$) all four candidate $F$-functions approach $F_{\textrm{IR}}= 8 \pi^2 (M \ell_{\textrm{IR}})^2$.
\item Most importantly, all four candidate $F$-functions interpolate monotonically between $F_{\textrm{UV}}$ and $F_{\textrm{IR}}$, that is we observe
\begin{align}
\frac{\partial \mathcal{F}_i(\mathcal{R})}{\partial \mathcal{R}} \geq 0 \, , \quad i =1,2,3,4 \, .
\end{align}
Hence every one of $\mathcal{F}_{1,2,3,4}$ is a good $F$-function.
\item We can also understand the qualitative differences between the
 plots for $\mathcal{F}_{1,2,3,4}$. By equations \eqref{eq:F1largeR}--\eqref{eq:F4largeR}, for $\Delta_-=1.2$ the functions $\mathcal{F}_{1,2}$ behave as $\mathcal{F}_{1,2} = F_{\textrm{UV}} + \mathcal{O}(\mathcal{R}^{-1.2})$ for large $\mathcal{R}$, while $\mathcal{F}_{3,4}$ behave as $\mathcal{F}_{3,4} = F_{\textrm{UV}} + \mathcal{O}(\mathcal{R}^{-0.5})$. As a result, we expect $\mathcal{F}_{3,4}$ to fall off faster than $\mathcal{F}_{1,2}$ when decreasing $\mathcal{R}$. This is exactly what we observe in fig.~\ref{fig:F0to4Delta1p2}.
\item The REE of Liu and Mezei $\mathcal{F}_0(\mathcal{R})$ is also a good $F$-function, interpolating monotonically between $F_{\textrm{UV}}$ and $F_{\textrm{IR}}$. Note, however, that it does not coincide with any of our $F$-functions $\mathcal{F}_{1,2,3,4}(\mathcal{R})$, which are manifestly different as functions of $\mathcal{R}$. In other words, the formulation of the $F$-function in $d=3$ in terms of a flat-space entanglement entropy across a ball with radius $\alpha$ and in terms of the free energy on $S^3$ with radius $\alpha$ differ as functions of $\alpha$.
\end{itemize}

These observations persist beyond the example with $\Delta_-=1.2$. In figure \ref{fig:F1Delta0p9and1p1and1p2and1p3} we plot $\mathcal{F}_{1}$ vs.~$\log(\mathcal{R})$ for a holographic model with dilaton potential \eqref{eq:potrepeat} and $\Delta_-=0.9$, $1.1$, $1.2$ $1.3$. In all cases $\mathcal{F}_{1}(\mathcal{R})$ is a good $F$-function, interpolating monotonically between $F_{\textrm{UV}}$ and $F_{\textrm{IR}}$. The same is true for $\mathcal{F}_{2,3,4}$, but we refrain from plotting the results explicitly.

\subsection{Alternative quantisation and the effective potential as an $F$-function}
\label{sec:altquant}
So far we have worked with holographic theories in what we referred to
as `standard quantisation' in section \ref{sec:holoRGreview}. That is,
we identified the dimension $\Delta$ of the operator $\mathcal{O}$
perturbing the UV CFT with $\Delta_+$. By doing so we restricted our
analysis to (in $d=3$)
\begin{align}
\frac{3}{2} < \Delta < 3 \, .
\end{align}
For such theories we found four potentially good $F$-functions $\mathcal{F}_{1,2,3,4} (\mathcal{R})$, which can be constructed from the free energy on $S^3$. The parameter along the RG flow is $\mathcal{R}$, which is the value of the curvature $R$ of the background space-time of the field theory in units of the operator source $\f_-$.

The question then arises, how the $F$-theorem is realised for theories
perturbed by an operator $\mathcal{O}$ with dimension $\Delta <
3/2$. In particular, how can one define good $F$-functions for such
theories? We can answer this question by switching to `alternative
quantisation'. This amounts to identifying the dimension $\Delta$ of
$\mathcal{O}$ with $\Delta_-$. In general dimension $d$,  this
possibility exists in the range $d/2 -1 < \Delta_- < d/2$. In this
range,  using the identification $\Delta=\Delta_-$ we can cover the region
\begin{align}
\frac{1}{2} < \Delta < \frac{3}{2} \, .
\end{align}

The main point to note is that by swapping the scheme of quantisation
none of the calculations and expressions we obtained so far  are changed in any way. All that changes is the interpretation of the various expressions. As we will argue presently, the functions $\mathcal{F}_{1,2,3,4} (\mathcal{R})$ defined in \eqref{eq:F1def}--\eqref{eq:F4def} will still be good $F$-functions for $\Delta < 3/2$. However, the interpretation in terms of field theory quantities will change when swapping the quantisation scheme.

For one, in alternative quantisation $\f_-$ is identified with the vev of $\mathcal{O}$ and hence $\mathcal{R} = R \, |\f_-|^{-2 / \Delta_-}$ is now the \emph{boundary curvature in units of the operator vev}. It is this quantity which is now the parameter describing the RG flow.

Secondly, note that while in \eqref{eq:F1def}--\eqref{eq:F4def} the
functions $\mathcal{F}_{1,2,3,4} (\mathcal{R})$ were constructed from
what we referred to as  the free energy $F^{(\textrm{ren})}$, this
language was tacitly  assuming standard quantisation (as we have
specified at the beginning of section \ref{sec:collectF}). In fact,
the functions $\mathcal{F}_{1,2,3,4} (\mathcal{R})$ can also be
defined in the more general terms of the Euclidean on-shell action,
\begin{align}
\label{eq:F1fromS} \mathcal{F}_1 (R) & \equiv \mathcal{D}_{1/2} \, \mathcal{D}_{3/2} \, S_{\textrm{on-shell},E}^{\textrm{ren}} (\mathcal{R} | B_{ct}, C_{ct}) \, , \\
\label{eq:F2fromS} \mathcal{F}_2 (R) & \equiv \mathcal{D}_{1/2} \, \hphantom{\mathcal{D}_{3/2}} \, S_{\textrm{on-shell},E}^{\textrm{ren}} (\mathcal{R} | B_{ct}, C_{ct,0}) \, , \\
\label{eq:F3fromS} \mathcal{F}_3 (R) & \equiv \hphantom{\mathcal{D}_{1/2}} \, \mathcal{D}_{3/2} \, S_{\textrm{on-shell},E}^{\textrm{ren}} (\mathcal{R} | B_{ct,0}, C_{ct}) \, , \\
\label{eq:F4fromS} \mathcal{F}_4 (R) & \equiv \hphantom{\mathcal{D}_{1/2} \, \mathcal{D}_{3/2}} \, S_{\textrm{on-shell},E}^{\textrm{ren}} (\mathcal{R} | B_{ct,0}, C_{ct,0}) \, .
\end{align}
This  does not require  a distinction between standard and
alternative quantisation.

In standard quantisation one has $S_{\textrm{on-shell},E}^{(\textrm{ren})} = F^{(\textrm{ren})}$ and \eqref{eq:F1fromS}--\eqref{eq:F4fromS} reduce to \eqref{eq:F1def}--\eqref{eq:F4def}.

In alternative quantisation the on-shell action corresponds to the \emph{quantum effective potential} $\Gamma^{\textrm{ren}}$, i.e.~the Legendre transform of the free energy. To illustrate this, we momentarily separate $\mathcal{R}$ into $R$ and $\f_-$. Then, in alternative quantisation, one has
\begin{align}
S_{\textrm{on-shell},E}^{\textrm{ren}} (R, \f_-) &= \Gamma^{\textrm{ren}} (R, \f_-) \, ,
\end{align}
with $\Gamma^{\textrm{ren}}$ a function of $R$ and the vev $\f_-$. The free energy is denoted by $F^{\textrm{ren}}(R, j)$, where $j$ is the source. This is then related to $\Gamma^{\textrm{ren}}(R , \f_-)$ as
\begin{align}
\Gamma^{\textrm{ren}}(R, \f_- ) &= F^{\textrm{ren}} \big(R , j(\f_-) \big)  - \int d^3x \sqrt{\gamma^{(0)}} \, j(\f_-) \, \f_- \, ,
\end{align}
where $j(\f_-)$ is defined by
\begin{align}
\left. \frac{\delta F (R , j)}{\delta j}\right|_{j(\f_-)} - \f_- =0 \, .
\end{align}

The key observation is that the functions defined in
\eqref{eq:F1fromS}--\eqref{eq:F4fromS} are good
$F$-functions (in the examples we considered in the previous section), regardless what quantisation is chosen. In
standard quantisation they correspond to $F$-functions for
theories where the perturbing operator has dimension $\tfrac{3}{2} <
\Delta < 3$. Using alternative quantisation, the same expressions now give $F$-functions for theories where the perturbing operator has dimension $\tfrac{1}{2} < \Delta < \tfrac{3}{2}$.

This for example implies that the $F$-functions for $\Delta_- =1.2$ plotted in fig~\ref{fig:F1to4Delta1p2} have two interpretations. Using standard quantisation, they can be understood as $F$-functions for a theory with $\Delta = 3 - \Delta_-=1.8$ with the parameter $\mathcal{R}$ the curvature in units of the source. Using alternative quantisation they become $F$-functions for a theory with $\Delta = \Delta_-=1.2$ with $\mathcal{R}$ the curvature in units of the vev.

There is another interesting consequence of the observations in this section. Our results imply that, depending on the operator dimension $\Delta$, we have to construct $F$-functions differently in terms of field-theoretic quantities. In particular, our findings suggest that for $\Delta > 3/2$ it is the free energy $F$ that acts as a $F$-functions while for $\Delta < 3/2$ it is the quantum effective potential $\Gamma$ that should be used. In section \ref{sec:freeboson} we will find that this indeed solves long-standing puzzles regarding the $F$-theorem for the free massive boson.

\section{De Sitter entanglement entropy and the $F$-theorem}
\label{sec:ent}
In this section we make the connection between the quantities
introduced so far (namely the various versions of the UV-finite free
energy) and the entanglement entropy across a spherical surface in de Sitter
space. The latter quantity has been discussed earlier in field theoretical
context \cite{1504.00913} as well as in holography
\cite{1102.0440,Maldacena:2012xp}. Here, both the free energy and the entanglement entropy are determined as functionals of corresponding holographic RG flows for theories in curved space-time. In this setting we will observe that the de
Sitter entanglement entropy corresponds to one of the contributions to the free
energy. In the de Sitter static patch, this relation translates
into the relation between free energy and the {\em thermodynamic}
entropy, computed by the area law. This is very interesting as it suggests that standard QFT on a fixed de Sitter background, and non-dynamical gravity satisfy thermodynamics equations that relate the on-shell action to the entanglement entropy.

As it was observed in \cite{1504.00913},  starting from the
(unrenormalized)  entanglement entropy  and performing a similar
subtraction as the one proposed by Liu and Mezei in flat space, one
can in principle obtain new candidate $F$-functions. We will show that
 the resulting $F$-functions are already part of  the set we
defined in section \ref{sec:Ffunc}.

\subsection{Entanglement entropy for a spherical  surface in
de Sitter space}
\label{sec:ent1}
Here we derive an expression for the entanglement entropy across a
spherically symmetric surface for a theory in dS$_d$. While we will be
once more mainly interested in expressions for $d=3$, we will work in
general $d$ when possible. In a holographic setup, this amounts to
computing the holographic entanglement entropy across a
$(d-2)$-dimensional spherical surface on the dS$_d$ boundary of our $(d+1)$-dimensional space-time.

To this end, consider our metric ansatz \eqref{eq:metric} for a field theory on dS$_d$ with the following choice of (global) coordinates on dS$_d$:
\begin{equation}
\label{eq:metricdS} ds^2=du^2+e^{2 A(u)}\left[-dt^2+\alpha^2 \cosh^2(t/\alpha) \left(d\theta^2 +\sin^2\theta d\Omega^{2}_{d-2}  \right) \right]
\end{equation}
where $d\Omega^{2}_{d-2} $ is the metric on a $(d-2)$-dimensional unit sphere.
In global coordinates the de Sitter metric describes a $S^{d-1}$ that starts infinite in the infinite past, decreases size until a minimum size of the order of the de Sitter curvature, and then increases again and becomes of infinite size in the infinite future.

We now wish to calculate the static entanglement entropy for an entangling
surface on the boundary QFT given by $\theta|_{u \rightarrow - \infty} =
\tfrac{\pi}{2}$ and $t=0$. This splits the spatial $S^{d-1}$ into the two hemispheres that touch at the entangling surface that is the equator (which is a $S^{d-2}$).

To calculate this in our holographic setting, we use the prescription of Ryu and Takayanagi \cite{RT}. We hence need to find the minimal surface in the bulk which has the entangling surface as the boundary. The entanglement entropy is then given by
\begin{align}
S_{\textrm{EE}}=\frac{\gamma}{4 G_{d+1}}
\end{align}
where $\gamma$ is the area of the minimal surface whose boundary is
the entangling surface at $u\to -\infty$,  and $G_{d+1}$ is Newton's constant in $(d+1)$ dimensions.

\begin{figure}[t]
\centering
\begin{overpic}
[scale=.65]{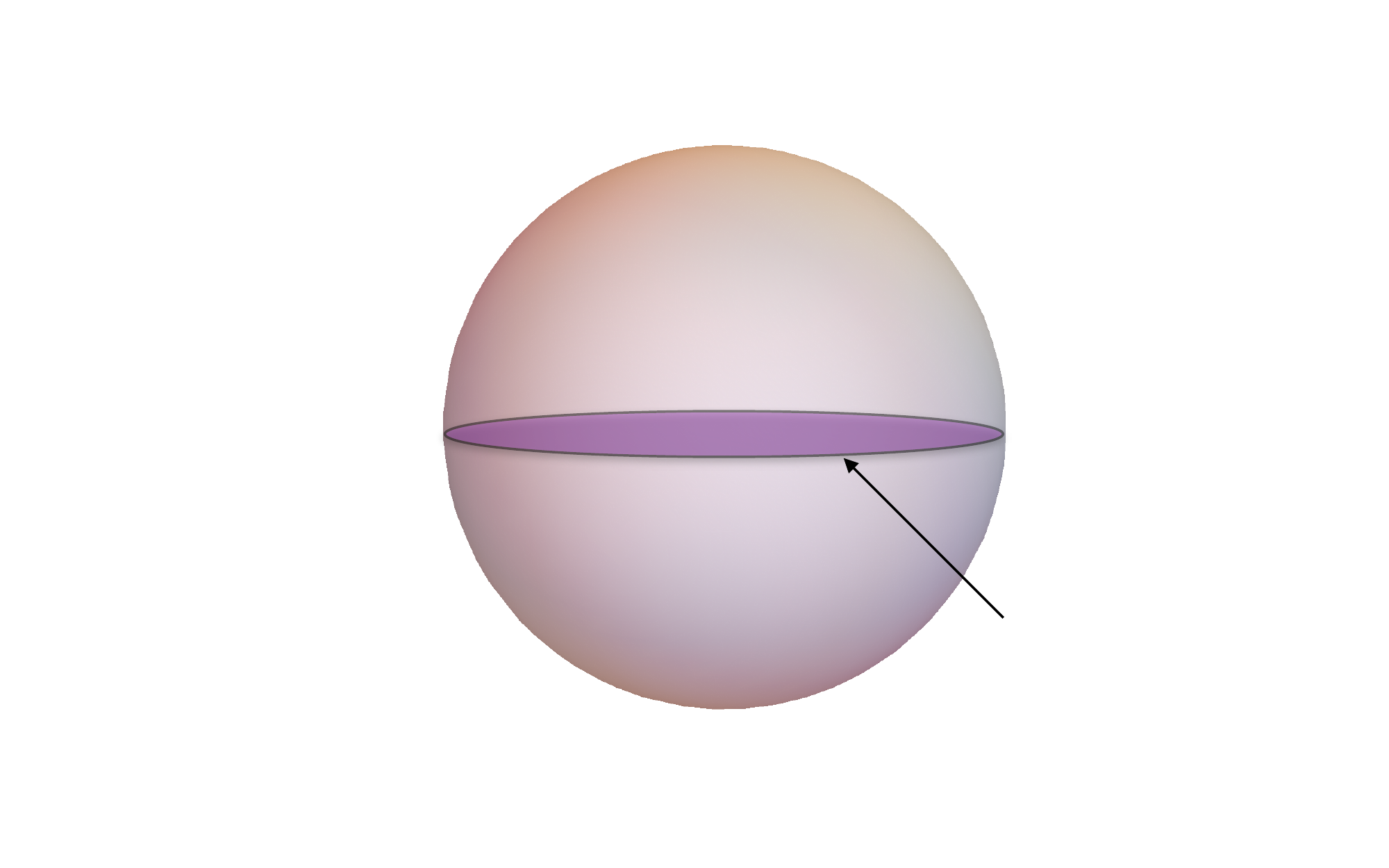}
\put(67,14){Entangling surface: $S^{d-2}$}
\put(45,30.5){\small{minimal surface}}
\end{overpic}
\caption{Sketch of the minimal surface which has the spherical surface
  $\theta=\pi/2$ as its boundary.}
\label{fig:EE}
\end{figure}

In  the Euclidean signature of the calculation, the
$t=0$ slice is the $S^{d-1}$ corresponding to the  equator of the
$S^{d}$ slice. The entangling surface  is then an $S^{d-2}$ at fixed
$\theta=\pi/2$, dividing the $S^{d-1}$ in two halves. One can show (see app.~\ref{app:SEEcalculation}) that the minimal Ryu-Takayanagi surface is described by the curve $\theta(u) = \pi/2$. The geometric
setup is  shown in figure \ref{fig:EE} (for $d=3$ and fixed $t$).

The detailed calculation is presented  in appendix
\ref{app:SEEcalculation} and here we just quote the result, which can
also be found in \cite{Maldacena:2012xp}:
\begin{equation}
\label{eq:SEE}
S_{\textrm{EE}}= M^{d-1} \frac{2 R}{d} V_d \int_{\textrm{UV}}^{\textrm{IR}} du \, e^{(d-2)A(u)} \, ,
\end{equation}
where $V_d$ is the volume of the $d$-dimensional sphere of radius
$\alpha$.

We can now make the following observation. The expression \eqref{eq:SEE} for the entanglement entropy is identical to the 2nd term of the free energy \eqref{eq:Son}. As a result, we find that here the entanglement entropy $S_{\textrm{EE}}$ is related to the free energy $F$ on $S^d$ as
\begin{align}
\label{eq:SEE_Frelation} S_{\textrm{EE}} = 2(d-1) M_p^{d-1} V_d \, {\big[e^{dA} \dot{A} \big]}_{\textrm{UV}} - F \, .
\end{align}
We will exploit this connection frequently in the following. We can also show that, just as $F$, the entanglement entropy $S_{\textrm{EE}}$ only depends on the UV curvature $R$ only through the dimensionless combination $\mathcal{R}$.

\subsection{Thermal interpretation}
\label{sec:ent2}
Equation (\ref{eq:SEE_Frelation}) is suggestive that there should be a thermal
interpretation to the entanglement entropy, and that the first term on
the right hand side should have the interpretation of an internal
energy, to reproduce a  relation of the form\footnote{Recall that our ``free
  energy'' is related to the usual thermodynamic free energy by a
  factor $\beta$, i.e. $F = \beta F_{\textrm{th}}$. See footnote 12.}
\be \label{thermo1}
S = \beta U_{\textrm{th}} - \beta F_{\textrm{th}}
\ee
for some appropriate definition of the inverse temperature $\beta$ and
internal energy $U_{\textrm{th}}$.

In fact, as shown in \cite{1102.0440}, the
thermal interpretation becomes manifest  if we go to the static patch
of the de Sitter slice, by a coordinate transformation which does not
involve the radial coordinate. Writing the $d$-dimensional de Sitter
slices in  static coordinates,  the bulk metric reads
\be \label{thermo2}
ds^2 =du^2+e^{2A(u)}\left[-\left(1-\frac{r^2}{\alpha^2}\right)d\tau^2 +\left(1-\frac{r^2}{\alpha^2}\right)^{-1}dr^2+r^2 d\Omega^2_{d-2}  \right] \ .
\ee
where $\alpha$ is the de Sitter radius and $0< r < \alpha$. The details and the explicit coordinate
transformation is given in Appendix \ref{app:ADM}.

The metric (\ref{thermo2}) has a horizon at $r=\alpha$, whose associated
temperature is
\be \label{thermo3}
T = {1\over 2\pi \alpha}
\ee
and whose  entropy, computed via the horizon area,
coincides with the expression for $S_{\textrm{EE}}$ in equation
(\ref{eq:SEE}),
\be\label{thermo4}
S_{\textrm{th}} \equiv {\textrm{Area} \over 4G_{d+1}} =  S_{\textrm{EE}} \, ,
\ee
as shown in appendix \ref{app:ADM}.

The final ingredient is the identification of $U_{\textrm{th}}$ in equation
(\ref{thermo1}): for static metrics, this corresponds to the ADM mass of
the solution.  A simple computation (see again appendix \ref{app:ADM})
shows that
\be\label{thermo5}
\beta U_{\textrm{th}} =  2 (d-1)M_P^{d-1} \left[e^{dA(u)}\dot{A}(u) \right]_{\textrm{UV}}V_d.
\ee

Using the identification \eqref{thermo4}--\eqref{thermo5}, equation
(\ref{eq:SEE_Frelation}) takes the first-law form (\ref{thermo1}).
From this relation, as shown in appendix \ref{app:thermo},  one can derive an identity
relating the scalar-vev and  curvature-vev parameters  $C({\mathcal R})$ and
$B({\mathcal R})$,
\be \label{thermo9}
 C'(\mathcal{R}) = \frac{1}{2} B(\mathcal{R}) - \mathcal{R} B'(\mathcal{R}) \, ,
 \ee
 with $\prime$ denoting a derivative with respect to
 $\mathcal{R}$. In the limit $R\to 0$ of this equation one obtains
 a relation between the leading term in $B$ and the first curvature
 correction to $C$ in the small $R$ limit, as defined in equations
 (\ref{eq:CsmallR}-\ref{eq:BsmallR}): 
\be
C_1 = {1\over 2} B_0
\ee

\subsection{Renormalized entanglement entropies and associated $F$-functions}
The entanglement entropy given in \eqref{eq:SEE} is UV-divergent and
we hence regulate it. Subsequently, we also define a renormalized entanglement entropy by adding appropriate counterterms.

The procedure parallels the one used in the case of the free
energy, using the dimensionless cutoff $\Lambda$ defined in equation  \eqref{eq:dimlesscutoff}. The entanglement entropy can then be written as a function of $\Lambda$ and $\mathcal{R}$. To be specific, we now again restrict to $d=3$ to find in the limit $\Lambda \rightarrow +\infty$:
\begin{align}
\label{eq:SEEd3cutoff} S_{\textrm{EE}}^{d=3} (\Lambda,\mathcal{R}) = - (M \ell)^2 \tilde{\Omega}_3 \, \Bigg\{ \mathcal{R}^{-1/2} \bigg[ \frac{2}{3} \Lambda \Big(1 + \mathcal{O} \big(\Lambda^{-2 \Delta_-} \big) + B(R) \Big) \bigg] + \mathcal{O} \Big( \mathcal{R}^{1/2} \Lambda^{-1} \Big) \Bigg\} \, ,
\end{align}
where $\tilde{\Omega}_3$ was defined in \eqref{eq:Volsphere}.
The function $B(R)$ is the same that appeared before in expressions for the free energy. Note that here all divergent terms have the same curvature dependence, i.e.~they are accompanied by a factor $\mathcal{R}^{-1/2}$.

We can now also define a renormalized entanglement entropy by adding
appropriate counterterms to \eqref{eq:SEEd3cutoff} and taking the
limit $\Lambda \rightarrow +\infty$. Note that the quantity we  define
this way differs from and should not be confused with the `Renormalized Entanglement Entropy' (REE) of Liu and Mezei defined in \cite{1202.2070}. Here we will proceed as in the case of the free energy and add appropriate counterterms.

Counterterms for the entanglement entropy should be defined in terms of an integral over the entangling surface or a related surface (see e.g.~\cite{1604.06808}). In particular, on the boundary, our entangling `surface' is given by $S^{d-2}$ with radius $\alpha$. However, using the relation \eqref{eq:SEE_Frelation}, note that for our setup the entanglement entropy \eqref{eq:SEE} can be written in terms of quantities which are proportional to the volume of $S^d$ (with radius $\alpha$) rather than the volume of the entangling surface $S^{d-2}$. As a result, we will be able to write down counterterms as integrals over $S^d$. These expressions will however only hold for the setup discussed here, but this is all that we need.

Given the relation \eqref{eq:SEE_Frelation} between $S_{\textrm{EE}}$ and the free energy, the analysis is similar to the one performed in section \ref{sec:Sren} for the free energy. We will hence be brief and give the main results. As a first step we rewrite the integral appearing in \eqref{eq:SEE} in terms of the functions $T(\f)$ and $U(\f)$, where $U(\f)$ was introduced before in \eqref{eq:Uequation0}:
\begin{align}
S_{\textrm{EE}} &= M^{d-1}  \frac{2 R}{d} \textrm{Vol}(S^d) \int_{\textrm{UV}}^{\textrm{IR}} du \, e^{(d-2)A(u)} \, \\
\nonumber &= M^{d-1} \tilde{\Omega}_d \, {\big[ T^{-\frac{d}{2}+1}(\f) \, U (\f) \big]}_{\textrm{UV}} \, ,
\end{align}
and with $\tilde{\Omega}_d$ defined in \eqref{eq:Volsphere}. Then, as in the case of the free energy, we can cancel divergences with the help of an appropriately defined function $\tilde{U}_{ct}(\f)$. In $d=3$ this will remove all the divergences. Hence, in $d=3$ we can write the renormalized entanglement entropy as follows:
\begin{align}
\label{eq:SEEren1} S_{\textrm{EE}}^{d=3, \textrm{ren}} = M_p^{2} \tilde{\Omega}_3 \, {\big[T^{-\frac{1}{2}}(\f) \big( U(\f) - \tilde{U}_{ct}(\f) \big) \big]}_{\textrm{UV}} \, ,
\end{align}
where $\tilde{U}_{ct}(\f)$ has to satisfy
\begin{align}
W_{ct}' \, \tilde{U}_{ct}' - \frac{d-2}{2 (d-1)} \, W_{ct} \, \tilde{U}_{ct} = -\frac{2}{d} \, ,
\end{align}
and $W_{ct}$ a solution to \eqref{eq:Wcteq}.
The function $\tilde{U}_{ct}(\f)$ will contain an integration constant which we will denote by $\tilde{B}_{ct}$. Fixing a value for $\tilde{B}_{ct}$ is equivalent to choosing a renormalization scheme.

Finally, inserting the near-boundary expansion for $T$, $U$ and $\tilde{U}_{ct}$ this becomes\footnote{Note that $\tilde{U}_{ct}$ satisfies the same equation as $U_{ct}$, \eqref{eq:Ucteq}, rescaled by a constant factor $2/d$. As a result, the near-boundary expansion of $\tilde{U}_{ct}$  will be given by the rescaled expression for $U_{ct}$ in \eqref{Uct}, but with $B_{ct}$ replaced by $\tilde{B}_{ct}$.}
\begin{align}
\label{eq:SEEren2} S_{\textrm{EE}}^{d=3, \textrm{ren}} (\mathcal{R} | \tilde{B}_{ct})= (M \ell)^2 \tilde{\Omega}_3 \, \mathcal{R}^{-1/2} \big( B(\mathcal{R}) - \tilde{B}_{ct}  \big) \, ,
\end{align}
where we also indicated that $S_{\textrm{EE}}^{d=3, \textrm{ren}}$ is a function of $\mathcal{R}$ that further depends on our choice for the parameter $\tilde{B}_{ct}$. In the following, we will work exclusively in $d=3$. Thus, the superscript $d=3$ on $S_{\textrm{EE}}^{d=3, \textrm{ren}}$ is henceforth obsolete and will be dropped to remove clutter.

As for the renormalized free energy, we can obtain large-curvature and
small curvature asymptotics for the renormalized entanglement
entropy (see appendix \ref{sec:largesmallR} for details):
\begin{align}
\label{eq:SEElargeR} S_{\textrm{EE}}^{\textrm{ren}}  & \underset{\mathcal{R}
  \rightarrow \infty}{=} - (M \ell)^2 \left(8 \pi^2 + \tilde{\Omega}_3
  \tilde{B}_{ct} \mathcal{R}^{-1/2}  +
  \mathcal{O}(\mathcal{R}^{-\Delta_-}) \right) \,
\\
& \nonumber \\
\label{eq:SEEsmallR} S_{\textrm{EE}}^{\textrm{ren}} & \underset{\mathcal{R} \rightarrow 0}{=} - (M \ell)^2 \tilde{\Omega}_3 \big(B_0 -\tilde{B}_{ct} \big) \mathcal{R}^{-1/2} - 8 \pi^2 (M \ell_{\textrm{IR}})^2 \left(1 + \mathcal{O}(\mathcal{R}^{- \Delta_-^{\textrm{IR}}}) \right) + \mathcal{O}(\mathcal{R}^{1/2}) \, ,
\end{align}

 The leading IR (i.e.~small-$\mathcal{R}$) divergence in
 $S_{\textrm{EE}}^{\textrm{ren}}$ scales as
 $\mathcal{R}^{-1/2}$. The reason is that the entanglement entropy
 scales with the volume of the entangling surface, which is given by
 $\textrm{Vol}(S^1) \sim \mathcal{R}^{-1/2}$ for fixed $\f_-$. Hence, the
 divergence of $S_{\textrm{EE}}^{\textrm{ren}}$ can also be understood as a
 volume-divergence, but this time of the entangling surface. Note that
 the same IR divergence also occurs in the unrenormalized quantity
 $S_{\textrm{EE}}(\Lambda, \mathcal{R})$.

The finite term for $\mathcal{R}\to 0$ is the same for
$F^{\textrm{ren}}$ and $S_{\textrm{EE}}^{\textrm{ren}}$ up to an overall sign
and is given by $\pm 8 \pi^2 (M \ell_{\textrm{IR}})^2$. The same is
true for  the UV limit as $\mathcal{R} \to +\infty$. This suggests
  that one can also construct candidate $F$-functions starting from the
  entanglement entropy, once the IR divergence is
  eliminated. Similarly to the case of the free energy discussed in
  section (\ref{sec:Ffunc})  we can either use the derivative
  operator ${\cal D}_{1/2}$ defined in \eqref{eq:Ddef} (since only the $\mathcal{R}^{-1/2}$ appears)
  acting on equation (\ref{eq:SEEren2}), or choose an appropriate
  scheme such that the first term in the IR expansion
  (\ref{eq:SEEsmallR}) vanishes. This gives rise to the following two candidate $F$-functions:

\begin{align}
\label{eq:F5def} \mathcal{F}_{5} (\mathcal{R}) & \equiv
-\mathcal{D}_{1/2} \ S_{EE}^{\textrm{ren}} (\mathcal{R}|
\tilde{B}_{ct}) = - \mathcal{D}_{1/2} \, S_{\textrm{EE}} (\Lambda, \mathcal{R})
\, , \\
& \nonumber \\
\label{eq:F6def} \mathcal{F}_{6} (\mathcal{R}) & \equiv  \hphantom{\mathcal{D}_{1/2}} - S_{\textrm{EE}}^{\textrm{ren}} (\mathcal{R}| \tilde{B}_{ct,0}) \, ,
\end{align}
where
\be \label{eq:Btildect0def}
\tilde{B}_{ct,0}  \equiv B(0)= B_0.
\ee
Note that $\mathcal{F}_5$ is the analogue of Liu and Mezei's
`Renormalized Entanglement Entropy' (REE) defined in \cite{1202.2070},
but for a theory defined on dS$_3$. It can either be defined in terms
of the renormalized or the unrenormalized entanglement entropy. As one can check from \eqref{eq:SEElargeR} and \eqref{eq:SEEsmallR}, both $\mathcal{F}_{5}$ and
$\mathcal{F}_{6}$ reduce to $F_{\textrm{UV}}$ and $F_{\textrm{IR}}$ in
the UV and IR, respectively. From equation \eqref{eq:SEEren2} it follows that we can write $\mathcal{F}_{5,6}$ in terms of $B(\mathcal{R})$ as follows:
\begin{align}
\label{eq:F5def2} \mathcal{F}_{5} (\mathcal{R}) & = -(M \ell)^2 \, \tilde{\Omega}_3 \ 2 \, \mathcal{R}^{1/2} \, B'(R) \, , \\
\label{eq:F6def2} \mathcal{F}_{6} (\mathcal{R}) & = -(M \ell)^2 \, \tilde{\Omega}_3 \ \mathcal{R}^{-1/2} \big( B(R) - \tilde{B}_{ct,0} \big) \, ,
\end{align}
where again ${}^{\prime}$ denotes an $\mathcal{R}$-derivative.

As it  turns out, equations (\ref{eq:F5def2}) and   (\ref{eq:F6def2})
do not give rise to new $F$-functions compared to those defined from the
free energy. Rather, as we will show below
\be \label{eq:F56areF13}
\mathcal{F}_5 (\mathcal{R}) \equiv \mathcal{F}_1 (\mathcal{R}), \quad  \mathcal{F}_6 (\mathcal{R}) \equiv \mathcal{F}_3 (\mathcal{R}),
\ee
where $\mathcal{F}_1 $ and $\mathcal{F}_3$ are defined in equations
(\ref{eq:F1def}) and (\ref{eq:F3def}).

The relations (\ref{eq:F56areF13})  follow from the
thermodynamic relation discussed in the previous subsection,
\begin{align}
\label{eq:thermorelmaintext} S_{\textrm{EE}} = \beta U_{\textrm{th}}  - \beta F_{\textrm{th}}
\end{align}
if $\beta U_{\textrm{th}}$ is identified with the space-time
integral of the boundary stress tensor.
As we show in detail in appendix \ref{app:thermo}, the relation \eqref{eq:thermorelmaintext} is equivalent to the following two identities
\begin{align}
\label{eq:SEEfromF} \mathcal{D}_{3/2} \, F(\Lambda, \mathcal{R}) &= -S_{EE} (\Lambda, \mathcal{R}) \, , \\
\label{eq:SEEfromFren} \mathcal{D}_{3/2} \, F^{\textrm{ren}}(\mathcal{R} | B_{ct}, C_{ct}) &= -S_{EE}^{\textrm{ren}}\big(\mathcal{R} \big| \tfrac{2}{3} \, B_{ct} \big) \, ,
\end{align}
together with the relation
\begin{align}
\label{eq:BintermsofBtilde} B_{ct,0} = \frac{3}{2} \tilde{B}_{ct,0} \, ,
\end{align}
between the counterterm parameters $B_{ct,0}$ and $\tilde{B}_{ct,0}$.

These in turn imply the identities \eqref{eq:F56areF13}. For example, starting with \eqref{eq:F5def} and using \eqref{eq:SEEfromF} and \eqref{eq:F1def3} one finds
\begin{align}
\mathcal{F}_5 (\mathcal{R}) \overset{\eqref{eq:F5def}}{=} - \mathcal{D}_{1/2} \, S_{\textrm{EE}} (\Lambda, \mathcal{R}) \overset{\eqref{eq:SEEfromF}}{=} \mathcal{D}_{1/2} \, \mathcal{D}_{3/2} \, F(\Lambda, \mathcal{R}) \overset{\eqref{eq:F1def3}}{=} \mathcal{F}_1(\mathcal{R}) \, .
\end{align}
Similarly, starting with \eqref{eq:F6def} and using \eqref{eq:SEEfromFren}, \eqref{eq:BintermsofBtilde} and \eqref{eq:F3def} one obtains
\begin{align}
\nonumber \mathcal{F}_6 (\mathcal{R}) \overset{\eqref{eq:F6def}}{=} - S_{\textrm{EE}}^{\textrm{ren}} (\mathcal{R}| \tilde{B}_{ct,0}) & \overset{\eqref{eq:SEEfromFren}}{=} \mathcal{D}_{3/2} \, F^{\textrm{ren}}(\mathcal{R} | \tfrac{3}{2} \tilde{B}_{ct,0}, C_{ct}) \\
& \overset{\eqref{eq:BintermsofBtilde}}{=} \mathcal{D}_{3/2} \, F^{\textrm{ren}}(\mathcal{R} | B_{ct,0}, C_{ct}) \overset{\eqref{eq:F3def}}{=} \mathcal{F}_3(\mathcal{R}) \, .
\end{align}

\section{Free field theories}
\label{sec:freefields}
Here we will check whether our four proposals for
$F$-functions also work more generally beyond a holographic
setting. Therefore, we turn to free field theories where many results can
be obtained analytically. We find that all our $F$-functions are
monotonic  in $d=3$ for a massive fermion (corresponding to
$\Delta_{\textrm{UV}}=2$) and for a massive scalar ($\Delta_{\textrm{UV}}=1$). In the latter case the UV dimension of the deforming operator $\phi^2$ is less than $d/2$ and therefore, according to our prescription, the
$F$-functions must be constructed from the quantum effective potential,
rather than from the free energy. This leads indeed to
monotonic $F$-functions, contrary to what one has been observed  using
either the free energy on the sphere (see \cite{1105.4598}) or the
entanglement entropy  on dS$_3$ (see \cite{1504.00913}).

\subsection{Free fermion on $S^3$}
\label{sec:freefermion}
Here we  consider the theory of a free massive fermion on $S^3$ with action given by
\begin{align}
\label{eq:fermionaction} S_D= \int d^3 \, \sqrt{\zeta} \, \left[ i \psi^{\dagger} \slashed{D} \psi - i m \psi^{\dagger} \psi \right] \, ,
\end{align}
where again $\zeta_{\mu \nu}$ is a metric on $S^3$ with curvature $R = 6 / \alpha$. This is a conformal theory perturbed by the operator $\psi^{\dagger} \psi$ with source $m$. The dimension of the perturbing operator is therefore
\begin{align}
\Delta [\psi^{\dagger} \psi] =2 > \frac{3}{2} \, .
\end{align}
According to our observations from holography (see sec.~\ref{sec:altquant}), we expect that the free energy can be used for constructing good $F$-functions.

Following \cite{1105.4598} the free energy can be written as
\begin{align}
\label{eq:FD} F_D = - \sum_{n=1}^{\infty} n(n+1) \, \log \left[ \left( n + \frac{1}{2} \right)^2  +(\alpha m)^2 \right] \, ,
\end{align}
which only depends on $m$ and $\alpha$ through the dimensionless combination $(\alpha m)$. To make contact with the notation in the previous sections, here we identify
\begin{align}
\label{eq:Rfermion} \mathcal{R} = (\alpha m)^{-2} \, ,
\end{align}
which is (proportional to) the curvature in units of the source.\footnote{The curvature in units of the source is given by $6 (\alpha m)^{-2}$, which differs from the expression in \eqref{eq:Rfermion} by a factor of 6. By defining $\mathcal{R}$ as in \eqref{eq:Rfermion} we can avoid a proliferation of factors of $\sqrt{6}$ in the following expressions without affecting the monotonicity properties.}

For $m \rightarrow 0$ ($\mathcal{R} \rightarrow \infty$ for fixed $\alpha$) the theory given by \eqref{eq:fermionaction} becomes conformal. In this case the free energy \eqref{eq:FD} was evaluated explicitly in \cite{1105.4598}. Using zeta-function renormalization one finds
\begin{align}
\label{eq:FDUV} F_{D,\textrm{UV}} = \frac{\log 2}{4} + \frac{3 \zeta(3)}{8 \pi^2} \, ,
\end{align}
which we will refer to as the `UV' value.

On the other hand, for $m \rightarrow \infty$ ($\mathcal{R} \rightarrow 0$  for fixed $\alpha$), the theory becomes non-dynamical and empty. We hence expect the corresponding value of the free energy to be
\begin{align}
F_{D,\textrm{IR}} = 0 \, .
\end{align}
A good $F$-function should then interpolate monotonically between the values $F_{D,\textrm{UV}}$ and $F_{D,\textrm{IR}}$ when going from the UV ($\mathcal{R} \rightarrow \infty$) to the IR ($\mathcal{R} \rightarrow 0$).

To this end, we now evaluate the free energy for an arbitrary value of $\mathcal{R}$. We start from the following observation in \cite{1105.4598}. There it was shown that, upon zeta-function renormalization, the free energy satisfies
\begin{align}
\frac{\partial F^{\textrm{ren}}_D}{\partial (\alpha m)^2} = \frac{4 (\alpha m)^2 +1}{\alpha m} \, \pi \, \tanh \big( \pi \alpha m \big) \, .
\end{align}
With the help of this, we can then write the zeta-function renormalized free energy in terms of an integral as
\begin{align}
\label{eq:FDren} F^{\textrm{ren}}_D(\mathcal{R}) = F_{D,\textrm{UV}} + \int_0^{1/\mathcal{R}} d x \, \frac{4 x +1}{\sqrt{x}} \, \pi \, \tanh \big( \pi \sqrt{x} \big) \, .
\end{align}
This expression will be sufficient for both the analytical and numerical evaluations in this section. By construction, $F^{\textrm{ren}}_D(\mathcal{R})$ reduces to $F_{D,\textrm{UV}}$ in the UV, i.e.~for $\mathcal{R} \rightarrow \infty$. However, in the IR ($\mathcal{R} \rightarrow 0$) the expression \eqref{eq:FDren} does not reproduce $F_{D,\textrm{IR}}$, but rather diverges. In particular, one finds
\begin{align}
F^{\textrm{ren}}_D(\mathcal{R}) \underset{\mathcal{R} \rightarrow 0}{=} \frac{\pi}{3} \mathcal{R}^{-3/2} + \frac{\pi}{4} \mathcal{R}^{-1/2} + (\textrm{vanishing for } \mathcal{R} \rightarrow 0) \, .
\end{align}
These can again be understood as volume-divergences, with the leading divergent term proportional to the dimensionless volume $\mathcal{R}^{-3/2}$.

Here we used zeta function renormalization to arrive at a finite expression for $F_D^{\textrm{ren}}$. However, we could have equally renormalized the free energy by adding specific covariant counterterms to the action \eqref{eq:fermionaction}. In this case our expression for $F_D^{\textrm{ren}}$ would be scheme-dependent, with different renormalization schemes parameterised by two real numbers $c_{ct}$ and $b_{ct}$. These parameters are the coefficients of the two \emph{finite} (i.e.~UV-cutoff-independent) counterterms
\begin{align}
\label{eq:finitect} F_{ct, 1}^{\textrm{finite}} &= c_{ct} \int d^3 x \sqrt{\zeta} \, m^3 = 2 \pi^2 \, c_{ct} \, \mathcal{R}^{-3/2} \, , \\
\nonumber F_{ct, 2}^{\textrm{finite}} &= b_{ct} \int d^3 x \sqrt{\zeta} \, R m = 12 \pi^2 \, b_{ct} \, \mathcal{R}^{-1/2} \, .
\end{align}
As we show in appendix \ref{app:zetafunctionrenorm},\footnote{In app.~\ref{app:zetafunctionrenorm}, we work with the theory of a free massive scalar on $S^3$. However, our findings can be easily generalised for the case of the Dirac fermion.} zeta-function renormalization is equivalent to renormalization via counterterms, with a particular choice of $c_{ct}$ and $b_{ct}$. However, we can always change renormalization scheme by adding terms of the form \eqref{eq:finitect} to $F_D^{\textrm{ren}}$. In the following, we will now exploit this to construct the equivalents of the $F$-functions  discussed in the context of
holography in section \ref{sec:Ftheorem}.

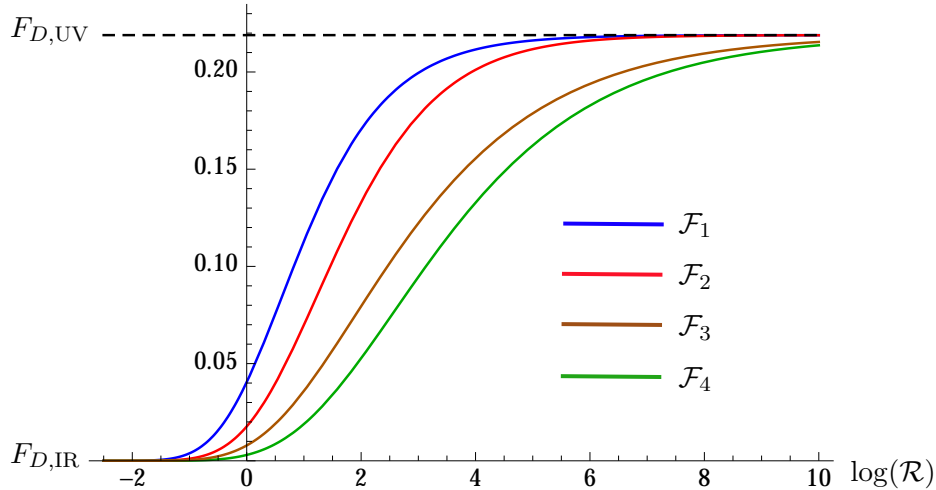
\begin{figure}[t]
\centering
\begin{overpic}
[width=0.65\textwidth]{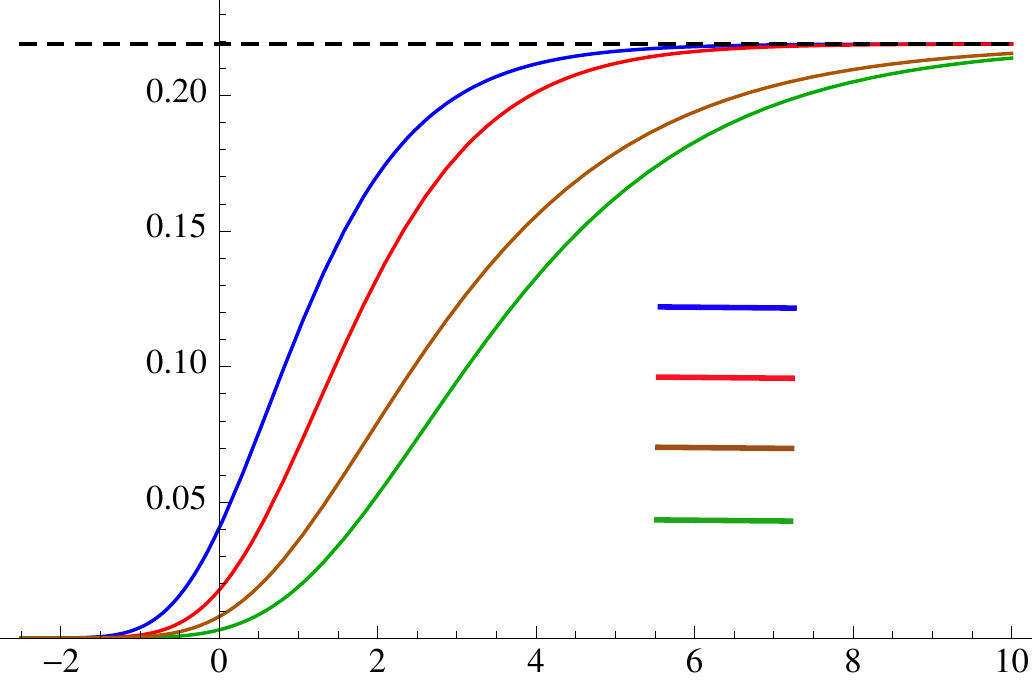}
\put(102,2){$\log(\mathcal{R})$}
\put(-10.5,62){$F_{D,\textrm{UV}}$}
\put(-10.5,4.5){$F_{D,\textrm{IR}}$}
\put(79,35.5){$\mathcal{F}_1$}
\put(79,28.6){$\mathcal{F}_2$}
\put(79,21.7){$\mathcal{F}_3$}
\put(79,14.8){$\mathcal{F}_4$}
\end{overpic}
\caption{$\mathcal{F}_{1,2,3,4}$ given in \protect\eqref{eq:F1fermion}--\protect\eqref{eq:F4fermion} vs.~$\log(\mathcal{R})$ for a theory of a free fermion on $S^3$. The black dashed line indicates the value of $F_{D,\textrm{UV}}$ given in \protect\eqref{eq:FDUV}.}
\label{fig:F1to4fermion}
\end{figure}

As explained in Section \ref{sec:Ffunc}, we can remove the two divergent terms $\sim \mathcal{R}^{-3/2}$ and $\sim \mathcal{R}^{-1/2}$ either with the help of a differential operator, or by subtracting them with the help of counterterms.\footnote{Here this amounts to adding the counterterms \eqref{eq:finitect} with $c_{ct} = - 1/ (6 \pi)$ and $b_{ct} = - 1/ (48 \pi)$.} In analogy with the functions $\mathcal{F}_{1,2,3,4}$ in \eqref{eq:F1def}--\eqref{eq:F4def}, here we can define four $F$-functions as follows:\footnote{The function $\mathcal{F}_4$ given in \eqref{eq:F4fermion} has already been confirmed to be a good $F$-function for the free fermion in \cite{1105.4598}.}
\begin{align}
\label{eq:F1fermion} \mathcal{F}_1 (\mathcal{R}) &= \mathcal{D}_{1/2} \, \mathcal{D}_{3/2} \, \, \, \, F^{\textrm{ren}}_D(\mathcal{R}) \, , \\
\mathcal{F}_2 (\mathcal{R}) &= \mathcal{D}_{1/2} \, \hphantom{\mathcal{D}_{3/2}} \, \Big( F^{\textrm{ren}}_D(\mathcal{R}) - \frac{\pi}{3} \mathcal{R}^{-3/2} \Big) \, , \\
\mathcal{F}_3 (\mathcal{R}) &= \hphantom{\mathcal{D}_{1/2}} \, \mathcal{D}_{3/2} \, \Big( F^{\textrm{ren}}_D(\mathcal{R}) - \frac{\pi}{4} \mathcal{R}^{-1/2} \Big) \, , \\
\label{eq:F4fermion} \mathcal{F}_4 (\mathcal{R}) &= \hphantom{\mathcal{D}_{1/2}} \, \hphantom{\mathcal{D}_{3/2}} \, \Big( F^{\textrm{ren}}_D(\mathcal{R}) - \frac{\pi}{3} \mathcal{R}^{-3/2} - \frac{\pi}{4} \mathcal{R}^{-1/2} \Big) \, ,
\end{align}
with $\mathcal{D}_{1/2}$ and $\mathcal{D}_{3/2}$ given in \eqref{eq:Ddef}. As one can check explicitly, the functions $\mathcal{F}_{1,2,3,4}$ reduce to the values $F_{D, \textrm{UV}}$ and $F_{D, \textrm{IR}}$ for $\mathcal{R} \rightarrow \infty$ and $\mathcal{R} \rightarrow 0$, respectively. In addition, by evaluating them numerically, one finds that all four functions $\mathcal{F}_{1,2,3,4}$ interpolate monotonically between the UV and IR values (see fig.~\ref{fig:F1to4fermion}). Therefore, all four functions $\mathcal{F}_{1,2,3,4}(\mathcal{R})$ in \eqref{eq:F1fermion}--\eqref{eq:F4fermion} are good $F$-functions for the free fermion. We find that our proposals for constructing $F$-functions also hold beyond the context of holographic theories.

\subsection{Free boson on $S^3$}
\label{sec:freeboson}
We now turn to the case of a free boson on $S^3$. The action is given by
\begin{align}
S_S = \frac{1}{2} \int d^3 x \, \sqrt{\zeta} \, \left[ (\nabla \phi)^2 + \frac{R}{8} \, \phi^2 + m^2 \phi^2  \right] \, ,
\end{align}
where $R= 6/ \alpha^2$ is again the scalar curvature of the $S^3$. This is a CFT perturbed by the operator $\phi^2$ with source $\tfrac{1}{2} m^2$. The dimension of the perturbing operator is given by
\begin{align}
\Delta[\phi^2] = 1 < \frac{3}{2} \, .
\end{align}
If our findings from holographic theories are correct, then a good $F$-function can be constructed from the quantum effective potential for the scalar (i.e.~the Legendre transform of the free energy with respect to the source). We will check this explicitly by first examining the suitability of the free energy as a $F$-function  before turning to its Legendre transform.

\subsubsection*{The free energy as a candidate $F$-function}
As shown in  \cite{1105.4598}, the free energy can be written as the following infinite sum:
\begin{align}
\label{eq:FSsum} F_S = \frac{1}{2} \sum_{n=1}^{\infty} n^2 \, \log \left[n^2 - \frac{1}{4} + (\alpha m)^2 \right] \, ,
\end{align}
which depends on $\alpha$ and $m$ only through the dimensionless combination $(\alpha m)$. In the following, it will also be useful to define
\begin{align}
\chi= (\alpha m)^{-2} \, ,
\end{align}
which is proportional to the curvature in units of the source. We do not use the label $\mathcal{R}$ for consistency with our holographic results earlier. There, for theories with $\Delta < 3/2$, the quantity $\mathcal{R}$ denoted the curvature in units of the vev, not the source. Here we maintain this convention.

We can now work in analogy of the free fermion discussed above. For $m \rightarrow 0$ ($\chi \rightarrow \infty$ at fixed $\alpha$) the theory becomes conformal. For this case the sum in \eqref{eq:FSsum} was evaluated \cite{1105.4598}. Using zeta-function renormalization one finds
\begin{align}
\label{eq:FSUV} F_{S, \textrm{UV}} = \frac{1}{16} \left(2 \log 2 -\frac{3 \zeta(3)}{\pi^2} \right) \, .
\end{align}
For $m \rightarrow \infty$ ($\chi \rightarrow 0$ at fixed $\alpha$) the theory becomes empty and we expect.
\begin{align}
F_{S, \textrm{IR}} = 0 \, .
\end{align}
A good $F$-function should reduce to these values in the limits $\chi \rightarrow \infty$ and $\chi \rightarrow 0$, respectively.

Using zeta-function renormalization, it was also shown in \cite{1105.4598} that the renormalized free energy satisfies
\begin{align}
\label{eq:FSderivative} \frac{\partial F^{\textrm{ren}}_S}{\partial (\alpha m)^2} = - \frac{\pi}{4}  \sqrt{(\alpha m)^2 - \frac{1}{4}} \, \coth \left(\pi   \sqrt{(\alpha m)^2 - \frac{1}{4}} \right) \, .
\end{align}
Therefore, we can again write the zeta-function renormalized free energy as an integral. Here one finds
\begin{align}
\label{eq:FSren} F^{\textrm{ren}}_{S} (\chi)= F_{S, \textrm{UV}} - \frac{\pi}{4} \int_0^{1/ \chi} dx \, \sqrt{x^2 - \frac{1}{4}} \, \coth \left(\pi   \sqrt{x^2 - \frac{1}{4}} \right) \, ,
\end{align}
which by definition reduces to $F_{S, \textrm{UV}}$ for $\chi \rightarrow \infty$. The expression \eqref{eq:FSren} again has IR divergences, which come with powers of $\chi^{-3/2}$ and $\chi^{-1/2}$. In particular, one finds
\begin{align}
\label{eq:FSIRdiv} F^{\textrm{ren}}_S(\chi) \underset{\chi \rightarrow 0}{=} - \frac{\pi}{6} \chi^{-3/2} + \frac{\pi}{16} \chi^{-1/2} + (\textrm{vanishing for } \chi \rightarrow 0) \, .
\end{align}

\begin{figure}[t]
\centering
\begin{overpic}
[width=0.65\textwidth]{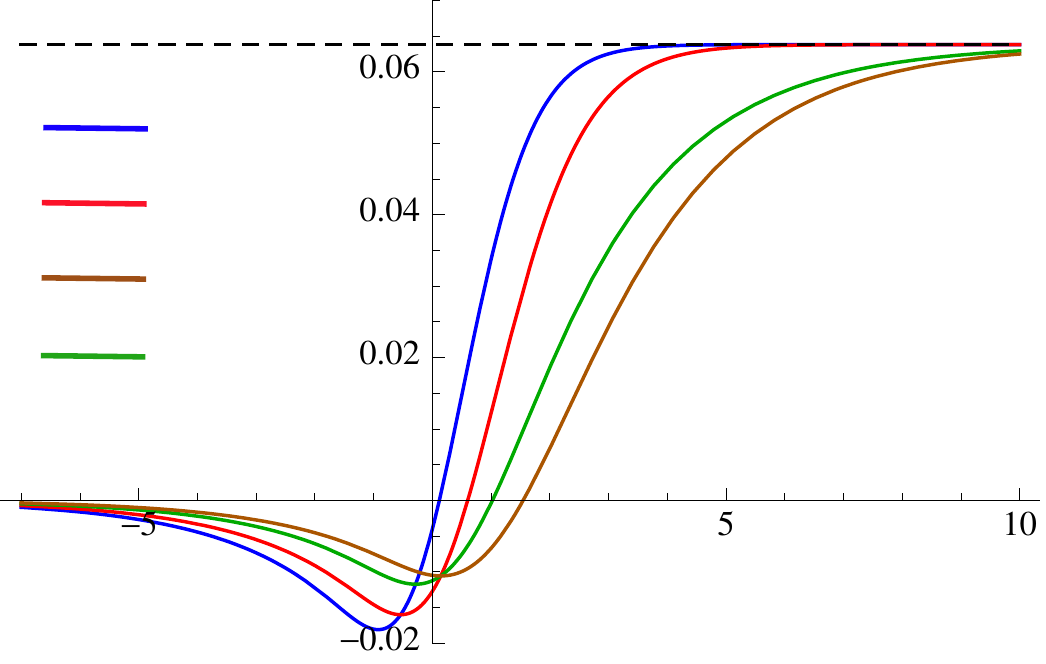}
\put(102,12){$\log(\chi)$}
\put(-10.5,58.25){$F_{S,\textrm{UV}}$}
\put(-10.5,14.75){$F_{S,\textrm{IR}}$}
\put(17,50.05){$\tilde{\mathcal{F}}_1$}
\put(17,42.7){$\tilde{\mathcal{F}}_2$}
\put(17,35.35){$\tilde{\mathcal{F}}_3$}
\put(17,28.0){$\tilde{\mathcal{F}}_4$}
\end{overpic}
\caption{$\tilde{\mathcal{F}}_{1,2,3,4}$ given in \protect\eqref{eq:F1bad}--\protect\eqref{eq:F4bad} vs.~$\log (\chi)$ with $\chi = (\alpha m)^{-2}$ for a theory of a free massive scalar on $S^3$. The black dashed line indicates the value of $F_{S,\textrm{UV}}$ given in \protect\eqref{eq:FDUV}.}
\label{fig:F1to4scalarbad}
\end{figure}
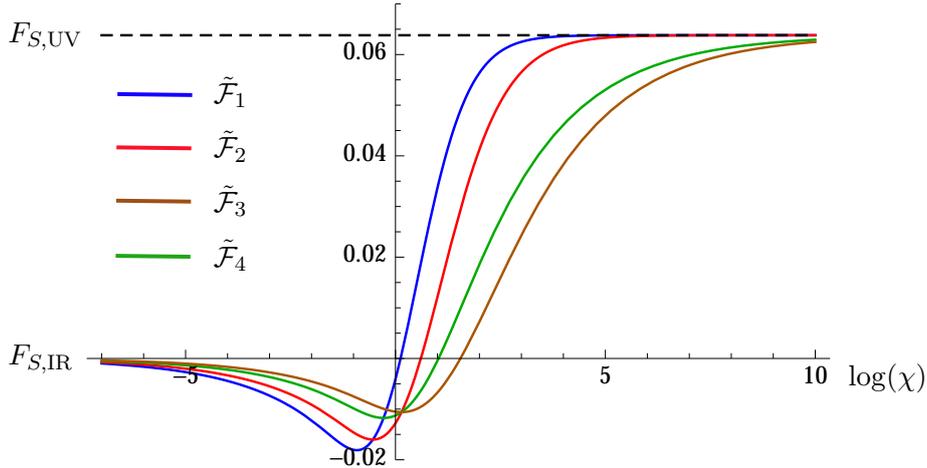

We can  again define finite candidate $F$-functions, where the divergent terms are removed by differentiation or subtraction or a combination thereof. By analogy with \eqref{eq:F1def}--\eqref{eq:F4def} we hence define the following for candidate $F$-functions constructed from the free energy \eqref{eq:FSren}:
\begin{align}
\label{eq:F1bad} \tilde{\mathcal{F}}_1 (\chi) &= \mathcal{D}^{\chi}_{1/2} \, \mathcal{D}^{\chi}_{3/2} \, \, \, \, F^{\textrm{ren}}_S(\chi) \, , \\
\tilde{\mathcal{F}}_2 (\chi) &= \mathcal{D}^{\chi}_{1/2} \, \hphantom{\mathcal{D}^{\chi}_{3/2}} \, \Big( F^{\textrm{ren}}_S(\chi) + \frac{\pi}{6} \chi^{-3/2} \Big) \, , \\
\tilde{\mathcal{F}}_3 (\chi) &= \hphantom{\mathcal{D}^{\chi}_{1/2}} \, \mathcal{D}^{\chi}_{3/2} \, \Big( F^{\textrm{ren}}_S(\chi) - \frac{\pi}{16} \chi^{-1/2} \Big) \, , \\
\label{eq:F4bad} \tilde{\mathcal{F}}_4 (\chi) &= \hphantom{\mathcal{D}^{\chi}_{1/2}} \, \hphantom{\mathcal{D}^{\chi}_{3/2}} \, \Big( F^{\textrm{ren}}_S(\chi) + \frac{\pi}{6} \chi^{-3/2} - \frac{\pi}{16} \chi^{-1/2} \Big) \, ,
\end{align}
where $\mathcal{D}^{\chi}_{1/2}$ and $\mathcal{D}^{\chi}_{3/2}$ are defined as in \eqref{eq:Ddef}, but with $\mathcal{R}$ replaced by $\chi$. By construction all four functions $\tilde{\mathcal{F}}_{1,2,3,4}$ reduce to the values $F_{S,\textrm{UV}}$ and $F_{S,\textrm{IR}}$ for $\chi \rightarrow \infty$ and $\chi \rightarrow 0$, respectively. However, a numerical evaluation shows that they do \emph{not} interpolate monotonically between the UV and IR. As shown in figure \ref{fig:F1to4scalarbad} all four functions $\tilde{\mathcal{F}}_{1,2,3,4}$ fail to exhibit monotonicity. Therefore, they are not good $F$-functions for the free scalar.

The fact that the free energy does not straightforwardly give rise to a good $F$-function for the free scalar has been observed before. In \cite{1105.4598} it was already found that the function $\tilde{\mathcal{F}}_4 (\chi)$ defined in \eqref{eq:F4bad} is not monotonic in $\chi$. To overcome this, a different subtraction of IR-divergent pieces was suggested. While this indeed solved the problem, the necessary subtraction is introduced ad hoc leaving the question unanswered how a good $F$-function can be constructed systematically.

Similar problems for constructing a good $F$-function for the free
massive scalar were found in \cite{1504.00913}. There the authors
examined the equivalent of the `Renormalized Entanglement Entropy'
(REE) of Liu and Mezei \cite{1202.2070} for the theory of a free
massive scalar on dS$_3$, again observing non-monotonicity. As can be deduced from the calculations in \cite{1504.00913}, the relations \eqref{eq:SEEfromF} and \eqref{eq:SEEfromFren} between the dS$_3$ entanglement entropy and the free energy on $S^3$ also hold for the free massive scalar. As a result, our function $\tilde{\mathcal{F}}_1 (\chi)$ defined in \eqref{eq:F1bad} is nothing but the REE on dS$_3$ studied in \cite{1504.00913}. The failure of finding monotonicity of the REE on dS$_3$ can then be understood as part of the more general failure of the free energy as an $F$-function for the free scalar on $S^3$.

Faced with this obstacle, we will now follow the intuition gained from our holographic analyses and consider the quantum effective potential rather than the free energy as an $F$-function for the free scalar. Interestingly, we will find that this will indeed give rise to a monotonic function interpolating between the UV and IR, thus answering the question how a good $F$-function for the free scalar on $S^3$ or dS$_3$ can be constructed.

\subsubsection*{The quantum effective potential as a candidate $F$-function}
We define the quantum effective potential as the Legendre transformation of the free energy with respect to the source $m^2$. While we found that the free energy only depends on the radius $\alpha$ and the source $m^2$ through the combination $(\alpha m)^2$, it will be convenient to write it in the following as $F_S^{\textrm{ren}}(\alpha, m^2)$, with radius and source appearing as separate arguments.

We begin by defining a quantity $G_S^{\textrm{ren}}(\alpha, \psi, m^2)$ as the Legendre transformation of the free energy with respect to $m^2$ as follows:
\begin{align}
\nonumber G_S^{\textrm{ren}}(\alpha, \psi, m^2) & = F_S^{\textrm{ren}}(\alpha, m^2) + \int d^3x \, \sqrt{\zeta} \, m^2 \psi \\
\label{eq:GammaLegendre} & = F_S^{\textrm{ren}}(\alpha, m^2) +2 \pi^2 \, (\alpha m)^2 (\alpha \psi) \, ,
\end{align}
where we have used that $\alpha$ and $m^2$ are constant in space-time and $\int d^3x \sqrt{\zeta} = 2 \pi^2 \alpha^3$. Here we introduced the variable $\psi$ `dual' to the source $m^2$, and which will be proportional to the vev of the operator $\phi^2$. Extremising with respect to $m^2$ gives
\begin{align}
\label{eq:m2ofpsiimplicitdef} \frac{\partial F^{\textrm{ren}}_S}{\partial (\alpha m)^2} + 2 \pi^2 \alpha \psi =0 \, ,
\end{align}
which can be inverted to find $m^2(\psi)$. The quantum effective potential $\Gamma_S^{\textrm{ren}}(\alpha, \psi)$ is then given by
\begin{align}
\label{eq:Gammadef}  \Gamma_S^{\textrm{ren}}(\alpha, \psi) \equiv G_S^{\textrm{ren}}(\alpha, \psi, m^2(\psi)) \, .
\end{align}

 Note that in \eqref{eq:GammaLegendre} and \eqref{eq:m2ofpsiimplicitdef} the source $m^2$ only appears in the combination $(\alpha m)^2$. Similarly $\psi$ only appears in the combination $(\alpha \psi)$. We already defined $\chi = (\alpha m)^{-2}$ as the dimensionless curvature in units of the source. Now, we also define
\begin{align}
\mathcal{R} \equiv (\alpha \psi)^{-2}
\end{align}
as the curvature in units of the vev, in analogy with our holographic
analysis for theories with $\Delta < 3/2$. Notice that we can write \eqref{eq:GammaLegendre} and \eqref{eq:m2ofpsiimplicitdef} using only the dimensionless combinations $\chi$ and $\mathcal{R}$ as variables without ever having to refer to $\alpha$, $m^2$ or $\psi$ individually. Also using the fact that the free energy is a function of $\chi$ only, $F_S^{\textrm{ren}} (\alpha, m^2)=F_S^{\textrm{ren}} (\chi)$, this implies that the quantum effective potential \eqref{eq:Gammadef} is only a function of $\mathcal{R}$, i.e.
\begin{align}
\Gamma_S^{\textrm{ren}}(\alpha, \psi) = \Gamma_S^{\textrm{ren}}(\mathcal{R}) \, .
\end{align}

The inversion required for finding $m^2(\psi)$ can only be done numerically, which, given expression \eqref{eq:FSderivative} is a straightforward exercise. However, for small and large $\psi$ we can also obtain analytical results. In particular, we find
\begin{align}
m^2(\psi) &\underset{\psi \rightarrow 0}{=} \frac{16 \psi}{\alpha} \Big(1+ \mathcal{O}(\psi) \Big) \, , \\
\label{eq:m2psilarge} m^2(\psi) &\underset{\psi \rightarrow \infty}{=} 64 \pi^2 \psi^2 \Big(1+ \mathcal{O}(\psi^{-2}) \Big) \, .
\end{align}
Therefore, for $\psi \rightarrow 0$ we find $m^2 (\psi) \rightarrow 0$ and for $\psi \rightarrow \infty$ we observe $m^2 (\psi) \rightarrow \infty$.

We can then make the following observations. For $\psi \rightarrow 0$ ($\mathcal{R} \rightarrow \infty$ at fixed $\alpha$) the second term in \eqref{eq:GammaLegendre} vanishes and correspondingly
\begin{align}
\Gamma_S^{\textrm{ren}}(\mathcal{R}\rightarrow \infty) = F_S^{\textrm{ren}} (\chi \rightarrow \infty) \rightarrow F_{S, \textrm{UV}} \, .
\end{align}
Therefore, in the UV limit $\mathcal{R} \rightarrow \infty$ the quantum effective potential reduces to the value of the free energy of a conformal scalar on $S^3$.

For $\psi \rightarrow \infty$ ($\mathcal{R} \rightarrow 0$ at fixed $\alpha$) the quantum effective action exhibits divergences which come with powers $\mathcal{R}^{-3/2}$ and $\mathcal{R}^{-1/2}$. These can be found by inserting \eqref{eq:FSIRdiv} and \eqref{eq:m2psilarge} into \eqref{eq:GammaLegendre} and expressing everything in terms of $\mathcal{R}$. One obtains
\begin{align}
\Gamma_S^{\textrm{ren}} (\mathcal{R}) \underset{\mathcal{R} \rightarrow 0}{=} \frac{128 \pi^4}{3} \mathcal{R}^{-3/2} +\frac{\pi^2}{2} \mathcal{R}^{-1/2} \, .
\end{align}

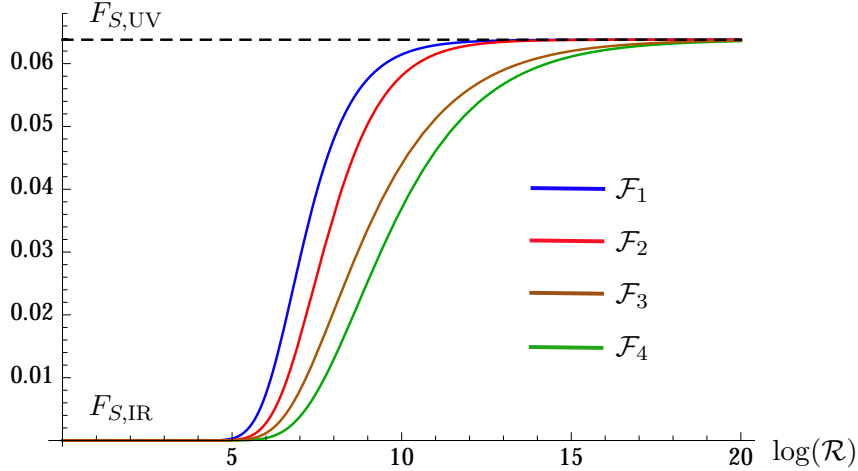
\begin{figure}[t]
\centering
\begin{overpic}
[width=0.65\textwidth]{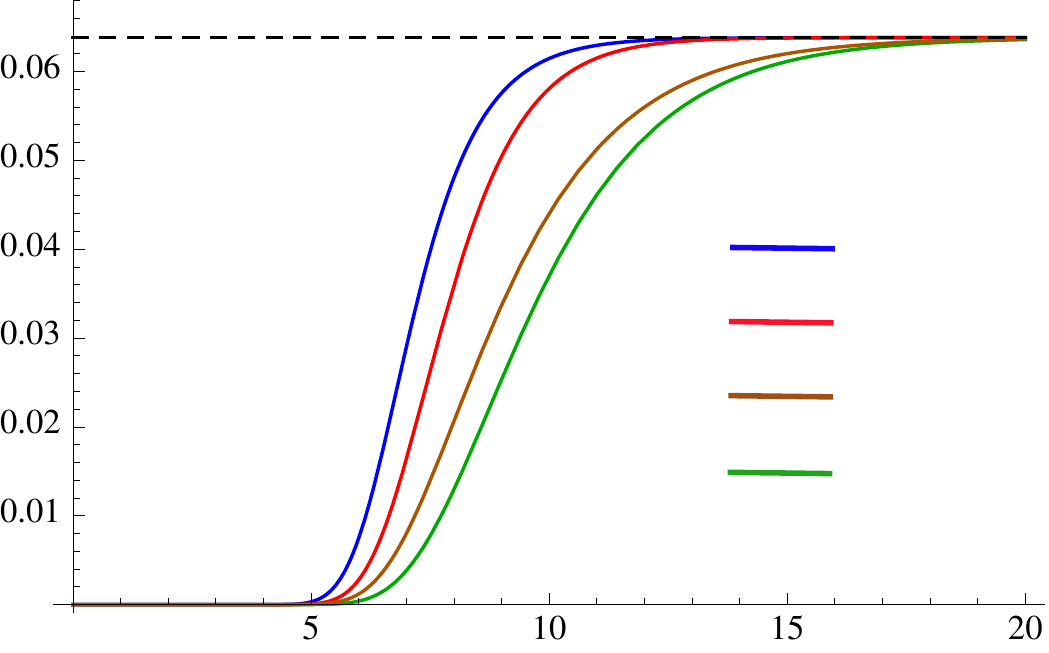}
\put(102,2){$\log(\mathcal{R})$}
\put(10.5,61){$F_{S,\textrm{UV}}$}
\put(10.5,8){$F_{S,\textrm{IR}}$}
\put(81,37.3){$\mathcal{F}_1$}
\put(81,30.2){$\mathcal{F}_2$}
\put(81,23.1){$\mathcal{F}_3$}
\put(81,16.0){$\mathcal{F}_4$}
\end{overpic}
\caption{$\mathcal{F}_{1,2,3,4}$ given in \protect\eqref{eq:F1scalar}--\protect\eqref{eq:F4scalar} vs.~$\log(\mathcal{R})$ for a theory of a free massive scalar on $S^3$. The black dashed line indicates the value of $F_{S,\textrm{UV}}$ given in \protect\eqref{eq:FSUV}.}
\label{fig:F1to4scalar}
\end{figure}

We now have all the ingredients to construct the analogues of the $F$-functions in \eqref{eq:F1fromS}--\eqref{eq:F4fromS} for the free scalar on $S^3$. Here these are given by
\begin{align}
\label{eq:F1scalar} \mathcal{F}_1 (\mathcal{R}) &= \mathcal{D}_{1/2} \, \mathcal{D}_{3/2} \, \, \, \, \Gamma_S^{\textrm{ren}}(\mathcal{R}) \, , \\
\mathcal{F}_2 (\mathcal{R}) &= \mathcal{D}_{1/2} \, \hphantom{\mathcal{D}_{3/2}} \, \Big( \Gamma_S^{\textrm{ren}}(\mathcal{R}) - \frac{128 \pi^4}{3} \mathcal{R}^{-3/2} \Big) \, , \\
\mathcal{F}_3 (\mathcal{R}) &= \hphantom{\mathcal{D}_{1/2}} \, \mathcal{D}_{3/2} \, \Big( \Gamma_S^{\textrm{ren}}(\mathcal{R}) - \frac{\pi^2}{2} \mathcal{R}^{-1/2} \Big) \, , \\
\label{eq:F4scalar} \mathcal{F}_4 (\mathcal{R}) &= \hphantom{\mathcal{D}_{1/2}} \, \hphantom{\mathcal{D}_{3/2}} \, \Big( \Gamma_S^{\textrm{ren}}(\mathcal{R}) - \frac{128 \pi^4}{3} \mathcal{R}^{-3/2} - \frac{\pi^2}{2} \mathcal{R}^{-1/2} \Big) \, ,
\end{align}
with $\mathcal{D}_{1/2}$ and $\mathcal{D}_{3/2}$ given in \eqref{eq:Ddef} and $\Gamma_S^{\textrm{ren}}(\mathcal{R})$ defined in \eqref{eq:Gammadef}. These are four candidate $F$-functions constructed from the quantum effective potential (i.e.~the Legendre transform of the free energy with respect to the source $m^2$). They are functions of $\mathcal{R}$, the (dimensionless) curvature in units of the vev $\langle \phi^2 \rangle$.

The candidate $F$-functions pass the test with flying colours. As one can check explicitly, they reduce to  the values $F_{S,\textrm{UV}}$ and $F_{S,\textrm{IR}}$ in the limits $\mathcal{R} \rightarrow \infty$ and $\mathcal{R} \rightarrow 0$, respectively. In fig.~\ref{fig:F1to4scalar} we then plot $\mathcal{F}_{1,2,3,4}$ vs.~$\log(\mathcal{R})$. The main observation is that all four functions \eqref{eq:F1scalar}--\eqref{eq:F4scalar} interpolate monotonically between $F_{S,\textrm{UV}}$ and $F_{S,\textrm{IR}}$. Therefore, the functions $\mathcal{F}_{1,2,3,4}(\mathcal{R})$ defined in \eqref{eq:F1scalar}--\eqref{eq:F4scalar} are indeed good $F$-functions for the free scalar on $S^3$.   

To conclude, for the case of the free massive boson on $S^3$ good $F$-functions can be constructed from the quantum effective potential, as suggested by our holographic analysis. The free energy fails to interpolate monotonically between $F_{S,\textrm{UV}}$ and $F_{S,\textrm{IR}}$ as observed in \cite{1105.4598} and reviewed at the beginning of this section. For the massive fermion on $S^3$ our holographic findings imply that a good $F$-function can be constructed from the free energy, which we confirmed explicitly in sec.~\ref{sec:freefermion}. To test our $F$-function proposals further one can check if anything goes wrong if one uses the quantum effective potential to construct an $F$-function for the free massive fermion. The calculation is very similar to the analysis presented in this section and we hence do not include it in this work and only quote the result. Interestingly, one finds that the analogues of \eqref{eq:F1scalar}--\eqref{eq:F4scalar} are \emph{not} monotonic in the fermionic case. This suggests that the choice between free energy and quantum effective potential is exclusive in the sense that only one leads to a good $F$-function for a given theory. We leave further investigations of this matter to future work.

\section*{Acknowledgements}\label{ACKNOWL}
\addcontentsline{toc}{section}{Acknowledgements}

We thank Oscar Dias, Rob Myers, Leandro Silva Pimenta, Kostas Skenderis for
useful discussions and Igor Klebanov, Dario Martelli and Marika Taylor
for correspondence.
We also thank Rob Myers for helpful comments on the draft of this work.

\noindent This work was supported in part  by the Advanced ERC grant SM-grav, No 669288.

\appendix
\renewcommand{\theequation}{\thesection.\arabic{equation}}
\addcontentsline{toc}{section}{Appendix\label{app}}
\section*{Appendix}

\section{Perturbative expansion near the maximum of the potential}
\label{app:nearmaximum}
In particular, in the vicinity of a maximum located at $\f_{\textrm{max}}=0$ the solutions $W_{C,\mathcal{R}}$ and $S_{C,\mathcal{R}}$ can be expanded in powers of $\f$ as follows:
\begin{align}
\nonumber W_{C,\mathcal{R}} (\f) & = \frac{1}{\ell} \left[2(d-1) + \frac{\Delta_-}{2} \f^2 + \mathcal{O}(\f^3) \right] + \frac{\mathcal{R}}{d \ell} \, |\f|^{\frac{2}{\Delta_-}} \ [1+ \mathcal{O}(\f) + \mathcal{O}(\mathcal{R})] \\
\label{eq:Wmsol} & \hphantom{=} \,  + \frac{C}{\ell} \, |\f|^{\frac{d}{\Delta_-}} \ [1+ \mathcal{O}(\f)+ \mathcal{O}(C) + \mathcal{O}(\mathcal{R})] \, , \\
\label{eq:Smsol} S_{C,\mathcal{R}}(\f) & = \frac{\Delta_-}{\ell} \f \ [1+ \mathcal{O}(\f)] + \frac{Cd}{\Delta_- \ell} \, |\f|^{\frac{d}{\Delta_-}-1} \ [1+ \mathcal{O}(\f) + \mathcal{O}(C)] \, , \\
\nonumber & \hphantom{=} \,  + \frac{1}{\ell} \mathcal{O}\left( \mathcal{R} |\f|^{\frac{2}{\Delta_-}+1} \right) + \frac{1}{\ell} \mathcal{O}\left(\mathcal{R} C |\f|^{\frac{2+d}{\Delta_-}-1} \right) \, .
\end{align}
with
\begin{align}
\Delta_{\pm} = \frac{1}{2} \left(d \pm \sqrt{d^2 + 4 \ell^2 V''(\f_{\textrm{max}})} \right) \, ,
\end{align}
and where $\mathcal{O}(\mathcal{R})$ and $\mathcal{O}(C)$ is shorthand for
\begin{align}
\mathcal{O}(\mathcal{R}) = \mathcal{O}\left(\mathcal{R} |\f|^{2/ \Delta_-}\right) \, , \qquad \mathcal{O}(C) = \mathcal{O}\left(\mathcal{R} |\f|^{d/ \Delta_-}\right) \, .
\end{align}
For completeness, we also present the corresponding expression for $T_{C,\mathcal{R}}$:
\begin{align}
\label{eq:Tnearmax} T_{C,\mathcal{R}} (\f)= \ell^{-2} \mathcal{R} \, |\f|^{\frac{2}{\Delta_-}} \, \left[1 - \frac{2Cd}{\Delta_-^2(d-2 \Delta_-)} |\f|^{\frac{d}{\Delta_-}-2} + \mathcal{O}(\f) + \mathcal{O}(\mathcal{R}) + \mathcal{O}(C^2)  \right] \, .
\end{align}
Given this solution to the 1st order system, we can integrate to obtain solutions for $A(u)$ and $\f(u)$. As a result, we obtain the following expansion for $A(u)$ and $\f(u)$ for $u \rightarrow - \infty$:
\begin{align}
\label{eq:Amsol} A(u) &= \bar{A} -\frac{u}{\ell} - \frac{\f_-^2 \, \ell^{2 \Delta_-}}{8(d-1)} e^{2\Delta_- u / \ell}  -\frac{\mathcal{R}|\f_-|^{2/\Delta_-} \, \ell^2}{4d(d-1)} e^{2u/\ell} \\
\nonumber & \hphantom{=} \ - \frac{\Delta_+ C |\f_-|^{d/\Delta_-} \, \ell^d}{d(d-1)(d-2 \Delta_-)}e^{du/\ell} +\ldots \, , \\
\label{eq:phimsol} \f(u) &= \f_- \ell^{\Delta_-}e^{\Delta_-u / \ell} \left[ 1+ \mathcal{O} \left(\mathcal{R} |\f_-|^{2/\Delta_-} e^{2u/\ell} \right) + \ldots \right] \\
\nonumber & \hphantom{=} \ + \frac{C d \, |\f_-|^{\Delta_+ / \Delta_-}}{\Delta_-(d-2 \Delta_-)} \, \ell^{\Delta_+} e^{\Delta_+ u \ell} \left[ 1+ \mathcal{O} \left(\mathcal{R}|\f_-|^{2/\Delta_-} e^{2u/\ell} \right) + \ldots \right] + \ldots \, ,
\end{align}
where $\bar{A}$ and $\f_-$ are integration constants. We can now make the following observation.
As $u \rightarrow - \infty$ we expect $A(u)$ to reproduce the scale factor for AdS$_{d+1}$ given in \eqref{eq:Aasymp}. We find that our result is consistent with as long as we make two identifications. For one, we need to set the integration constant $\bar{A}=0$. This is equivalent to our choice of setting $\alpha = \frac{\ell}{2} e^{-c / \ell}$. In addition, by comparing the coefficient of $e^{2u / \ell}$ we confirm that
\begin{align}
\mathcal{R} = R |\f_-|^{-2 / \Delta_-} \, .
\end{align}

Another quantity of interest is the function $U(\f)$, which is defined via equation \eqref{eq:Uequation0}. Its near-boundary expansion can be found from the corresponding results for $W$ and $S$ displayed above. In particular, we find
\begin{align}
\label{eq:Unearmax} U_{B, \mathcal{R}}(\f) = \ell \left[ \frac{2}{d(d-2)} + B \, |\f|^{(d-2)/ \Delta_-} + \mathcal{O} \big(\mathcal{R} \, |\f|^{2/ \Delta_-} \big)  \right] \, ,
\end{align}
where $B$ is a constant of integration.

\section{Calculation of the on-shell action}
\label{app:freeenergy}
We begin with on-shell action in the case of a field theory on a Lorentzian manifold. The starting point is the action \eqref{eq:action} which we proceed to rewrite as a functional of $A(u)$.
As a first step note that for our ansatz the curvature scalar $R^{(g)}$ of the $(d+1)$ space-time can be written as:
\begin{align}
\nonumber R^{(g)} &= - \left(2d \ddot{A} + d(d+1) \dot{A}^2 \right) + e^{-2 A} R^{(\zeta)} = \frac{1}{2} \dot{\f}^2 + \frac{d+1}{d-1} \, V \, .
\end{align}
where in the last step we used the equations of motion \eqref{eq:EOM1}--\eqref{eq:EOM3}. Inserting this into \eqref{eq:action} we find
\begin{align}
\label{eq:Son1app} S_{\textrm{on-shell}} = \frac{2}{d-1}M^{d-1} V_d \int_{\textrm{UV}}^{\textrm{IR}} du \, e^{dA} V + S_{GHY}\, ,
\end{align}
i.e.~we have successfully eliminated the explicit dependence on $R^{{g}}$ and $\dot{\f}^2$. Here we also defined
\begin{align}
V_d \equiv \int d^dx \sqrt{|\zeta|} \, .
\end{align}
In the next step we use \eqref{eq:EOM1}--\eqref{eq:EOM2} to replace the potential $V$ in \eqref{eq:Son1app} by
\begin{align}
V= - (d-1) \ddot{A} - d(d-1) \dot{A}^2 + \frac{d-1}{d} \, e^{-2A} R^{(\zeta)} \, .
\end{align}
Inserting and after some manipulations we obtain:
\begin{align}
\nonumber S_{\textrm{on-shell}} = \ & 2 M^{d-1}V_d \, {\big[e^{dA} \dot{A} \big]}_{\textrm{UV}} - 2 M^{d-1} V_d \, {\big[e^{dA} \dot{A} \big]}_{\textrm{IR}} \\
\label{eq:Son2app} & + \frac{2 M^{d-1} R^{(\zeta)}}{d} V_d \int_{\textrm{UV}}^{\textrm{IR}} du \, e^{(d-2)A} + S_{GHY} \, ,
\end{align}
where the subscript UV/IR denotes that the expression is to be evaluated at the UV/IR locus. One can check that, given the asymptotic form of $A(u)$ in the IR \eqref{eq:expAasympIR} the second term in \eqref{eq:Son2} vanishes (as long as $d \geq 2$). Last, we turn to the Gibbons-Hawking-York term, This is given by
\begin{align}
S_{GHY} = 2M^{d-1} {\left[ \int d^dx \sqrt{|\gamma|} \, K \right]}_{\textrm{UV}} = - 2 d M^{d-1}V_d \, {\big[e^{dA} \dot{A}\big]}_{\textrm{UV}} \, ,
\end{align}
where $K=-d \dot{A}$ is the extrinsic curvature of the boundary and the induced metric $\gamma_{\mu \nu}$ was defined in \eqref{eq:gammadef}. Putting everything together we obtain
\begin{align}
\label{eq:Sonapp} S_{\textrm{on-shell}} = -2(d-1) M^{d-1} V_d \, {\big[e^{dA} \dot{A} \big]}_{\textrm{UV}} + \frac{2 M^{d-1} R}{d} V_d \int_{\textrm{UV}}^{\textrm{IR}} du \, e^{(d-2)A} \, .
\end{align}
Here we also replaced $R^{(\zeta)}$ by $R$ as the two are identical in our conventions.
The expression \eqref{eq:Son} is the on-shell action for a holographic
RG flow on a Lorentzian manifold. For the
corresponding expression in the  Euclidean case we simply have to swap the sign, i.e.~we have
\begin{align}
\label{eq:SonE} S_{\textrm{on-shell}, E} = -S_{\textrm{on-shell}} \, .
\end{align}

\section{Calculation of the entanglement entropy}
\label{app:SEEcalculation}
In this appendix, we review the calculation of the entanglement entropy which will be used for constructing $F$-functions. Here we work with a field theory on dS$_d$, which in our holographic setting corresponds to a bulk space-time with dS$_d$ boundary. To calculate an entanglement entropy, we then need to specify an entangling surface on the boundary. To this end, we first specify the bulk metric. Here we will work with
\begin{equation}
\label{eq:metricdSapp} ds^2=du^2+e^{2 A(u)}\left[-dt^2+\alpha^2 \cosh^2(t/\alpha) \left(d\theta^2 +\sin^2\theta d\Omega^{2}_{d-2}  \right) \right]
\end{equation}
where $d\Omega^{2}_{d-2} $ is the metric on a $(d-2)$-dimensional unit sphere. To find the static entanglement entropy we set $t=0$ so that the bulk metric becomes:
\begin{equation}
ds^2=du^2+\alpha^2 e^{2A(u)}\left(d\theta^2 +\sin^2\theta d\Omega^{2}_{d-2}  \right)
\end{equation}
Our choice of entanglement surface is then given by $\theta|_{u \rightarrow - \infty} = \tfrac{\pi}{2}$. This corresponds to calculating the entanglement entropy between two cap-like regions as shown in fig.~\ref{fig:EE}.

The entanglement entropy in our holographic setting is then calculated following the prescription of Ryu and Takayanagi \cite{RT}. According to this we need to find the minimal surface in the bulk which has the entangling surface as the boundary. The entanglement entropy is then given by
\begin{align}
S_{\textrm{EE}}=\frac{\gamma}{4 G_{d+1}}
\end{align}
where $\gamma$ is the area of the minimal surface.
The equation for the surface is $\theta=\theta(u)$. Then the metric on the surface is
\begin{equation}
ds^2=\left[1+\alpha^2 e^{2 A(u)}\left(\frac{d\theta}{du}\right)^2 \right] du^2+\alpha^2 \sin^2 \theta e^{2 A(u)} d\Omega^{2}_{d-2} .
\end{equation}
From this the surface area functional is obtained as:
\begin{equation}
\gamma=\alpha^{d-2} \text{Vol}(S^{d-2})\int du \left[1+\alpha^2 e^{2 A(u)}\left(\frac{d\theta}{du}\right)^2 \right]^{1/2}\sin^{d-2}\theta\  e^{(d-2)A(u)}
\end{equation}
where $\text{Vol}(S^{d-2})$ is the volume of a unit radius $(d-2)$-dimensional sphere. To minimise the area the surface $\theta(u)$ has to satisfy the following equation:
\begin{align}
\alpha ^2 e^{2 A(u)} \left[\dot{A}(u) \dot{\theta}(u) \sin (\theta (u)) \left\{\alpha ^2 (d-1) e^{2 A(u)} (\dot{\theta}(u))^2+d\right\}-(d-2) (\dot{\theta}(u))^2 \cos (\theta (u))\right. \nonumber \\
\left. +\ddot{\theta}(u) \sin (\theta (u))\right]-(d-2) \cos (\theta (u))=0 \label{RT}
\end{align}
subject to the boundary condition
\begin{equation}
\lim_{u\to -\infty }\theta(u)=\frac{\pi}{2}. \label{bc}
\end{equation}
We also impose regularity on the surface. The solution of the equation \eqref{RT} subject to  the boundary condition \eqref{bc} is:
\begin{equation}
\theta(u)=\dfrac{\pi}{2}.
\end{equation}
Then the minimal surface area is
\begin{align}
\gamma=\alpha^{d-2} \Omega_{d-2} \int_{\textrm{UV}}^{\textrm{IR}} du \, e^{(d-2)A(u)} \, , \qquad
\textrm{with} \qquad
\Omega_n = \frac{2 \pi^{\frac{n+1}{2}}}{\Gamma(\tfrac{n+1}{2})} \, .
\end{align}
Using $\Omega_d=\frac{2\pi}{d-1} \alpha^2 \Omega_{d-2}$ and $R=\frac{d(d-1)}{\alpha^2}$ the minimal surface area can be written as
\begin{equation}
\gamma= \frac{2 R}{d} \frac{1}{4\pi} \textrm{Vol}(S^{d}) \int_{\textrm{UV}}^{\textrm{IR}} du \, e^{(d-2)A(u)} \, .
\end{equation}
The holographic entanglement entropy is then
\begin{equation}
\label{eq:SEEapp} S_{EE}=\frac{\gamma}{4 G_{d+1}}= M^{d-1}  \frac{2 R}{d} \textrm{Vol}(S^{d}) \int_{\textrm{UV}}^{\textrm{IR}} du \, e^{(d-2)A(u)} \, ,
\end{equation}
where we also rewrote Newton's constant as $G_{d+1} = 1/ (16 \pi M_p^{d-1})$.

\section{Analytical results for large and small boundary curvature}
\label{sec:largesmallR}
\subsection{Large curvature expansion}
\label{app:largeR}
RG flows with large dimensionless curvature $\mathcal{R}$ are found when the IR end point is very close to the corresponding UV fixed point $\f_{\textrm{UV}}$. In particular, when $\f_0 \rightarrow \f_{\textrm{UV}}$ we find $\mathcal{R} \rightarrow \infty$. In this regime we can find solutions analytically by solving perturbatively in $\f_{\star} \equiv |\f_0 - \f_{\textrm{UV}}|$.

UV fixed points are associated with extrema of the potential, so that in the vicinity of $\f_{\textrm{UV}}$ we can write the potential as\footnote{The calculation performed here only applies to UV fixed points at maxima of the potential, as these fixed points come with a family of RG flow solutions with $\f_0$ a continuous parameter over this family. In contrast, UV fixed points at minima only exist as individual solutions with a discrete set of IR end points $\f_0$. Hence the limit of taking $\f_0 \rightarrow \f_{\textrm{UV}}$ is ill-defined in this case.}
\begin{align}
V(\f) = - \frac{d(d-1)}{\ell^2} - \frac{\Delta_-(d-\Delta_-)}{2 \ell^2} (\f-\f_{\textrm{UV}})^2 + \mathcal{O} \big((\f-\f_{\textrm{UV}})^3 \big) \, .
\end{align}

The solutions for $A(u)$ and $\f(u)$ can then be organised as an expansion in $\f_{\star}$ about the solution associated with the UV fixed point $\f_{\textrm{UV}}$, which we will refer to as $A_0(u)$ and $\f_0(u)$. At the fixed point, the scale factor $A(u)$ is that of AdS$_{d+1}$ space-time given in \eqref{eq:expAasymp} and $\f(u)$ is constant:
\begin{align}
\label{eq:A0solphi0sol} A_0(u) = \ln \left(- \frac{\ell}{\alpha} \sinh \frac{u+c}{\ell} \right) \, , \qquad \f_0(u)= \f_{\textrm{UV}} =const.
\end{align}
We then find that the system of equations \eqref{eq:EOM1}--\eqref{eq:EOM3} can be solved self-consistently by expanding about ($A_0(u)$, $\f_0(u)$) as follows:
\begin{align}
\label{eq:AlargeRansatz} A(u) &= A_0(u) + \mathcal{O}(\f_{\star}^2) \, , \\
\label{eq:philargeRansatz} \f(u) &= \f_0 + \f_1(u) + \mathcal{O}(\f_{\star}^2) \, ,
\end{align}
and where the subscript indicates the order in the expansion in $\f_{\star}$. This has been used before and analytical results for $\f(u)$ in $d=3$ and $d=4$ have been obtained in \cite{Taylor} and \cite{curvedRG}, respectively. Here we continue for general $d$.

In the following, we will solve explicitly for $\f_1(u)$. Also, for simplicity, we will set $\f_{\textrm{UV}}=0$ in the following. By inserting the ansatz \eqref{eq:AlargeRansatz}, \eqref{eq:philargeRansatz} together with \eqref{eq:A0solphi0sol} into the equation of motion \eqref{eq:EOM3}, we obtain the following differential equation for the function $\f_1(u)$:
\begin{align}
\label{eq:phi1equation} \ddot{\f}_1(u)+ \frac{d}{\ell} \coth \left(\frac{u+c}{\ell}\right) \dot{\f}_1(u)+ \Delta_- (d-\Delta_-) \f_1 (u) =0 \, .
\end{align}
which we need to solve subject to the boundary conditions
\begin{align}
\f(-c) =\f_0 \, , \quad \dot{\f}(-c)=0 \, , \quad \textrm{or, equivalently} \quad \f_1(-c) =\f_\star \, , \quad \dot{\f}_1(-c)=0 \, .
\end{align}
This can be solved analytically, which results in the following solution for $\f(u)$:
\begin{align}
\nonumber \f(u) = & \ \f_\star \, \frac{2^{\frac{d}{2}-1}\sqrt{\pi}}{\sin \left(\pi \left(\frac{d-2\Delta_-}{2} \right) \right)} \, \frac{\Gamma \left(\frac{d+1}{2} \right)}{\Gamma(\Delta_-) \Gamma (d-\Delta_-)} \, \times \\
\nonumber & \times \left[ (U-1)^{\frac{\Delta_-}{2}} (U+1)^{\frac{d-\Delta_-}{2}} \frac{\Gamma(\Delta_-)}{\Gamma \left( \frac{2-d+2 \Delta_-}{2} \right)} \, {_2}F_1 \left(\frac{2-d}{2}, \frac{d}{2}; \frac{2-d+2\Delta_-}{2}; \frac{1-U}{2} \right) \right. \\
\nonumber & \ \left. - (U-1)^{\frac{d-\Delta_-}{2}} (U+1)^{\frac{\Delta_-}{2}} \frac{\Gamma(d-\Delta_-)}{\Gamma \left( \frac{2+d-2 \Delta_-}{2} \right)} \, {_2}F_1 \left(\frac{2-d}{2}, \frac{d}{2}; \frac{2+d-2\Delta_-}{2}; \frac{1-U}{2} \right) \right] \\
\label{eq:phiexactlargeR} & + \mathcal{O}(\f_{\star}^2) \, ,
\end{align}
where we defined $U \equiv -\coth(\tfrac{u+c}{\ell})$.

\subsubsection*{Analytical relations between UV data}
Given the analytical solutions for $A(u)$ and $\f(u)$, we can then find how the UV data $(\f_-, R, B, C)$ are related to one another and also to the IR quantity $\f_0$.

For one, note that both $\f_-$ and $C$ appear in the near-boundary expansion of $\f(u)$ given in \eqref{eq:phimsol} and which we reproduce here for convenience:
\begin{align}
\f(u) = \f_- \ell^{\Delta_-} e^{\Delta_- u / \ell} + \frac{Cd |\f_-|^{\Delta_+/\Delta_-}}{\Delta_- (d- 2 \Delta_-)} \ell^{\Delta_+} e^{\Delta_+ u / \ell} + \ldots \, .
\end{align}
We can hence find relations involving $\f_-$ and $C$ by comparing with the near boundary behaviour of \eqref{eq:phiexactlargeR} above. The boundary is reached for $u \rightarrow -\infty$. In this limit the expression \eqref{eq:phiexactlargeR} becomes:
\begin{align}
\f(u) \underset{u \rightarrow -\infty}{=} & \ \f_\star \, \frac{2^{d-1}\sqrt{\pi}}{\sin \left(\pi \left(\frac{d-2\Delta_-}{2} \right) \right)} \, \frac{\Gamma \left(\frac{d+1}{2} \right)}{\Gamma(\Delta_-) \Gamma (d-\Delta_-)} \, \times \\
\nonumber & \times \left[ \frac{\Gamma(\Delta_-)}{\Gamma \left( \frac{2-d+2 \Delta_-}{2} \right)} \, e^{\Delta_- c / \ell} \, e^{\Delta_- u / \ell} \left(1 + \mathcal{O}(e^{2u/\ell}) + \mathcal{O}(e^{2\Delta_- u/\ell}) \right) \right. \\
\nonumber & \quad \left. + \frac{\Gamma(\Delta_+)}{\Gamma \left( \frac{2-d+2 \Delta_+}{2} \right)} \, e^{\Delta_+ c / \ell} \, e^{\Delta_+ u / \ell} \left(1 + \mathcal{O}(e^{2u/\ell}) + \mathcal{O}(e^{2\Delta_+ u/\ell}) \right)  \right] + \mathcal{O}(\f_{\star}^2) \, ,
\end{align}
where $c$ is related to the UV curvature as $\ell^2 R= 4d(d-1) e^{2c / \ell}$. By comparing the coefficients of $e^{\Delta_- u / \ell}$ and of $e^{\Delta_+ u / \ell}$ we can then find
\begin{align}
\label{eq:phi0largeR} \f_{\star} \sim \mathcal{R}^{- \frac{\Delta_-}{2}} \Big( 1 + \mathcal{O} \big( \mathcal{R}^{-\frac{\Delta_-}{2}} \big) \Big) \, ,
\end{align}
where we neglected extracting the exact numerical prefactor, as this will not be important. This confirms that an expansion in small $\f_{\star}$ is indeed an expansion for large $\mathcal{R}$. Similarly, we find
\begin{align}
C \sim \mathcal{O} \left(\mathcal{R}^{\frac{d}{2}- \Delta_-} \right) \, ,
\end{align}
where again we ignored numerical prefactors.

Finally, we wish to extract the behaviour of $B$ at large
$\mathcal{R}$. While the above analysis is valid for any $d$, here it
will be convenient to work with $d=3$, which is all we need. Here we
will extract $B$ by calculating the entanglement entropy $S_{\textrm{EE}}$
explicitly for large $\mathcal{R}$. Recall that the unrenormalized
entanglement entropy is given by (see e.g. equation~\eqref{eq:SEEd3cutoff})
\begin{align}
\label{eq:SEEappendix1} S_{\textrm{EE}}(\Lambda, \mathcal{R}) = (M \ell)^2 \tilde{\Omega}_3 \,  \mathcal{R}^{-1/2} \, \Big( \Lambda\textrm{-dependent part} + B \Big) \, .
\end{align}
That is the parameter $B$ corresponds to what we will call the `universal contribution' to the entanglement entropy, i.e.~the part of the entanglement entropy that does not depend on the cutoff $\Lambda$ defined in \eqref{eq:dimlesscutoff}. This observation will allow us to extract $B$ as follows. From \eqref{eq:SEE} recall that the entanglement entropy (in $d=3$) can also be written as
\begin{align}
\label{eq:SEEappendix2} S_{\textrm{EE}} (\Lambda, \mathcal{R}) = \frac{2}{3} M^2 \tilde{\Omega}_3 \, R^{-1/2} \int_{\log \epsilon}^{-c} du \, e^{A(u)} \, .
\end{align}
The idea is to insert our analytic solution $A(u) = A_0(u) + \mathcal{O}(\f_{\star}^2)$ into \eqref{eq:SEEappendix2} and evaluate the integral. After isolating the cutoff-independent part we can then read off $B$. One obtains
\begin{align}
S_{\textrm{EE}} (\Lambda, \mathcal{R}) &= - \frac{2}{3} M^2 \tilde{\Omega}_3 \, R^{-1/2} \int_{\log \epsilon}^{-c} du \, \frac{\ell}{\alpha} \, \sinh \left( \frac{u+c}{\ell} \right) \left(1 +  \mathcal{O}(\f_{\star}^2) \right) \\
&= -\frac{2}{3} (M \ell)^2 \tilde{\Omega}_3 \, \alpha^{-1} \, R^{-1/2} \left[ \cosh \left( \frac{u+c}{\ell} \right)\right]_{\log \epsilon}^{-c} - \frac{1}{\alpha} \, \mathcal{O}(\f_{\star}^2) \\
&= - 8 \pi^2 (M \ell)^2 \tilde{\Omega}_3 \left[ \cosh \left( \frac{u+c}{\ell} \right)\right]_{\log \epsilon}^{-c} - R^{1/2} \, \mathcal{O}(\f_{\star}^2) \, ,
\end{align}
where we have used $R = 6 / \alpha^2$ and $\tilde{\Omega}_3 = 12 \sqrt{6} \pi^2$.
Any contribution from the lower integration limit will depend on $\epsilon$ and hence $\Lambda$. The universal contribution  purely comes from the upper integration limit $-c$. Then, comparing with \eqref{eq:SEEappendix1} one finds
\begin{align}
B = - 8 \pi^2 \tilde{\Omega}_3^{-2} \mathcal{R}^{1/2} \Big(1 +  \mathcal{O}\big( \mathcal{R}^{- \Delta_-} \big) \Big) \, .
\end{align}
where we also used \eqref{eq:phi0largeR}.

\subsection{Small curvature expansion}
\label{app:smallR}
Here we will derive analytical expressions for \textsc{rg} flow solutions for small values of the dimensionless curvature $\mathcal{R}$. The results will be obtained by expanding about a \emph{flat} \textsc{rg} flow solution. We will derive two expansions, one valid in the vicinity of the UV fixed point, and one appropriate when close to the IR end point. With the help of these we will then find expressions for $C(\mathcal{R})$ and $B(\mathcal{R})$ valid for small $\mathcal{R}$.

\subsubsection*{Expansion in the vicinity of the UV fixed point}
This has been discussed to some extent in section \ref{sec:holoRGreview} and we can be brief. Consider a potential with a maximum at $\f= \f_{\textrm{UV}}$ which gives rise to a UV fixed point. In the vicinity of $\f_{\textrm{UV}}$ the potential can then be expanded as in \eqref{eq:Vnearmax}. We also choose the UV to be reached for $u \rightarrow - \infty$. An \textsc{rg} flow solution with $R=0$ then has the following well-known expansion in the vicinity of the UV (see e.g.~\cite{9905104,exotic}):
\begin{align}
A_{\textrm{flat}}(u) &= \bar{A} - \frac{u}{\ell} - \frac{\f_-^2 \ell^{2 \Delta_-}}{8 (d-1)} e^{2 \Delta_- u / \ell} + \ldots \, , \\
\f_{\textrm{flat}}(u) &= \f_{\textrm{UV}} + \f_- \ell^{\Delta_-} e^{\Delta_- u / \ell} + \f_+ \ell^{\Delta_+} e^{\Delta_+ u / \ell} + \ldots \, ,
\end{align}
with $\bar{A}$, $\f_-$ and $\f_+$ integration constants and $\Delta_{\pm}$ defined in \eqref{eq:Deltapmdef}.

Once we switch on a small finite value for $\ell^2 R$, we write the solution for $(A(u), \f(u))$ as an expansion in powers of $\ell^2 R$ about the flat solution. In particular, one can check that the equations of motion \eqref{eq:EOM1}--\eqref{eq:EOM3} can be solved self-consistently with the ansatz:
\begin{align}
\label{eq:AsmallRinUV} A(u) &= A_{\textrm{flat}}(u) + \ell^2 R \, A_1(u) + \mathcal{O}\big( (\ell^2 R)^2\big) \, , \\
\label{eq:phismallRinUV} \f(u) &= f_{\textrm{flat}}(u) + \ell^2 R \, \f_1(u) + \mathcal{O}\big( (\ell^2 R)^2\big) \, .
\end{align}

Instead of solving for $(A(u), \f(u))$ we could have equally considered the functions $W(\f)$ and $S(\f)$. Again, the solution in the vicinity of $\f_{\textrm{UV}}$ can be organised as an expansion about the flat solutions $W_{\textrm{flat}}(\f)$ and $S_{\textrm{flat}}(\f)$, but now expanding in powers of $\mathcal{R}$. Explicit solutions for $(A(u), \f(u))$ and $(W(\f), S(\f))$ in the vicinity of a UV fixed point at a maximum of the potential are given in appendix \ref{app:nearmaximum}.

However, in the vicinity of the IR end point $\f_0$ of an RG flow the above expansion is not sufficient. As can be seen from the results in section \ref{app:nearmaximum} a power of $\ell^2 R$ in $(A(u), \f(u))$ always comes together with $e^{2u / \ell}$, so that the effective expansion parameter is $\ell^2 R \, e^{2 u / \ell}$. Note that for small $R$ the IR end point will will be at $u =u_0 \gg 1$ with $u_0 \rightarrow + \infty$ for $R \rightarrow 0$. As a result, in the vicinity of the IR the combination $\ell^2 R \, e^{2 u / \ell}$ ceases to be a good expansion parameter. Hence, in the following, we will find a different expansion for $(A(u), \f(u))$ valid in the vicinity of the IR end point.

\subsubsection*{Expansion in the vicinity of the IR end point}
Here, we will again expand about the solution for a flat flow. Recall that a flow with $\mathcal{R}=0$ has its IR end point $\f_{\textrm{IR}}$ at a minimum of the potential. In contrast, a flow with finite $\mathcal{R}$ can never reach the minimum and ends at a generic point $\f_0$ (see \cite{curvedRG} for details). For $\mathcal{R} \rightarrow 0$ the IR end point will approach the minimum, i.e.~$\f_0 \rightarrow \f_{\textrm{IR}}$. Here, we will be interested in solutions valid in the vicinity of $\f_0$, and hence close to $\f_{\textrm{IR}}$. As we will confirm later, a good expansion parameter at small $\mathcal{R}$ for obtaining solutions perturbatively is then $\f_* \equiv |\f_{\textrm{IR}} -\f_0|$.

To begin, we record the expressions for the scale factor $A_{\textrm{flat}}(u)$ and the dilaton $\f_{\textrm{flat}}(u)$ for a flat \textsc{rg} flow solution. The IR fixed point $\f_{\textrm{IR}}$ is reached for $u \rightarrow + \infty$ and in its vicinity one finds \cite{exotic}:
\begin{align}
\label{eq:AflatinIR} A_{\textrm{flat}}(u) &= \bar{\bar{A}} - \frac{u}{\ell_{\textrm{IR}}} + \mathcal{O} \big( e^{2 \Delta_-^{\textrm{IR}} u / \ell_{\textrm{IR}}} \big) \, , \\
\label{eq:phiflatinIR} \f_{\textrm{flat}}(u) &= \f_{\textrm{IR}} + \bar{\f}_- \, e^{\Delta_-^{\textrm{IR}} u / \ell_{\textrm{IR}}} + \mathcal{O} \big( e^{2 \Delta_-^{\textrm{IR}} u / \ell_{\textrm{IR}}} \big) \, ,
\end{align}
where $\bar{\bar{A}}$ and $\bar{\f}_-$ are integration constants,
$\ell_{\textrm{IR}}$ and  $\Delta_-^{\textrm{IR}}$  are given in \eqref{eq:LIRdef}.
On the contrary, for finite $\mathcal{R}$ the IR end point $\f_0$ is reached for $u \rightarrow u_0$ with $u_0$ finite. The scale factor and dilaton in the vicinity of the IR are given by (see \cite{curvedRG}):
\begin{align}
\label{eq:Aphicurvedapp} A(u) \underset{u \rightarrow u_0} = \ln \left(- \frac{\ell_0}{\alpha} \sinh \frac{u-u_0}{\ell_0} \right)  \, , \qquad \f(u) = \f_0 + \mathcal{O} \big( (u-u_0)^2 \big)  \, ,
\end{align}
and
\begin{align}
\label{eq:l0def} \ell_0^2 \equiv -\frac{d(d-1)}{V(\f_0)} \, .
\end{align}

We are now in a position to set up the small-curvature expansion. The idea is to expand about a solution of the form \eqref{eq:Aphicurvedapp} but with $\ell_0 \rightarrow \ell_{\textrm{IR}}$ and $\f_0 \rightarrow \f_{\textrm{IR}}$. That is, we expand about a curved ansatz in the flat limit. Also, $u_0$ is taken to be large and positive, i.e.~$u_0 \gg 1$. The full solution can then be constructed in a perturbative expansion in $\f_* \equiv |\f_{\textrm{IR}} -\f_0| \ll 1$ about the leading order expressions. In particular, the equations of motion \eqref{eq:EOM1}--\eqref{eq:EOM3} can be solved self-consistently with the following expansion:
\begin{align}
\label{eq:AsmallRexpsystematic} A(u) &= A_{\textrm{IR}}(u) + \mathcal{O}(\f_*^2) \, , \quad \textrm{with} \quad A_{\textrm{IR}}(u) = \ln \left(- \frac{\ell_\textrm{IR}}{\alpha} \sinh \frac{u-u_0}{\ell_\textrm{IR}} \right)  \, , \\
\label{eq:phismallRexpsystematic} \f(u) &= \f_{\textrm{IR}} + \f_1(u) + \mathcal{O} (\f_*^2) \, ,
\end{align}
subject to the boundary conditions $\dot{A}(u)|_{u \rightarrow u_0} \rightarrow \frac{1}{(u - u_0)}$, $\f(u_0) = \f_0$ and $\dot{\f}(u_0)=0$. The subscript on $\f_1$ indicates that this term is of order $\mathcal{O}(\f_*)$. Furthermore, note that in the regime of interest, the potential can be expanded as
\begin{align}
\label{eq:Vminexp} V = -\frac{d(d-1)}{\ell_{\textrm{IR}}^2} + \frac{1}{2} m_{\textrm{IR}}^2 (\f - \f_{\textrm{IR}})^2 + \mathcal{O} \big( (\f - \f_{\textrm{IR}})^3 \big) \, .
\end{align}

We can now make the following observation. Consider a point $u=u_1$ towards the IR end of the flow, but not  close to $u_0$, that is
\begin{align}
\label{eq:u1properties} u_1 \gg 1, \, , \quad \textrm{with} \quad u_1 < u_0 \, , \quad \textrm{and} \quad |u_0 - u_1 | \gg 1 \, .
\end{align}
We choose $u_1$ sufficiently large such that our expansion \eqref{eq:AsmallRexpsystematic} is expected to hold. At the same time, any flow that reaches $u_1 \gg 1$ must be a small correction to a flat flow and hence the expression \eqref{eq:AflatinIR} should also be a good approximation at $u=u_1$. We can then write two expressions for $A(u_1)$. From \eqref{eq:AsmallRexpsystematic} we find
\begin{align}
\nonumber A(u_1) &= \left. \ln \left(- \frac{\ell_\textrm{IR}}{\alpha} \sinh \frac{u_1-u_0}{\ell_\textrm{IR}} \right) \right|_{\substack{u_0 \gg 1 \\ u_1 \ll u_0}} + \mathcal{O} \big( \f_*^2 \big) \\
\label{eq:AIRu1expansion} & = \ln \Big(\frac{\ell_\textrm{IR}}{2 \alpha} \Big) + \frac{u_0}{\ell_\textrm{IR}} - \frac{u_1}{\ell_\textrm{IR}} + \mathcal{O} \big( e^{-2 (u_0-u_1) / \ell_\textrm{IR}} \big) + \mathcal{O} \big( \f_*^2 \big) \, .
\end{align}
while from \eqref{eq:AflatinIR} one obtains:
\begin{align}
A(u_1) = \bar{\bar{A}} - \frac{u_1}{\ell_{\textrm{IR}}} + \mathcal{O} \big( e^{2 \Delta_-^{\textrm{IR}} u_1 / \ell_{\textrm{IR}}} \big) \, ,
\end{align}
At leading order, the two expressions for $A(u_1)$ for are consistent if we make the following identification:
\begin{align}
\label{eq:u0Rrelation} \frac{u_0}{\ell_{\textrm{IR}}} = \ln \left(\frac{2 \alpha}{\ell_{\textrm{IR}}} \right) + \bar{\bar{A}} = \frac{1}{2} \ln \left(\frac{4 d (d-1)}{\ell^{2}_{\textrm{IR}} \, R} \right) + \bar{\bar{A}} \, ,
\end{align}
i.e.~the value of $u_0$ is related to the UV curvature $R$. It is this observation which will be useful in the following. Note that, as expected, we find that $u_0 \rightarrow + \infty$ for $R \rightarrow 0$.

We can then make one more observation. For $u_0 \gg 1$ we also expect that the flat solution for $\f(u)$ given in \eqref{eq:phiflatinIR} should be a good approximation to the full result. In particular, at $u=u_0$ the exact result is given by $\f(u_0) = \f_0$. Comparing this to the flat expression at $u=u_0$ one finds
\begin{align}
\nonumber \f_0 &= \f_{\textrm{IR}} + \bar{\f}_- \, e^{\Delta_-^{\textrm{IR}} u_0 / \ell_{\textrm{IR}}} + \mathcal{O} \big( e^{2 \Delta_-^{\textrm{IR}} u_0 / \ell_{\textrm{IR}}} \big) \, .
\end{align}
Using the relation \eqref{eq:u0Rrelation} between $u_0$ and $R$  this can be rewritten as
\begin{align}
\label{eq:phistarRrelation} \f_* \sim (\ell_{\textrm{IR}}^2 R)^{- \frac{\Delta_-^{\textrm{IR}}}{2}} + \mathcal{O} \big( (\ell_{\textrm{IR}}^2 R)^{-\Delta_-^{\textrm{IR}}} \big)  \, .
\end{align}
This confirms that an expansion in $\f_*$ is indeed an expansion for small UV curvature $R$ (recalling that $\Delta_-^{\textrm{IR}} <0$).

\subsubsection*{Expressions for $B(\mathcal{R})$ and $C(\mathcal{R})$ at small $\mathcal{R}$}
While above we have been working with general $d$ here we again specialise to $d=3$. To extract $B$ we use our insight developed in the previous section \ref{app:largeR}. There we used the fact that the parameter $B(\mathcal{R})$ is given by the universal (i.e.~cutoff-independent) contribution to the following integral:
\begin{align}
\label{eq:Bfromint} B(\mathcal{R}) = \frac{2}{3} \ell^{-2} \f_-^{1/ \Delta_-} \int_{\log \epsilon}^{u_0} du \, e^{A(u)} - (\Lambda\textrm{-dependent terms}) \, .
\end{align}
Here we wish to determine $B(\mathcal{R})$ for small $\mathcal{R}$. In particular, we want to write $B(\mathcal{R})$ in terms of an expansion about $B_0 =B(0)$, which in terms of an integral is given by
\begin{align}
\label{eq:B0fromint}B_0 = \frac{2}{3} \ell^{-2} \f_-^{1/ \Delta_-} \int_{\log \epsilon}^{\infty} du \, e^{A_{\textrm{flat}}(u)} - (\Lambda\textrm{-dependent terms}) \, .
\end{align}

The main difficulty in determining $B$ is to calculate the integral appearing in \eqref{eq:Bfromint}. For small $R$ this can be done perturbatively by using the expansions developed above. The main idea is to to split the integration interval $[\log \epsilon, u_0]$ into two and integrate over $[\log \epsilon, u_1]$ and $[u_1, u_0]$ separately. Here $u_1$ is an auxiliary parameter and can take any value as long as it satisfies \eqref{eq:u1properties}. We therefore write the integral in \eqref{eq:Bfromint} as:
\begin{align}
I &= \int_{\log \epsilon}^{u_0} du \, e^{A(u)} = \int_{\log \epsilon}^{u_1} du \, e^{A(u)} + \int_{u_1}^{u_0} du \, e^{A(u)} \, .
\end{align}
Then, on the interval $[\log \epsilon, u_1]$ we can use the expansion \eqref{eq:AsmallRinUV} for $A(u)$ in powers of $R$. On the interval $[\log \epsilon, u_1]$ close to the IR we instead use the expansion given in \eqref{eq:AsmallRexpsystematic}. This is summarised below:
\begin{align}
\label{eq:Acases} A(u) = \left\{
                \begin{array}{ll}
                  A_{\textrm{flat}}(u) + \mathcal{O}(R) \, , & \quad u < u1 \, ,  \\
                  A_{\textrm{IR}}(u) + \mathcal{O}(\f_*^2) \, , & \quad u > u_1 \, ,
                \end{array}
              \right.
\end{align}
with the two solutions matched at $u=u_1$ up to the required order. Inserting this, the integral becomes:
\begin{align}
\nonumber I & = \int_{\log \epsilon}^{u_1} du \, \exp \left(A_{\textrm{flat}}(u) + \mathcal{O}(R) \right) + \int_{u_1}^{u_0} du \, \exp \left( A_{\textrm{IR}}(u) + \mathcal{O}(\f_*^2) \right) \\
\label{eq:I2} & = \int_{\log \epsilon}^{u_1} du \, e^{A_{\textrm{flat}}(u)}  + \int_{\log \epsilon}^{u_1} du \, \mathcal{O}(R) + \int_{u_1}^{u_0} du \, e^{A_{\textrm{IR}}(u) + \mathcal{O}(\f_*^2)} \, .
\end{align}
In the next step, we split the first integral in \eqref{eq:I2} into two as follows:
\begin{align}
\label{eq:I3a} I & = \int_{\log \epsilon}^{\infty} du \, e^{A_{\textrm{flat}}(u)} - \int_{u_1}^{\infty} du \, e^{A_{\textrm{flat}}(u)} + \int_{\log \epsilon}^{u_1} du \, \mathcal{O}(R) + \int_{u_1}^{u_0} du \, e^{A_{\textrm{IR}}(u) + \mathcal{O}(\f_*^2)} \, .
\end{align}
Also, as the two solutions for $u < u_1$ and $u > u_1$ are matched at $u_1$, the contributions to the integrals in \eqref{eq:I3a} from the limit $u_1$ will vanish. Hence we need to evaluate
\begin{align}
\label{eq:I3} I & = \int_{\log \epsilon}^{\infty} du \, e^{A_{\textrm{flat}}(u)} - \int^{\infty} du \, e^{A_{\textrm{flat}}(u)} + \int_{\log \epsilon} du \, \mathcal{O}(R) + \int^{u_0} du \, e^{A_{\textrm{IR}}(u) + \mathcal{O}(\f_*^2)} \, \\
\nonumber &= I_1 + I_2 + I_3 + I_4 \, .
\end{align}
The first term $I_1$ in the above is then the same integral that appears in the expression for $B_0$ given in \eqref{eq:B0fromint}. Therefore, the first term will indeed contribute a term $B_0$ to $B(\mathcal{R})$ while the remaining terms will give rise to corrections.

We will then proceed as follows. In the next step we will insert the appropriate expressions for $A_{\textrm{flat}}$ and $A_{\textrm{IR}}$ into the 2nd ($I_2$) and 4th term ($I_4$) in \eqref{eq:I3} and perform the integrations. In particular, in the 2nd integral we replace $A_{\textrm{flat}}$ by its near IR-expansion given in \eqref{eq:AflatinIR}. In the 4th term we insert expression \eqref{eq:AsmallRexpsystematic}. To remove clutter we will set $\bar{\bar{A}}=0$ in the following. Then these two integrals become:
\begin{align}
\nonumber I_2 + I_4 = \, & - \int^{\infty} du \, e^{A_{\textrm{flat}}(u)} + \int^{u_0} du \, e^{A_{\textrm{IR}}(u) + \mathcal{O}(\f_*^2)} \\
\nonumber = \, & - \int^{\infty} du \, e^{- \frac{u}{\ell_{\textrm{IR}}}} \left( 1+ \mathcal{O} \big( e^{2 \Delta_-^{\textrm{IR}} u / \ell_{\textrm{IR}}} \big) \right) - \frac{\ell_{\textrm{IR}}}{\alpha} \int^{u_0} du \, \sinh \left( \tfrac{u-u_0}{\ell_{\textrm{IR}}} \right) \left( 1 + \mathcal{O}(\f_*^2) \right) \, \\
\nonumber = \, & \ell_{\textrm{IR}} \left[ e^{- \frac{u}{\ell_{\textrm{IR}}}} + \mathcal{O} \big( e^{(2 \Delta_-^{\textrm{IR}} -1) u / \ell_{\textrm{IR}}} \big) \right]^{\infty} - \frac{\ell_{\textrm{IR}}^2}{\alpha} \left[ \cosh \left( \frac{u-u_0}{\ell_{\textrm{IR}}} \right) \right]^{u_0} + \frac{\ell_{\textrm{IR}}}{\alpha} \, \mathcal{O}\big(\f_*^2\big)  \, \\
\nonumber = \, & - \frac{\ell_{\textrm{IR}}^2}{\alpha} \left( 1+ \mathcal{O}(\f_*^2) \right)  \, \\
\label{eq:I4} = \, & - \frac{\ell_{\textrm{IR}}^2}{\sqrt{6}} \, R^{1/2} \left( 1+ \mathcal{O}\big(R^{-\Delta_-^{\textrm{IR}}}\big) \right) \, ,
\end{align}
where in the last step we have used $R= 6 / \alpha^2$ and \eqref{eq:phistarRrelation}.
Then, putting everything together, we are left with
\begin{align}
\label{eq:I7} I = \int_{\log \epsilon}^{\infty} du \, e^{A_{\textrm{flat}}(u)} + \mathcal{O}(R) - \frac{\ell_{\textrm{IR}}^2}{\sqrt{6}} \, R^{1/2} \left( 1+ \mathcal{O}\big(R^{-\Delta_-^{\textrm{IR}}}\big) \right) \, .
\end{align}
Inserting this back into \eqref{eq:Bfromint} we obtain
\begin{align}
B(\mathcal{R}) = \, & \frac{2}{3} \ell^{-2} \f_-^{1/ \Delta_-} \int_{\log \epsilon}^{\infty} du \, e^{A_{\textrm{flat}}(u)} + \mathcal{O}(R) \\
\nonumber & - 8 \pi^2 \tilde{\Omega}_3^{-2} \frac{\ell_{\textrm{IR}}^2}{\ell^2} \mathcal{R}^{1/2} \left( 1+ \mathcal{O}\big(\mathcal{R}^{-\Delta_-^{\textrm{IR}}}\big) \right) - (\Lambda\textrm{-dependent terms}) \, ,
\end{align}
where we also replaced $R$ by $\mathcal{R}$, as $B$ only depends on $R$ via $\mathcal{R}$. As stated before, the first term contributes $B_0$. Therefore, overall we find
\begin{align}
\label{eq:BsmallRapp} B(\mathcal{R}) \underset{\mathcal{R} \rightarrow 0}{=} B_0 + \mathcal{O}(R)  - 8 \pi^2 \tilde{\Omega}_3^{-2} \frac{\ell_{\textrm{IR}}^2}{\ell^2} \mathcal{R}^{1/2} \left( 1+ \mathcal{O}\big(\mathcal{R}^{-\Delta_-^{\textrm{IR}}}\big) \right) \, .
\end{align}

We can perform a similar analysis for extracting $C(\mathcal{R})$ for small $\mathcal{R}$. In particular, we can determine $C(\mathcal{R})$ from the first term in \eqref{eq:Son}. After introducing a cutoff $\Lambda$ as in \eqref{eq:dimlesscutoff} this term gives rise to the first line in expression \eqref{eq:Fd3cutoff}. From this it follows that $C(\mathcal{R})$ can be determined as
\begin{align}
\label{eq:CsmallRboundary} C(\mathcal{R}) = -4 \ell^{-2} |\f_-|^{-3 / \Delta_-} \left.e^{3 A(u)} \dot{A}(u) \right|_{u= \log \epsilon} - (\Lambda\textrm{-dependent terms}) \, .
\end{align}
We define $C_0=C(0)$ as the value of $C(\mathcal{R})$ for $\mathcal{R}=0$. This can be calculated as
\begin{align}
\label{eq:C0smallRboundary} C_0 = -4 \ell^{-2} |\f_-|^{-3 / \Delta_-} \left.e^{3 A_{\textrm{flat}}(u)} \dot{A}_{\textrm{flat}}(u) \right|_{u= \log \epsilon} - (\Lambda\textrm{-dependent terms}) \, .
\end{align}
In the following, it will be useful to write $C(\mathcal{R})$ in terms of an integral. In particular note that
\begin{align}
\int_{\log \epsilon}^{u_0} du \, e^{3A} (\ddot{A} + 3 \dot{A}^2) = \left[ e^{3 A} \dot{A} \right]_{\log \epsilon}^{u_0} = - \left. e^{3 A} \dot{A}(u) \right|_{u= \log \epsilon} \, ,
\end{align}
where we observe that the integral does not receive any contributions from its IR limit. Hence we can write $C(\mathcal{R})$ as
\begin{align}
\label{eq:CsmallRasint} C(\mathcal{R}) = 4 \ell^{-2} |\f_-|^{-3 / \Delta_-} \int_{\log \epsilon}^{u_0} du \, e^{3A} (\ddot{A} + 3 \dot{A}^2)  - (\Lambda\textrm{-dependent terms}) \, .
\end{align}
As before, we again split the integration range into the two intervals $[\log \epsilon, u_1]$ and $[u_1,u_0]$ with $u_1$ satisfying \eqref{eq:u1properties}. We then insert for $A(u)$ with the appropriate expansions as detailed in \eqref{eq:Acases}. Then the integral in \eqref{eq:CsmallRasint} becomes
\begin{align}
\nonumber \mathcal{I} &= \int_{\log \epsilon}^{u_0} du \, e^{3A} (\ddot{A} + 3 \dot{A}^2) \\
\nonumber &= \int_{\log \epsilon}^{u_1} du \, e^{3A_{\textrm{flat}}} (\ddot{A}_{\textrm{flat}} + 3 \dot{A}_{\textrm{flat}}^2) \\
\nonumber & \hphantom{=} \ + \ell^2_{\textrm{IR}} R \int_{\log \epsilon}^{u_1} du \, e^{3A_{\textrm{flat}}} \left(\ddot{A}_1 + 3 \dot{A}_1^2 + 3A_1(\ddot{A}_{\textrm{flat}} + 3 \dot{A}_{\textrm{flat}}^2) \right) \\
\label{eq:J1} & \hphantom{=} \ + \int_{\log \epsilon}^{u_1}  du \, \mathcal{O}(R^2) +\int_{u_1}^{u_0} du \, e^{3A_{\textrm{IR}}} (\ddot{A}_{\textrm{IR}} + 3 \dot{A}_{\textrm{IR}}^2) \Big(1 + \mathcal{O}(\f_*^2) \Big)
\end{align}
To make contact with the expression for $C_0$ we rewrite the first integral in \eqref{eq:J1} as
\begin{align}
\nonumber & \int_{\log \epsilon}^{u_1} du \, e^{3A_{\textrm{flat}}} (\ddot{A}_{\textrm{flat}} + 3 \dot{A}_{\textrm{flat}}^2) \\
= \, & \int_{\log \epsilon}^{\infty} du \, e^{3A_{\textrm{flat}}} (\ddot{A}_{\textrm{flat}} + 3 \dot{A}_{\textrm{flat}}^2) - \int_{u_1}^{\infty} du \, e^{3A_{\textrm{flat}}} (\ddot{A}_{\textrm{flat}} + 3 \dot{A}_{\textrm{flat}}^2) \, .
\end{align}
By ensuring that the solutions for $u < u_1$ and $u > u_1$ match at $u=u_1$ to the appropriate order, we can ensure that $u_1$ does not appear in the final expression and can therefore be deleted from \eqref{eq:J1}. We are hence left with
\begin{align}
\nonumber \mathcal{I} &= \int_{\log \epsilon}^{\infty} du \, e^{3A_{\textrm{flat}}} (\ddot{A}_{\textrm{flat}} + 3 \dot{A}_{\textrm{flat}}^2) - \int^{\infty} du \, e^{3A_{\textrm{flat}}} (\ddot{A}_{\textrm{flat}} + 3 \dot{A}_{\textrm{flat}}^2) \\
\nonumber & \hphantom{=} \ + \ell^2_{\textrm{IR}} R \int_{\log \epsilon} du \, e^{3A_{\textrm{flat}}} \left(\ddot{A}_1 + 3 \dot{A}_1^2 + 3A_1(\ddot{A}_{\textrm{flat}} + 3 \dot{A}_{\textrm{flat}}^2) \right) \\
\nonumber & \hphantom{=} \ + \int_{\log \epsilon}  du \, \mathcal{O}(R^2) +\int^{u_0} du \, e^{3A_{\textrm{IR}}} (\ddot{A}_{\textrm{IR}} + 3 \dot{A}_{\textrm{IR}}^2) \Big(1 + \mathcal{O}(\f_*^2) \Big) \\
\nonumber &= \left[ e^{3 A_{\textrm{flat}}} \dot{A}_{\textrm{flat}} \right]_{\log \epsilon} + \ell^2_{\textrm{IR}} R \int_{\log \epsilon} du \, e^{3A_{\textrm{flat}}} \left(\ddot{A}_1 + 3 \dot{A}_1^2 + 3A_1(\ddot{A}_{\textrm{flat}} + 3 \dot{A}_{\textrm{flat}}^2) \right) \\
\label{eq:J2} & \hphantom{=} \ + \mathcal{O}(R^2) + \left[ e^{3 A_{\textrm{IR}}} \dot{A}_{\textrm{IR}} \right]^{u_0} - \frac{\ell_{\textrm{IR}}^3}{\alpha^3} \, \mathcal{O}(\f_*^2) \, .
\end{align}
Inserting for $A_{\textrm{IR}}$ from \eqref{eq:AsmallRexpsystematic} one finds that
\begin{align}
\left[ e^{3 A_{\textrm{IR}}} \dot{A}_{\textrm{IR}} \right]^{u_0} = 0 \, .
\end{align}
Hence we are left with
\begin{align}
\nonumber \mathcal{I} &= \left[ e^{3 A_{\textrm{flat}}} \dot{A}_{\textrm{flat}} \right]_{\log \epsilon} + \ell^2_{\textrm{IR}} R \int_{\log \epsilon} du \, e^{3A_{\textrm{flat}}} \left(\ddot{A}_1 + 3 \dot{A}_1^2 + 3A_1(\ddot{A}_{\textrm{flat}} + 3 \dot{A}_{\textrm{flat}}^2) \right) \\
\label{eq:J3} & \hphantom{=} \ + \mathcal{O}(R^2) + \mathcal{O} \Big( R^{\frac{3}{2} - \Delta_-^{\textrm{IR}}} \Big) \, .
\end{align}
where we have also used $R= 6 /\alpha^2$ and \eqref{eq:phistarRrelation}. Inserting this back into \eqref{eq:CsmallRboundary} we then find the following. We identify the first term in \eqref{eq:J3} with $C_0$, while the second term gives a contribution $\sim \mathcal{R} C_1$. Overall we hence find that for small $\mathcal{R}$ we can write
\begin{align}
C(\mathcal{R}) = \Big(C_0 + \mathcal{R} C_1 + \mathcal{O}(\mathcal{R}^2) \Big) + \mathcal{O} \Big( \mathcal{R}^{\frac{3}{2} - \Delta_-^{\textrm{IR}}} \Big) \, ,
\end{align}
where we also used that $C(\mathcal{R})$ depends on $R$ only via $\mathcal{R}$.

\section{Holographic entanglement entropy of a spherical region in flat space} \label{FEE}
In this appendix, we will calculate the holographic entanglement entropy of a spherical entangling region in flat space. In this regard, we write the $(d+1)$-dimensional bulk in flat slicing as
\begin{equation}
ds^2=du^2+e^{2 A(u)} \left(-dt^2+dr^2+r^2 d\Omega^2_{d-2}\right).
\end{equation}
 At constant time slicing we can write the metric in conformal coordinates as
 \begin{equation}
 ds^2=\rho^2(z)\left(dz^2+dr^2+r^2 d\Omega^2_{d-2} \right)
 \end{equation}
 where the coordinate $z$ and the scale factor $a(z)$ are defined as
 \begin{equation}
 e^{-A(u)}du=dz \ , \quad \rho(z)=e^{A(u(z))} \ .
 \end{equation}
 We are interested in computing the entanglement entropy between a
 ball of radius $\alpha$ , $ r \leq \alpha $, and the rest on the
 boundary. That means the entangling ``surface" is a
 $(d-2)$-dimensional sphere of radius $\alpha$ on the boundary. To
 compute the entanglement entropy holographically, we need to find the
 minimal $(d-1)$-dimensional ``surface" in the bulk which coincides
 with  a $(d-2)$-dimensional sphere of radius $\alpha$ on the
 boundary. To find this, we take the ansatz: $r=r(z)$ while the angular
 coordinates coincide. The induced metric on this ``surface" becomes
 \begin{equation}
  ds^2_{ind}=\rho^2(z) \left[\left(1+r'^2(z) \right) dz^2 +r^2(z) d\Omega^2_{d-2}\right] \ .
  \end{equation}
From this induced the metric we can find the area functional
\begin{equation}
 \mathcal{S}[r(z)]= \Omega_{d-2} \int dz \ \rho^{d-1}(z) r^{d-2}(z) \sqrt{1+r'^2(z)} .
 \end{equation}
which needs to be minimized subject to the boundary condition that
$r(\epsilon)=\alpha$, where $z=\epsilon$ is the boundary. Variation of
the area functional leads to the equation
\begin{equation}\label{eq:RTflat}
\rho(z) \left((d-2) r'(z)^2+d-2-r(z) r''(z)\right)-(d-1) r(z) \rho'(z) r'(z) \left(r'(z)^2+1\right) =0 \ .
\end{equation}
This equation needs two boundary conditions. One is fixed by asking
that the surface intersects the  (regulated) boundary at $z=\epsilon$
on a circle  of radius $\alpha$;  The second condition is that the
surface is regular where it closes off:  if the minimal surface ends $z=z_0$, we must impose \begin{equation}
z(0)=z_0 \ , \quad z'(0)=0 \ .
\end{equation}
for regularity.  Near  $r=0$ we then have
\begin{equation}
z(r)\approx z_0+\kappa r^2 \ ,  \label{initialflat}
\end{equation}
where $\kappa$ can be taken to 1 which is equivalent to a coordinate rescaling. Eq.~(\ref{initialflat}) sets the initial condition for $r(z_0)$ and $r'(z_0)$ and then we solve the eq.~(\ref{eq:RTflat}) numerically.

Denoting the solution of this equation as $r=r_0(z)$, we find the minimal area is
\begin{equation}
\mathcal{A}= \Omega_{d-2} \int dz \ \rho^{d-1}(z) r_0^{d-2}(z)
\sqrt{1+r_0'^2(z)} \ .
\end{equation}

The entanglement entropy calculated by this way is divergent near the
UV boundary $z=\epsilon$ and requires regularisation. The holographic entanglement entropy is then
\begin{equation}
S_{\textrm{FEE}}=\dfrac{\mathcal{A}}{4 G_{d+1}}= 4 \pi M^{d-1} \mathcal{A} \ .
\end{equation}

\section{De Sitter entanglement entropy and thermodynamics} \label{app:thermo0}
In this appendix we show that the entanglement entropy computed in
section \ref{sec:ent} is the same as the bulk gravitational  entropy of the
space-time which one obtains  by writing the slice metric in de Sitter static
coordinates. This gives rise to a  horizon in the bulk, with an
associate temperature and entropy. The internal energy is identified
with the ADM mass associated to the timelike killing vector.

Using the thermodynamic relation between free energy, internal energy
and entropy, it is possible to derive  relation (\ref{thermo9})  relating  the
functions $B(\mathcal{R})$ and $C(\mathcal{R})$ which appear in the
F-functions.

\subsection{The de Sitter static patch: thermal entropy, and the ADM mass} \label{app:ADM}
The metric of dS in the expanding patch is
\begin{equation}
\zeta_{\mu\nu}dx^\mu dx^\nu=-dt^2+\alpha^2 \cosh^2(t/\alpha)\left(d\theta^2 + \sin^2\theta d\Omega^2_{d-2} \right) \ .
\end{equation}
On the other hand the dS metric in the static patch is
\begin{equation}
\zeta_{\mu\nu}dx^\mu dx^\nu=-\left(1-\frac{r^2}{\alpha^2}\right)d\tau^2 +\left(1-\frac{r^2}{\alpha^2}\right)^{-1}dr^2+r^2 d\Omega^2_{d-2} \ .
\end{equation}
The coordinate transformations from the expanding patch to the static patch is
\begin{align}
\tau & = \alpha \sinh^{-1}\left\{\frac{\sinh(t/ \alpha)}{\sqrt{1-\cosh^2(t/\alpha)\sin^2\theta}} \right\} , \\
r & =\alpha\cosh(t/\alpha) \sin\theta \ ,
\end{align}
and all the other angular coordinates are the same.

The bulk metric in the static patch coordinates is
\begin{equation} \label{eq:static-app}
ds^2=du^2+e^{2A(u)}\left[-\left(1-\frac{r^2}{\alpha^2}\right)d\tau^2 +\left(1-\frac{r^2}{\alpha^2}\right)^{-1}dr^2+r^2 d\Omega^2_{d-2}  \right] \ .
\end{equation}
We can see that there is a horizon at $r=\alpha$, parametrized by the
coordinates $(u, \Omega_{d-2})$. The associated temperature  is
\begin{equation}
T=\frac{1}{2\pi\alpha} \ .
\end{equation}
as can be easily seen by going to Euclidean time and  imposing the
right periodicity to demand
regularity at $r=\alpha$, i.e. $\tau \sim \tau + i 2\pi \alpha$.

\subsubsection*{Entropy}
The thermodynamic entropy  associated to the horizon is given by
\be
S_{\textrm{th}}  = {\textrm{Area} \over 4G_{d+1}} = 4\pi M^{d-1} \textrm{Vol}(S^{d-2}) \, ,
\int_{\textrm{UV}}^{\textrm{IR}} du \, e^{(d-2)A(u)}
\ee
where we have also used $G_{d+1} = (16\pi M^{d-1})^{-1}$. Using the fact that
$\alpha^2 = d(d-1)/R$ and the geometric relation $\textrm{Vol}(S^{d-2}) =
\tfrac{d-1}{2 \pi \, \alpha^2} \textrm{Vol}(S^{d})$, we obtain
\be
S_{\textrm{th}} = 2 M^{d-1}{R\over d} V_d \int_{\textrm{UV}}^{\textrm{IR}} du \, e^{(d-2)A(u)}
=  S_{\textrm{EE}},
\ee
where we have used the explicit expression for $S_{\textrm{EE}}$ established in
equation (\ref{eq:SEE}) for the last identification, and the definition $V_d
= \textrm{Vol}(S^d)$.

\subsubsection*{ADM Mass}
For a static metric such as the one in equation (\ref{eq:static-app})
the internal energy is identified with the   ADM mass, defined as
\begin{align} \label{eq:ADMmass}
M_{\textrm{ADM}}& =-2 M_P^{d-1} \int_{\textrm{UV}} d^{d-1}x \sqrt{h} N K_{\textrm{ADM}} \nonumber \\
& = 2 (d-1)M^{d-1} \left[e^{dA(u)}\dot{A}(u) \right]_{\textrm{UV}}\Omega_{d-2} \int_0^\alpha dr r^{d-2} \ .
\end{align}
On the fixed time slice and on boundary the metric is
\begin{equation}
h_{ab}dx^a dx^b= \left[ e^{2A(u)}\right]_{\textrm{UV}} \left[ \left(1-\frac{r^2}{\alpha^2}\right)^{-1}dr^2+r^2 d\Omega^2_{d-2}  \right]
\end{equation}
where $ \left[ e^{2A(u)}\right]_{\textrm{UV}}$ means $e^{2A(u)}$ evaluated at the UV boundary. The extrinsic curvature of this hypersurface of codimension 2 is
\begin{equation}
K_{\textrm{ADM}}=-(d-1)[\dot{A}(u)]_{\textrm{UV}} \ .
\end{equation}
and the  lapse function is
\begin{equation}
N=e^{A(u)} \left(1-\frac{r^2}{\alpha^2}\right)^{1/2} \ .
\end{equation}
Using the relation $\Omega_d=\frac{2\pi}{d-1}\Omega_{d-2}$ and
evaluating the integral in equation (\ref{eq:ADMmass})  we find,
\begin{equation}
M_{\textrm{ADM}}=2 (d-1)M^{d-1} \left[e^{dA(u)}\dot{A}(u) \right]_{\textrm{UV}} \Omega_d \frac{\alpha^{d-1}}{2\pi } \ .
\end{equation}
Using the relations $\beta=2\pi\alpha$ and $V_d = \alpha^d \Omega_d$, we arrive  at
\begin{equation}
\beta M_{\textrm{ADM}}=2 (d-1)M^{d-1} \left[e^{dA(u)}\dot{A}(u) \right]_{\textrm{UV}}V_d
\end{equation}

Identifying the ADM mass with the internal energy, $M_{\textrm{ADM}} = U_{\textrm{th}}$,
equation (\ref{eq:Son}) becomes the (integrated) first law.
\be \label{eq:thermorelation}
\beta F_{\textrm{th}}  = \beta U_{\textrm{th}} - S_{\textrm{th}}
\ee

\subsection{Identities from thermodynamic relations}
\label{app:thermo}
We begin with the renormalized stress-tensor, which is computed holographically by
\begin{align}
\label{eq:Tmunudef} \langle T_{\mu \nu}^{(\textrm{ren})} \rangle = - \frac{2}{\sqrt{\zeta}} \frac{\delta S_{\textrm{on-shell}}^{(\textrm{ren})}}{\delta \, \zeta^{\mu \nu}} \, ,
\end{align}
where $\zeta_{\mu \nu}$ denotes the metric on $S^3$. Expressions for
both the cutoff-regulated and the renormalized free energy $F =
-S_{\textrm{on-shell}}$ are
collected in \eqref{eq:Fsummary1} and \eqref{eq:Fsummary3}. Using
these expressions we obtain
\begin{align}
\langle T_{\mu \nu}^{(\textrm{ren})} \rangle = - \frac{1}{3} \ R^{3/2} \, \mathcal{C}^{(\textrm{ren})} \, \zeta_{\mu \nu} \, ,
\end{align}
with
\begin{align}
\label{eq:Ccaldef} \mathcal{C} &= - (M \ell)^2 \tilde{\Omega}_3 \Big(3 \mathcal{R}^{-3/2} \big[4 \Lambda^3  (1 + \ldots) + C \big] + \mathcal{R}^{-1/2} \big[\Lambda (1 + \ldots) + B -2 C' \big] -2 \mathcal{R}^{1/2} B' \Big) \, , \\
\label{eq:Ccalrendef} \mathcal{C}^{\textrm{ren}} &= - (M \ell)^2 \tilde{\Omega}_3 \Big(3 \mathcal{R}^{-3/2}(C-C_{ct}) + \mathcal{R}^{-1/2}(B-B_{ct}) -2 \mathcal{R}^{-1/2} C' -2 \mathcal{R}^{1/2} B'  \Big) \, ,
\end{align}
where $(\ldots)$ contain all remaining terms that depend on the cutoff $\Lambda$ explicitly. Furthermore, starting with \eqref{eq:Fsummary1} and \eqref{eq:Fsummary3} one can also show that
\begin{align}
\label{eq:Ferivativeid} \mathcal{R} \frac{\partial}{\partial \mathcal{R}} F^{(\textrm{ren})} = - \frac{1}{2} \mathcal{C}^{(\textrm{ren})} \, .
\end{align}

We now use the thermodynamic identifications discussed in the first part of
this appendix,
\begin{align}
\label{eq:Fthrelation} \beta F_{\textrm{th}} &= F^{(\textrm{ren})} \,
, \\
\label{eq:Sthrelation} S_{\textrm{th}} &= S_{\textrm{EE}}^{(\textrm{ren})} \, , \\
\label{eq:Uthrelation} \beta U_{\textrm{th}} &= \int d^3 x \,
\sqrt{\zeta} \, \langle T_{0}^{\,0} \rangle = \frac{1}{3}
\mathcal{C}^{(\textrm{ren})} \, .
\end{align}

Then the thermodynamic relation \eqref{eq:thermorelation} implies:
\begin{align}
\nonumber S_{\textrm{EE}}^{(\textrm{ren})} & = \frac{1}{3} \mathcal{C}^{(\textrm{ren})} - F^{(\textrm{ren})} \\
\nonumber &= - \frac{2}{3} \mathcal{R} \frac{\partial}{\partial \mathcal{R}} F^{(\textrm{ren})} - F^{(\textrm{ren})} \\
\label{eq:SEEfromFapp} &= - \mathcal{D}_{3/2} \, F^{(\textrm{ren})}
\end{align}
where when going from the 1st to the 2nd line we used \eqref{eq:Ferivativeid} and where $\mathcal{D}_{3/2}$ is defined in \eqref{eq:Ddef}. Therefore,  starting with the thermodynamic relation \eqref{eq:thermorelation} we successfully reproduced the relations \eqref{eq:SEEfromF}--\eqref{eq:SEEfromFren}.

In addition, starting again with the thermodynamic relation (now in terms of the cutoff-regulated quantities only)
\begin{align}
\label{eq:thermorelation2} S_{\textrm{EE}} (\Lambda, \mathcal{R}) & = \frac{1}{3} \mathcal{C} - F(\Lambda, \mathcal{R}) \, ,
\end{align}
and inserting \eqref{eq:SEEd3cutoff}, \eqref{eq:Ccaldef} and \eqref{eq:Fsummary1} one can show that the above reduces to
\begin{align}
\label{eq:BCrelationapp} C'(\mathcal{R}) = \frac{1}{2} B(\mathcal{R}) - \mathcal{R} B'(\mathcal{R}) \, ,
\end{align}
which we also found to hold numerically in section \ref{sec:numerical}.

Note that to arrive at \eqref{eq:SEEfromFapp} and \eqref{eq:BCrelationapp} it was crucial that the the entanglement entropy $S_{\textrm{EE}}^{(\textrm{ren})}$ is identified with a thermal entropy as in \eqref{eq:Sthrelation}. Our numerical evidence for the validity of \eqref{eq:SEEfromFapp} and \eqref{eq:BCrelationapp} can therefore be seen as evidence for the validity of this assertion.

From \eqref{eq:BCrelationapp} we can make another observation. As shown in app.~\ref{app:smallR} for small $\mathcal{R}$ the functions $B(\mathcal{R})$ and $C(\mathcal{R})$ can be expanded as
\begin{align}
\nonumber B(\mathcal{R}) &= B_0 + B_{1/2} \mathcal{R}^{1/2} + \mathcal{O} (\mathcal{R}) + \mathcal{O}(\mathcal{R}^{1/2-\Delta_-^{\textrm{IR}}}) \, \\
\nonumber C(\mathcal{R}) &= C_0 + C_1 \mathcal{R} + \mathcal{O}(\mathcal{R}^2) + \mathcal{O}(\mathcal{R}^{3/2-\Delta_-^{\textrm{IR}}}) \, .
\end{align}
Then \eqref{eq:BCrelationapp} implies that
\begin{align}
\label{eq:BasderivativeofC} B_0 = 2 C_1 \qquad \Rightarrow \qquad B(\mathcal{R}) \big|_{\mathcal{R} = 0} = 2 \left. \frac{\partial C(\mathcal{R})}{\partial \mathcal{R}} \right|_{\mathcal{R} = 0} \, .
\end{align}
This in turn gives the following relation between the renormalization scheme parameters $B_{ct,0}$ and $\tilde{B}_{ct,0}$ defined in \eqref{eq:Bct0Cct0def} and \eqref{eq:Btildect0def}:
\begin{align}
B_{ct, 0} = B_0 + C_1 \overset{\eqref{eq:BasderivativeofC}}{=} \frac{3}{2} B_0 = \frac{3}{2} \tilde{B}_{ct,0} \, .
\end{align}

\section{Comments on the renormalization scheme}
\label{app:renormalizationscheme}
In this section we will give further physical insight into the renormalization scheme employed in section \ref{sec:Ffunc}. There we found that for constructing good $\mathcal{F}$-function from the on-shell action for a theory on $S^3$, we need to choose the two counterterms $C_{ct,0}$ and $B_{ct,0}$ as
\begin{align}
\label{eq:renschemeapp} C_{ct,0} = C(0) = C_0 \, , \qquad B_{ct,0} = B(0) + C'(0) = B_0 + C_1 \, .
\end{align}
Here we will relate \eqref{eq:renschemeapp} to a renormalization condition for correlation functions of the stress tensor for the field theory on $S^3$. The key ingredient will be a set of identities already derived in \cite{1504.00913}. Here we we reproduce the relevant equations, rewriting them using our notation.

We start by collecting the relevant expressions. The (expectation value of the) renormalized stress tensor can be written in terms of the renormalized on-shell action as
\begin{align}
\label{eq:Trenmunudefapp} \langle T_{\mu \nu}^{\textrm{ren}} (x) \rangle = - \frac{2}{\sqrt{\zeta(x)}} \frac{\delta \, S_{\textrm{on-shell}}^{\textrm{ren}}(\mathcal{R}| B_{ct}, C_{ct})}{\delta \, \zeta^{\mu \nu}(x)} = - \frac{1}{3} |\f_-|^{3/ \Delta_-} \mathcal{R}^{3/2} \, \mathcal{C}^{\textrm{ren}} (\mathcal{R}| B_{ct}, C_{ct}) \, \zeta_{\mu \nu}(x) \, ,
\end{align}
with $\mathcal{C}^{\textrm{ren}}$ given in \eqref{eq:Ccalrendef}. We also have the two-point function
\begin{align}
\label{eq2ptfuncdef} \langle T_{\mu \nu}^{\textrm{ren}}(x) T_{\rho \sigma}^{\textrm{ren}}(y)  \rangle = \frac{4}{\sqrt{\zeta(x)} \sqrt{\zeta(y)}} \frac{\delta^2 \, S_{\textrm{on-shell}}^{\textrm{ren}}(\mathcal{R}| B_{ct}, C_{ct})}{\delta \, \zeta^{\mu \nu}(x) \ \delta \, \zeta^{\rho \sigma}(y)}  \, .
\end{align}
If we consider variations with respect to $\zeta_{\mu \nu}$ respecting the spherical symmetry of $S^3$ (i.e.~restricting to homogeneous Weyl rescalings,) the following holds
\begin{align}
\int d^3 x \, \zeta^{\mu \nu}(y) \, \frac{\delta}{\delta \, \zeta^{\rho \sigma}(y)} = \mathcal{R} \, \frac{\partial}{\partial \mathcal{R}} \, .
\end{align}
By applying the above operator to \eqref{eq:Trenmunudefapp} one can derive the following identity:
\begin{align}
\nonumber \int d^3 y \, & \sqrt{\zeta(y)} \, \langle T^{\textrm{ren}} (y) \, T_{\mu \nu}^{\textrm{ren}} (x) \rangle - 3 \langle T_{\mu \nu}^{\textrm{ren}} (x) \rangle \\
\label{eq:2ptidentity} = \ & \frac{2}{3} |\f_-|^{3/ \Delta_-} \mathcal{R} \frac{\partial}{\partial \mathcal{R}} \Big(\mathcal{R}^{3/2} \, \mathcal{C}^{\textrm{ren}} \Big) \, \zeta_{\mu \nu}(x) + \frac{2}{3} |\f_-|^{3/ \Delta_-} \, \mathcal{R}^{3/2} \, \mathcal{C}^{\textrm{ren}} \, \zeta_{\mu \nu(x)} \, ,
\end{align}
where we have also used \eqref{eq2ptfuncdef} and $T^{\textrm{ren}} \equiv \zeta^{\mu \nu} T^{\textrm{ren}}_{\mu \nu}$.

We are now in a position to rewrite the conditions \eqref{eq:renschemeapp} as a set of conditions on the 1pt and 2pt-functions of $T^{\textrm{ren}}_{\mu \nu}$. To this end we take expression \eqref{eq:Trenmunudefapp} and expand $\mathcal{C}^{\textrm{ren}}$ for small $\mathcal{R}$. Using our results from appendix \ref{app:smallR} one finds
\begin{align}
\nonumber  \langle T_{\mu \nu}^{\textrm{ren}} (x) \rangle & =  - \frac{1}{3} |\f_-|^{3/ \Delta_-} \mathcal{R}^{3/2} \,  \mathcal{C}^{\textrm{ren}} (\mathcal{R}| B_{ct}, C_{ct}) \, \zeta_{\mu \nu}(x) \\
\nonumber & \underset{\mathcal{R} \rightarrow 0}{=} (M \ell)^2 |\f_-|^{3/ \Delta_-} \Big( (C_0-C_{ct}) + \mathcal{O} (\mathcal{R}) + \mathcal{O} \big(\mathcal{R}^{\frac{3}{2}-\Delta_-^{\textrm{IR}}} \big) \Big) \, \zeta_{\mu \nu}(x) \, .
\end{align}
Rearranging this we find
\begin{align} \label{eq:C-T}
(C_0-C_{ct}) \, \zeta_{\mu \nu}(x) = (M \ell)^{-2}  |\f_-|^{- 3/ \Delta_-} \, \langle T_{\mu \nu}^{\textrm{ren}} (x) \rangle \Big|_{R \rightarrow 0} \, .
\end{align}
If we recall the identification of $\f_-$ with the source $j$ and  $C$ with the (dimensionless) vev of
the deforming operator, equation (\ref{eq:C-T}) is nothing but the
trace identity $\< T \> = \beta(j) \< O\>$ in the $R=0$ theory. 
Equation (\ref{eq:C-T}) implies that renormalizing with $C_{ct} =
C_{ct,0} = C_0$ is equivalent to the renormalization condition, that
the renormalized stress tensor $\langle T_{\mu \nu}^{\textrm{ren}} (x) \rangle$
(or, equivalently, the renormalized  operator vev $\< O\>$)  of the flat theory ($R=0$) vanishes.

For the 2nd renormalization condition, we start with expression \eqref{eq:2ptidentity}, adding $2\langle T_{\mu \nu}^{\textrm{ren}} (x) \rangle$ on both sides:
\begin{align}
\int d^3 y \, & \sqrt{\zeta(y)} \, \langle T^{\textrm{ren}} (y) \, T_{\mu \nu}^{\textrm{ren}} (x) \rangle - \langle T_{\mu \nu}^{\textrm{ren}} (x) \rangle  = \frac{2}{3} |\f_-|^{3/ \Delta_-} \mathcal{R} \frac{\partial}{\partial \mathcal{R}} \Big(\mathcal{R}^{3/2} \, \mathcal{C}^{\textrm{ren}} \Big) \, \zeta_{\mu \nu}(x) \, .
\end{align}
Inserting for $\mathcal{C}^{\textrm{ren}}$ with \eqref{eq:Ccalrendef} this becomes:
\begin{align}
\nonumber \int d^3 y \, & \sqrt{\zeta(y)} \, \langle T^{\textrm{ren}} (y) \, T_{\mu \nu}^{\textrm{ren}} (x) \rangle - \langle T_{\mu \nu}^{\textrm{ren}} (x) \rangle  \\
\nonumber = \ & - \frac{2}{3} (M \ell)^2 |\f_-|^{3/ \Delta_-} \Big( \mathcal{R}(B +C' - B_{ct}) - \mathcal{R}^2 B' - 2 \mathcal{R}^2 C'' - 2 \mathcal{R}^3 B'' \Big) \, \zeta_{\mu \nu}(x) \\
\underset{\mathcal{R} \rightarrow 0}{=} \ & - \frac{2}{3} (M \ell)^2 |\f_-|^{3/ \Delta_-} \Big( \mathcal{R}  (B_0 +C_1 - B_{ct}) +\mathcal{O} (\mathcal{R}^2) + \mathcal{O} \big(\mathcal{R}^{\frac{3}{2}-\Delta_-^{\textrm{IR}}} \big) \Big) \, \zeta_{\mu \nu}(x) \, .
\end{align}
This can be rearranged as follows:
\begin{align}
\nonumber & (B_0 +C_1 - B_{ct})  \, \zeta_{\mu \nu}(x) \\
\label{eq:2ndcondition} = \ & \frac{3}{2} (M \ell)^{-2} |\f_-|^{-1/ \Delta_-} \Big[ \frac{1}{R} \Big( \int d^3 y \, \sqrt{\zeta(y)} \, \langle T^{\textrm{ren}} (y) \, T_{\mu \nu}^{\textrm{ren}} (x) \rangle - \langle T_{\mu \nu}^{\textrm{ren}} (x) \rangle \Big) \Big]_{R \rightarrow 0} \, .
\end{align}
Therefore, the choice $B_{ct} = B_{ct,0} = B_0 +C_1$ is again related to a
vanishing condition on correlators involving $T_{\mu
  \nu}^{\textrm{ren}}$ for $R \rightarrow 0$.

\section{Zeta-function renormalization vs.~covariant counterterms}
\label{app:zetafunctionrenorm}
Here we calculate the free energy for conformally coupled massive boson on $S^3$ and renormalize it with the help of covariant counterterms. We then compare with the corresponding expression obtained via zeta-function renormalization.

The action for a conformally coupled massive scalar on $S^3$ is given by
\begin{align}
\label{eq:scalaractionapp} S= \frac{1}{2} \int d^d x \, \sqrt{\zeta} \, \left((\nabla \phi)^2 + \frac{d-2}{4(d-1)} R \phi^2 + m^2 \phi^2 \right) \, .
\end{align}
where $\zeta_{\mu \nu}$ is a metric on $S^3$ of radius $\alpha$. The free energy is then calculated as
\begin{align}
F_S = - \log |Z| = \frac{1}{2} \log \det \left(\mu_0^{-2} \mathcal{O}_S \right) \, , \quad \textrm{with} \quad \mathcal{O}_S = - \nabla^2 + \frac{d-2}{4(d-1)}R + m^2 \, ,
\end{align}
and $\mu_0$ is a scale introduced to make the functional determinant well-defined.

The determinant of $\mathcal{O}_S$ can be calculated as the product of its eigenvalues. In $d=3$ these are given by (see e.g.~\cite{1105.4598})
\begin{align}
\lambda_j = \frac{1}{\alpha^2} \left(j + \frac{3}{2} \right) \left(j + \frac{1}{2} \right) + m^2 \, , \qquad j \geq 0 \, .
\end{align}
The multiplicity of each level $n$ is
\begin{align}
m_j = (j+1)^2  \, .
\end{align}
Putting everything together and defining $n = j+1$, we arrive at the following expression for the free energy:
\begin{align}
F_S= \frac{1}{2} \sum_{n = 1}^{\infty} n^2  \, \log \left(\frac{n^2 - \frac{1}{4} + (\alpha m)^2}{(\alpha \mu_0)^2} \right) \, .
\end{align}

To regulate this expression, we cut off the sum at a maximum level $n_{\textrm{max}} =N$. Also, for $\mu_0$ we choose the corresponding eigenvalue at this level, i.e.
\begin{align}
\mu_0^2 = \lambda_{j_{\textrm{max}}} = \lambda_{N-1} = \frac{1}{\alpha^2} \left(N^2 - \frac{1}{4} + (\alpha m)^2 \right) \, .
\end{align}
Thus we arrive at an expression for the regulated free energy which is given by
\begin{align}
F_S^{\textrm{reg}} (N, \alpha m)= \frac{1}{2} \sum_{n = 1}^{N} n^2  \, \log \left(\frac{n^2 - \frac{1}{4} + (\alpha m)^2}{N^2 - \frac{1}{4} + (\alpha m)^2} \right) \, ,
\end{align}
where we made it manifest that it is a function of the cutoff $N$ and the dimensionless combination $\alpha m$.

We will be particularly interested in the divergent terms (i.e.~terms with positive powers of $N$) and finite terms ($\sim \mathcal{O}(N^0)$) in $F_S$ for $N \rightarrow \infty$. One can extract those explicitly by rewriting the sum in $F_S$ with the help of the Euler-Maclaurin formula:
\begin{align}
\sum_{n=a}^b f(n) = \int_a^b dx \, f(x) + \frac{f(a) + f(b)}{2} + \sum_{k=1}^{\lfloor p/2 \rfloor} \frac{B_{2k}}{(2k)!} \Big(f^{(2k-1)}(b) - f^{(2k-1)}(a) \Big) + R_p \, ,
\end{align}
with
\begin{align}
R_p = (-1)^{p+1} \int_a^b dx \, \frac{B_p(x - \lfloor x \rfloor)}{p!} \, f^{(p)}(x) \, ,
\end{align}
where $\lfloor x \rfloor$ is the largest integer that is not greater than $x$. Here $B_{2k}$ denote Bernoulli numbers and $B_p(x)$ is the $p$-th Bernoulli polynomial.

Then, in the limit $N \rightarrow \infty$ one finds that
\begin{align}
\label{eq:Fscalarregulatedapp} F_S^{\textrm{reg}} (N, \alpha m) & = - \frac{1}{9} N^3 +  \frac{(\alpha m)^2}{3}  N + F_S^{\textrm{finite}} (\alpha m) +\mathcal{O}(N^{-1}) \, ,
\end{align}
where we denoted the $\mathcal{O}(N^0)$-term by $F_S^{\textrm{finite}} (\alpha m)$. It is a function of $(\alpha m)$ and we can only evaluate it numerically.

Having arrived at a regulated expression for $F_S$, we now renormalize by adding appropriate counterterms to the action \eqref{eq:scalaractionapp}. As a fist step, we define a dimensionful cutoff $\Lambda$ as
\begin{align}
\Lambda \equiv \frac{N}{\alpha} \, .
\end{align}
We then add covariant counterterms of the form
\begin{align}
S_{ct,1} = \int d^3x \sqrt{\zeta} \, \Lambda^3 \, f_1(m / \Lambda) \, , \qquad S_{ct,2} = \int d^3x \sqrt{\zeta} \, R \, \Lambda \, f_2(m / \Lambda) \, ,
\end{align}
where the functions $f_1(m / \Lambda)$ and $f_2(m / \Lambda)$ are to be chosen such that one arrives at a finite expression for the free energy. Given the divergences in \eqref{eq:Fscalarregulatedapp}, the appropriate counterterms are
\begin{align}
S_{ct,1} &= \int d^3x \sqrt{\zeta} \, \left(\frac{\Lambda^3}{18 \pi^2} - \frac{m^2 \Lambda}{6 \pi^2} + c_{ct} m^3 \right) = \frac{1}{9} N^3 - \frac{(\alpha m)^2}{3} N + 2 \pi^2 c_{ct} (\alpha m)^3 \, , \\
S_{ct,2} &= \int d^3x \sqrt{\zeta} \, R \, b_{ct} m = 12 \pi^2 b_{ct} \, \alpha m \, .
\end{align}
Here $c_{ct}$ and $b_{ct}$ are two unspecified coefficients that
multiply finite counterterms, i.e.~UV-cutoff-independent
terms. Picking values for $c_{ct}$ and $b_{ct}$ modifies the finite part of $F_S$ and hence a choice of $c_{ct}$ and $b_{ct}$ amounts to choosing a renormalization scheme.

Hence, we arrive at an expression for the (counterterm-)renormalized free energy. This is given by
\begin{align}
\label{eq:Frenctscalarapp} F_S^{\textrm{ren}}(\alpha m \, | \,  b_{ct}, c_{ct}) = \lim_{N \rightarrow \infty} \Big( F^{\textrm{reg}}(N, \alpha m) + S_{ct,1} (\alpha m \, | \,  c_{ct})+ S_{ct,2} (\alpha m \, | \,  b_{ct}) \Big) \, ,
\end{align}
where we also indicated the dependence on the renormalization-scheme parameters $b_{ct}, c_{ct}$.

We are now in a position to compare the expression \eqref{eq:Frenctscalarapp} with the zeta-function-renormalized expression for $F_S$ given in \eqref{eq:FSren}. While we cannot do this analytically, a numerical evaluation shows that
\begin{align}
F_S^{\textrm{ren}} (\alpha m \, | \, \zeta\textrm{-function-renormalized}) = F_S^{\textrm{ren}}(\alpha m \, | \, b_{ct}=0, c_{ct}=0) \, .
\end{align}
Thus, in the case of the free massive scalar on $S^3$, zeta-function-renormalization is equivalent to adding counterterms with all finite counterterms chosen to vanish ($b_{ct}=0, c_{ct}=0$).

\section{Further monotonic functions\label{mono}}

We start from the expression of the unrenormalized (cutoff) Euclidean free energy
\begin{align}
\label{b1} S_{\textrm{on-shell}} = 2(d-1) M^{d-1} V_d \, {\big[e^{dA} \dot{A} \big]}_{\textrm{UV}} - \frac{2 M^{d-1} R}{d} V_d \int_{\textrm{UV}}^{\textrm{IR}} du \, e^{(d-2)A} \, .
\end{align}
where the UV cutoff is at $u=\epsilon$ and $R$ is the UV curvature of the slices.
We recall the first order evolution equations 
\begin{align}
\label{b2} W(\f) & \equiv -2 (d-1) \dot{A} \, , \\
\label{b3} S(\f) & \equiv \dot{\f} \, , \\
\label{b4}  T(\f) & \equiv R \, e^{-2A} = \frac{d(d-1)}{\alpha^2} \, e^{-2A} \, .
\end{align}
along with the non-linear equations for $W,S,T$,
\begin{align}
\label{b5} S^2 - SW' + \frac{2}{d} T &=0 \, , \\
\label{b6} \frac{d}{2(d-1)} W^2 -S^2 -2 T +2V &=0 \, , \\
\label{b7} SS' - \frac{d}{2(d-1)} SW - V' &= 0 \, .
\end{align}

We now consider the free-energy per unit boundary volume evaluated in UV units
\begin{align}
\label{b8} f \equiv { S_{\textrm{on-shell}} \over V_d~\Lambda^d} &= \hphantom{-} M^{d-1} \ell^{d} \left[2(d-1)\dot A+{2R\over d}e^{-dA(\e)}\int_{u_{\textrm{IR}}}^{\epsilon}du ~e^{(d-2)A}\right] \\
\nonumber & =-M^{d-1} \ell^{d} \left[W(\f(\e))+{2R\over d}e^{-dA(\e)}\int^{u_{\textrm{IR}}}_{\epsilon}du ~e^{(d-2)A}\right] \, ,
\end{align}
where
\be
\Lambda\equiv {1\over \ell}e^{A(\e)} \, ,
\ee
and $f$ is a dimensionless functional of two dimensionless numbers made out of the three dimensionful numbers of the problem: the cutoff $\e$, the relevant coupling $\f_-$ and the curvature $R$.

Consider now the derivative of $f$ with respect to $\e$:
\begin{align}
\label{b9} {df\over d\e} &=-M^{d-1} \ell^{d} \left[{dW(\f(\e))\over d\e}-{2R~\dot A(\e)}e^{-dA(\e)}\int^{u_{\textrm{IR}}}_{\epsilon}du ~e^{(d-2)A}-{2R\over d}~e^{-2A(\e)}\right] \\
\nonumber &=-M^{d-1} \ell^{d} \left[{dW(\f(\e))\over d\e}+{R\over d-1}~W(\e)~e^{-dA(\e)}\int^{u_{\textrm{IR}}}_{\epsilon}du ~e^{(d-2)A}-{2\over d}~T(\e)\right] \, .
\end{align}
We also use
\be
{dW(\f(\e))\over d\e}={dW(\f(\e))\over d\f(\e)}\dot \f(\e)=W'(\e)S(\e)
\label{b9b}\ee
and (\ref{b5})
to rewrite it as
\begin{align}
\label{b10}
{df\over d\e} &=-M^{d-1} \ell^{d} \left[
W'(\e)S(\e)+{R\over d-1}~W(\e)~e^{-dA(\e)}\int^{u_{\textrm{IR}}}_{\epsilon}du ~e^{(d-2)A}-{2\over d}T(\e)\right] \\
\nonumber &=-M^{d-1} \ell^{d} \left[
S^2(\e)+{R\over d-1}~W(\e)~e^{-dA(\e)}\int^{u_{\textrm{IR}}}_{\epsilon}du ~e^{(d-2)A}\right] \, .
\end{align}

Using (\ref{b10}) we can calculate the invariant dimensionless derivative
\begin{align}
\label{b11}
\Lambda {df\over d\Lambda } &={df\over dA(\e)}={1\over \dot A(\e)}{df\over d\e} \\
\nonumber & = {2(d-1)M^{d-1} \ell^{d} \over W(\e)}\left[
S^2(\e)+{R\over d-1}~W(\e)~e^{-dA(\e)}\int^{u_{\textrm{IR}}}_{\epsilon}du ~e^{(d-2)A}\right]~~\geq 0 \, .
\end{align}

We also have
\be
{df\over d\f(\e)}={df\over dA(\e)}{dA(\e)\over d\f(e)}=-{W(\e)\over 2(d-1)S(\e)}\Lambda {df\over d\Lambda} \, .
\label{b12}\ee

Equation (\ref{b11}) is a version of the $F$-theorem, as we can use $\Lambda$ to track the flow: $\Lambda\to \infty$ corresponds to the UV fixed point, while $\Lambda\to 0$ to the IR fixed point. Then the inequality
\be
\Lambda {df\over d\Lambda } \geq 0
\ee
implies that $f$ decreases along the RG flow.
Of course, for finite $R$, the IR limit $\Lambda\to 0$ does not correspond to minimum of the bulk potential as the theory is at finite curvature $R$.
 
When $\Lambda\to\infty$ we obtain
\be
\lim_{\Lambda\to\infty}f=-2(d-1)(M\ell)^{d-1} \, .
\ee
However, when $R\not=0$ the flow does not really fully asymptote to the IR fixed point and therefore the $\Lambda\to 0$ limit is ${\cal R}$-dependent.
If however we take the ${\cal R}\to 0$ limit, then
\be
f_{{\cal R}\to 0}=-(M\ell)^{d-1} W(\f(\e))
\ee
and
\be
\lim_{\Lambda\to 0}f_{{\cal R}\to 0}=-2(d-1)(M\ell)^{d-1}{\ell\over \ell_{\textrm{IR}}}
\ee
As expected, this is the wrong answer for the IR limit!

There maybe however a different definition that could lead to an $F$-function.

\addcontentsline{toc}{section}{References}
\bibliography{Ftheorem.bib}
\bibliographystyle{JHEP}

\end{document}